# Electron Dynamics in Open Quantum Systems: The Driven Liouville-von Neumann Methodology within Time Dependent Density Functional Theory

Annabelle Oz,[1] Abraham Nitzan,[1,2] Oded Hod,[1] Juan E. Peralta[3]

[1] *Department of Physical Chemistry, School of Chemistry, The Raymond and Beverly Sackler Faculty of Exact Sciences, and The Sackler Center for Computational Molecular and Materials Science, Tel Aviv University, Tel Aviv, IL 6997801*

[2] *Department of Chemistry, University of Pennsylvania, Philadelphia, PA, USA 19103*

[3] *Department of Physics, Central Michigan University, Mount Pleasant, MI, USA 48859*

## Abstract

A first-principles approach to describe electron dynamics in open quantum systems driven far from equilibrium via external time-dependent stimuli is introduced. Within this approach, the driven Liouville von Neumann methodology is used to impose open boundary conditions on finite model systems, whose dynamics is described using time-dependent density functional theory. As proof of concept, the developed methodology is applied to simple spin-compensated model systems including a hydrogen chain and a graphitic molecular junction. Good agreement between steady-state total currents obtained via direct propagation and from the self-consistent solution of the corresponding Sylvester equation indicates the validity of the implementation. The capability of the new computational approach to analyze, from first principles, non-equilibrium dynamics of open quantum systems in terms of temporally and spatially resolved current densities is demonstrated. Future extensions of the approach towards the description of dynamical magnetization and decoherence effects are briefly discussed.

# Introduction

The need to realize miniaturized electronic devices with optimal efficiency, has led to the extensive study of electronic transport in nanoscale constrictions over the past decades. Nowadays, various aspects of steady-state conductance are routinely explored,[1–4] including current switching[5–9] and rectification,[10–15] thermopower,[5–8,10–21] interference effects,[22–26] as well as the role of lead-molecule coupling[7,27–29] and chemical composition.[30] Despite the many advancements made in the field, the study of time-dependent phenomena in molecular junctions still poses scientific challenges with potentially significant technological merits, ranging from high-speed and quantum computing to opto-electronic devices operating at the nanoscale.

To model electron dynamics in molecular junctions, theoretical methods have been developed,[16,24,31–71] which can be broadly divided into methods that use model Hamiltonians (usually formulated in energy space) treating general transport phenomena while circumventing detailed descriptions of specific junctions,[63,64,72–78] and methods that explicitly take into account the chemical composition and structure of the system and thus allowing for a direct comparison with experimental findings.[31,32,56,60,61,71,79–85] The latter are often highly computationally demanding and thus limited to treating relatively small systems.

These two views have been bridged via the driven Liouville-von Neumann (DLvN) approach,[65,66,69,86–89] which enables the calculation of time-dependent electronic transport in fully atomistic junction models. Within the DLvN, a finite atomistic model system is coupled to implicit external Fermionic reservoirs by imposing appropriate nonequilibrium boundary conditions in energy space. The DLvN method can be used with any single-particle description of the system, such as tight-binding[65,66,87,88] (TB) and extended Hückel[69,89] (EH) Hamiltonians. However, simulations of electron dynamics in realistic molecular junctions, often require a more accurate description of the underlying electronic structure. To this end, time-dependent density functional theory (TDDFT)[90] offers a tractable and reliable framework for describing the electronic response of the fully interacting system to varying external stimuli, in terms of a fictitious single-particle system.[14,16,31,31–40,42–46,49,56,58–60,71,79,80,82,83,87–89,91–107] In this work, we present an implementation of the DLvN within the framework of TDDFT and demonstrate its performance for two systems, namely, a hydrogen chain "toy" model and a graphene nano-ribbon (GNR) based molecular junctions.

## 2. Methodology

### 2.1 The Driven Liouville-von Neumann Formalism

The DLvN formalism relies on a unitary transformation from a finite real-space representation of the junction model to its spectral state-representation,[65,66,68,69,87–89] where the states of left and right lead sections couple to those of an extended molecule. This allows one to apply non-equilibrium boundary conditions to absorb outgoing electrons and inject incoming electrons with an appropriate thermal distribution at the far edges of the finite model system. Notably, the DLvN equation of motion (see below) ensures positive semi-definiteness of the density matrix and prevents violation of Pauli's exclusion principle.[68]

The general DLvN scheme can be divided into four steps:

*(i) Spatial partitioning:* the molecular junction is represented by a fully atomistic finite model system that is formally partitioned into three sections: the left lead (L), the extended molecule (EM), and the right lead (R, see illustration in Figure 1). The EM includes the active molecule and its adjacent lead subsections that serve to buffer the molecule from the lead regions, where boundary conditions are applied. In a non-orthogonal, atom-centered basis-set representation of the Kohn-Sham (KS) molecular orbitals, the partitioned KS Hamiltonian and overlap matrices ($H_{KS}$ and $S$, respectively) can then be written in the following block form:

$$H_{KS} = \begin{pmatrix} H_L & V_{L,EM} & V_{L,R} \\ V_{EM,L} & H_{EM} & V_{EM,R} \\ V_{R,L} & V_{R,EM} & H_R \end{pmatrix} \text{ and } S = \begin{pmatrix} S_L & S_{L,EM} & S_{L,R} \\ S_{EM,L} & S_{EM} & S_{EM,R} \\ S_{R,L} & S_{R,EM} & S_R \end{pmatrix}. \quad (1)$$

For simplicity, in what follows we assume that the lead sections are spatially well separated such that the $V_{L,R}$ and $S_{L,R}$ blocks (and their conjugate counterparts) are negligible and can be safely replaced by zero matrix blocks of appropriate dimensions.

*(ii) Block orthogonalization:* when a non-orthogonal basis-set representation is used, one must ensure that the boundary conditions applied at the far edges of the systems do not directly affect the dynamics of the extended molecule region.[69] To this end, the block orthogonalization procedure of Kwok et al.[108] is adopted to transform the localized basis-functions of the EM section into new EM basis-functions that are *mutually orthogonal* to those of the finite L/R lead models. This block orthogonalization procedure involves a non-unitary transformation matrix of the form:[108]

$$U_b \equiv \begin{pmatrix} I_L & -S_L^{-1}S_{L,EM} & 0 \\ 0 & I_{EM} & 0 \\ 0 & -S_R^{-1}S_{R,EM} & I_R \end{pmatrix}; \quad U_b^{-1} = \begin{pmatrix} I_L & S_L^{-1}S_{L,EM} & 0 \\ 0 & I_{EM} & 0 \\ 0 & S_R^{-1}S_{R,EM} & I_R \end{pmatrix}, \tag{2}$$

where $\mathbf{0}$ and $\mathbf{I}$ are null and unit matrices of the relevant dimensions, respectively. Following this transformation, the overlap matrix becomes block diagonal:

$$\tilde{S} \equiv U_b^{-1} S U_b = \begin{pmatrix} S_L & 0 & 0 \\ 0 & S_{EM} - S_{EM,L}S_L^{-1}S_{L,EM} - S_{EM,R}S_R^{-1}S_{R,EM} & 0 \\ 0 & 0 & S_R \end{pmatrix}, \tag{3}$$

and the KS Hamiltonian matrix retains its block form (see supporting information (SI) sec. 1):

$$\widetilde{H}_{KS} \equiv U_b^{-1} H U_b = \begin{pmatrix} H_L & \widetilde{V}_{L,EM} & 0 \\ \widetilde{V}_{EM,L} & \widetilde{H}_{EM} & \widetilde{V}_{EM,R} \\ 0 & \widetilde{V}_{R,EM} & H_R \end{pmatrix}, \tag{4}$$

whose transformed blocks become:

$$\begin{cases} \widetilde{V}_{L,EM} = V_{L,EM} - H_L S_L^{-1} S_{L,EM}, \\ \widetilde{V}_{EM,L} = V_{EM,L} - S_{EM,L} S_L^{-1} H_L, \\ \widetilde{V}_{EM,R} = V_{EM,R} - S_{EM,R} S_R^{-1} H_R, \\ \widetilde{V}_{R,EM} = V_{R,EM} - H_R S_R^{-1} S_{R,EM}, \end{cases} \tag{5}$$

and

$$\widetilde{H}_{EM} = H_{EM} - S_{EM,L}S_L^{-1}V_{L,EM} - V_{EM,L}S_L^{-1}S_{L,EM} - V_{EM,R}S_R^{-1}S_{R,EM} - S_{EM,R}S_R^{-1}V_{R,EM}$$

$$+ S_{EM,L}S_L^{-1}H_L S_L^{-1}S_{L,EM} + S_{EM,R}S_R^{-1}H_R S_R^{-1}S_{R,EM}. \tag{6}$$

Since the transformation leaves the diagonal lead blocks, $S_L, S_R, H_L$ and $H_R$ unaffected, the procedure for applying boundary conditions does not change. We note that if an orthogonal basis-set is used, this step can be skipped.

*(iii) Site-to-state transformation:* To allow the application of open boundary conditions, a basis transformation is performed on the block-orthogonal matrices, shifting them from the real-space representation to the basis of eigenfunctions of the individual system sections. This site-to-state transformation is represented by the unitary matrix:

$$\widetilde{\mathbf{U}} \equiv \begin{pmatrix} \mathbf{U}_L & 0 & 0 \\ 0 & \mathbf{U}_{EM} & 0 \\ 0 & 0 & \mathbf{U}_R \end{pmatrix}, \quad \widetilde{\mathbf{U}}^{-1} = \widetilde{\mathbf{U}}^\dagger = \begin{pmatrix} \mathbf{U}_L^\dagger & 0 & 0 \\ 0 & \mathbf{U}_{EM}^\dagger & 0 \\ 0 & 0 & \mathbf{U}_R^\dagger \end{pmatrix}, \tag{7}$$

where $\mathbf{U}_{i=L,EM,R}$ are the unitary matrix blocks that transform the generalized eigenvalue equations:

$$\widetilde{\mathbf{H}}_i c^i = \varepsilon^i \widetilde{\mathbf{S}}_i c^i, \tag{8}$$

to their diagonal representation, where $\varepsilon^i$ and $c^i$ are the generalized eigenvalues and eigenvectors matrices of each block, respectively. Within this representation, $\widetilde{\widetilde{\mathbf{H}}}_i = \mathbf{U}_i^\dagger \widetilde{\mathbf{H}}_i \mathbf{U}_i$ is a diagonal matrix containing the eigenvalues of $\widetilde{\mathbf{H}}_i$, and $\widetilde{\widetilde{\mathbf{S}}}_i = \mathbf{U}_i^\dagger \widetilde{\mathbf{S}}_i \mathbf{U}_i = \mathbf{I}_i$. With this, the full overlap matrix becomes the identity:

$$\widetilde{\widetilde{\mathbf{S}}} = \mathbf{I}, \tag{9}$$

and the KS Hamiltonian adopts the following form:

$$\widetilde{\widetilde{\mathbf{H}}}_{KS} = \mathbf{U}^\dagger \widetilde{\mathbf{H}}_{KS} \mathbf{U} = \begin{pmatrix} \widetilde{\widetilde{\mathbf{H}}}_L & \widetilde{\widetilde{\mathbf{V}}}_{L,EM} & 0 \\ \widetilde{\widetilde{\mathbf{V}}}_{EM,L} & \widetilde{\widetilde{\mathbf{H}}}_{EM} & \widetilde{\widetilde{\mathbf{V}}}_{EM,R} \\ 0 & \widetilde{\widetilde{\mathbf{V}}}_{R,EM} & \widetilde{\widetilde{\mathbf{H}}}_R \end{pmatrix}, \tag{10}$$

where the off-diagonal $\widetilde{\widetilde{\mathbf{V}}}_{i,j}$ blocks represent couplings between the eigenstates of sections $i$ and $j$, and the lead-lead couplings remain zero. Following the site-to-state transformation, the atomistic representation of the junction is replaced by a state-representation, where the single-particle states of the EM section are coupled to the corresponding lead states (see Figure 2).

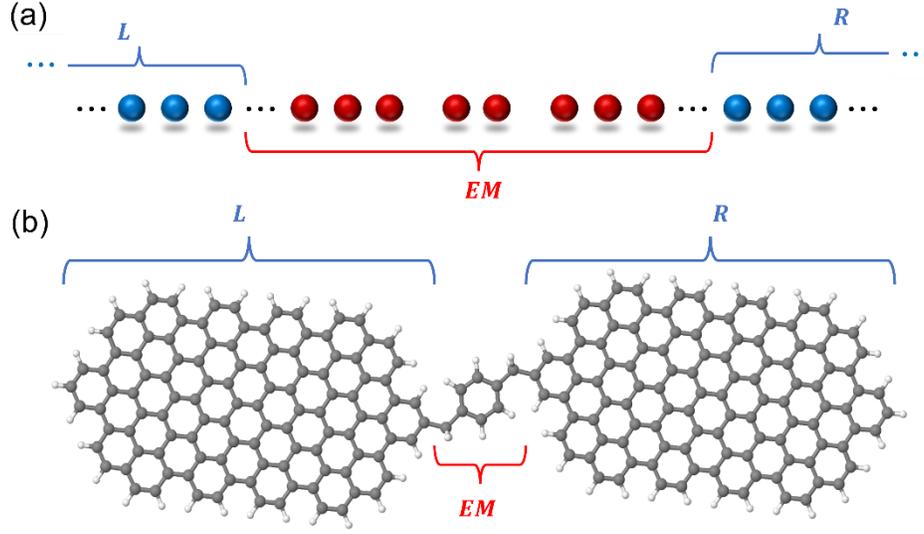

Figure 1: Real-space formal partitioning of a molecular junction model composed of (a) a hydrogen chain, and (b) two finite graphene nanoribbons bridged by a benzene molecule, into left (L) and right (R) lead sections and an extended molecule (EM) region.

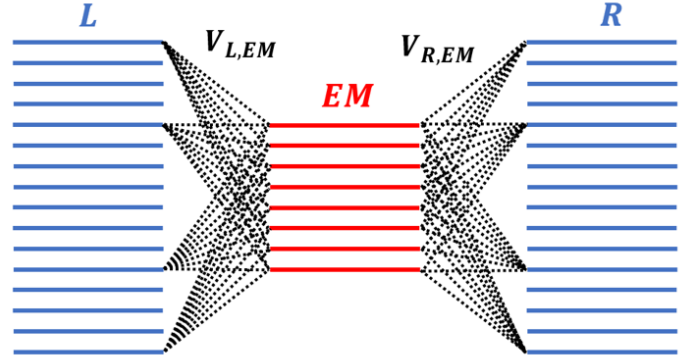

Figure 2: A schematic illustration of the state representation, where extended molecule single-particle states (EM, red lines) couple to left (L) and right (R) lead states (blue lines).

*(iv) Application of the open boundary conditions:* The final step in the DLvN scheme is enforcing the boundary conditions on the lead sections using the following equation of motion (given in atomic units (a.u.), see SI sec. 1 for a detailed derivation):[65,66,68,69,86–89]

$$\dot{\widetilde{\widetilde{\mathcal{P}}}} = -i\left[\overline{\widetilde{\mathcal{H}}}_{KS}, \widetilde{\widetilde{\mathcal{P}}}\right] - \Gamma \begin{pmatrix} \widetilde{\widetilde{\mathcal{P}}}_L - \widetilde{\widetilde{\mathcal{P}}}_L^0 & \frac{1}{2}\widetilde{\widetilde{\mathcal{P}}}_{L,EM} & \widetilde{\widetilde{\mathcal{P}}}_{LR} \\ \frac{1}{2}\widetilde{\widetilde{\mathcal{P}}}_{EM,L} & 0 & \frac{1}{2}\widetilde{\widetilde{\mathcal{P}}}_{EM,R} \\ \widetilde{\widetilde{\mathcal{P}}}_{RL} & \frac{1}{2}\widetilde{\widetilde{\mathcal{P}}}_{R,EM} & \widetilde{\widetilde{\mathcal{P}}}_R - \widetilde{\widetilde{\mathcal{P}}}_R^0 \end{pmatrix}. \quad (11)$$

Here, $\widetilde{\widetilde{\mathcal{P}}}$ is the single-particle density matrix of the entire system, given in the state-representation.

The first term of Eq. (11) represents the unitary dynamics according to the standard Liouville-von Neumann equation of motion, where $[.,.]$ denotes the commutator and $i = \sqrt{-1}$. The second term drives the lead sections toward the equilibrium state of the corresponding reservoirs by coupling them to implicit baths at a driving rate $\Gamma$. Here, the upper and lower diagonal blocks, $(\widetilde{\widetilde{\mathcal{P}}}_L - \widetilde{\widetilde{\mathcal{P}}}_L^0)$ and $(\widetilde{\widetilde{\mathcal{P}}}_R - \widetilde{\widetilde{\mathcal{P}}}_R^0)$, drive the lead state occupations toward diagonal target density matrices, $\widetilde{\widetilde{\mathcal{P}}}_{L/R}^0(\varepsilon_n^{L/R}) = \left[1 + e^{(\varepsilon_n^{L/R} - \mu_{L/R})/k_B T_{L/R}}\right]^{-1}$, which represent Fermi-Dirac occupation distributions of the manifolds of lead levels of energies $\varepsilon_n^{L/R}$, where $k_B$ is the Boltzmann constant. These target density matrices encode the electronic temperatures $(T_{L/R})$ and chemical potentials $(\mu_{L/R})$ of the equilibrium reservoirs, to which the driven leads are implicitly coupled. With this, a bias voltage, $V$, can be effectively applied by setting the target chemical potentials of the leads to $\mu_{L/R} = \varepsilon_F \pm 0.5 \times |e| \times V$, where $\varepsilon_F$ is the Fermi level of the unbiased full model system and $e$ is the electron charge ($e = -1$ in a.u.). The off-diagonal blocks serve to dampen the coherences of electrons that exit the extended molecule region into the driven lead sections. This scheme allows for the use of reservoirs that differ with respect to their material properties, chemical potentials, and electronic temperatures. In such cases, a non-equilibrium state is obtained, inducing a time-dependent current flow through the system.

The lead driving rate, $\Gamma$, appearing in Eq. (11), represents the inverse timescale at which thermal relaxation takes place in the leads due to their coupling to the implicit baths. While the value of $\Gamma$ can be extracted from the self-energy of the implicit semi-infinite bath models,[86] it is usually sufficient to approximate it as the typical lead level spacing in the vicinity of the Fermi energy of the lead models, $\Gamma \sim \Delta \varepsilon_{i=L/R}$. With this choice, the discrete density of states of the finite lead model is sufficiently broadened to represent that of the corresponding semi-infinite lead. In practice, the simulated electron dynamics weakly depends on the value of the driving rate over a wide parameter range (see SI sec. 2).[98–100,104–106,109]

**2.2 The DLvN scheme within TDDFT**

Unlike the simplified TB and EH electronic structure approximations previously used in conjunction with the DLvN equation of motion (EOM), within a TDDFT framework Eq. (11) becomes non-linear as the KS Hamiltonian depends explicitly on the electron density and thus implicitly on time. Moreover, the KS Hamiltonian is evaluated from the real-space single-particle density matrix, whereas the boundary conditions are applied in the state representation. Therefore, at each propagation time-step one needs to go back and forth between the site and the state representations. As discussed above, this involves the block-orthogonalization procedure of Eq. (2) and the site-to-state transformation of Eq. (7). The former remains constant in time as it depends only on the stationary atomic orbital overlaps (as long as the nuclei are kept fixed), whereas the latter depends on the KS Hamiltonian and thus has to be updated at every time-step to account for the varying eigenfunctions. Further details on this issue are provided in sec. 1 of the SI.

The first TDDFT implementation of the DLvN EOM circumvented repeated site-to-state transformations by using as reference the polarized state of the finite model junction under an axial electric field.[97,110] This resulted in considerable gain in computational efficiency at the expense of a less accurate representation of the equilibrium state of the Fermionic reservoirs, loss of a unique definition of the bias voltage and electronic temperatures, and possible violations of Pauli's exclusion principle.[102] Furthermore, approaches based on field-induced polarized boundary conditions are limited in practice to linear two-lead setups, where a uniform field is applied along the main axis of the junction model.

Therefore, we opt to pursue a full-fledged implementation of the DLvN scheme, where at each time-step the following workflow is followed:

(a) obtain the KS Hamiltonian, $\boldsymbol{H}_{KS}$, in a general non-orthogonal atomic orbital basis.

(b) transform to the state-representation to obtain $\widetilde{\widetilde{\boldsymbol{\mathcal{H}}}}_{KS}$.

(c) build the target density matrices $\widetilde{\widetilde{\boldsymbol{\mathcal{P}}}}_L^0$ and $\widetilde{\widetilde{\boldsymbol{\mathcal{P}}}}_R^0$ and construct the driving term.

(d) transform the driving term to the site representation.

(e) propagate the single-particle density matrix, $\boldsymbol{\mathcal{P}}$.

Note that to avoid the implicit time-dependence of the $\widetilde{\boldsymbol{U}}$ transformation (via $\widetilde{\boldsymbol{H}}[\widetilde{\boldsymbol{\mathcal{P}}}]$), the propagation of the density matrix in item (e) is performed in real-space (see SI sec. 1 for further details.). To this end, we use an implicit Euler propagation scheme (see SI sec. 3), where:

$$\boldsymbol{\mathcal{P}}(t+dt) = \boldsymbol{\mathcal{P}}(t) + dt \cdot f(t+dt, \boldsymbol{\mathcal{P}}(t+dt)), \qquad (12)$$

and $f(t+dt, \boldsymbol{\mathcal{P}}(t+dt))$ represents the right-hand-side of Eq. (11) in the site representation. To solve Eq. (12), an iterative fixed-point scheme is implemented at each time step, where $f(t+dt, \boldsymbol{\mathcal{P}}(t+dt))$ at a given iteration is evaluated using $\mathcal{P}(t + dt)$ of the previous iteration, keeping $\mathcal{P}(t)$ on the right-hand-side of Eq. (12) fixed. The fixed-point iterations proceed until the convergence criterion is met, such that $|\mathcal{P}_{i+1}(t + dt) - \mathcal{P}_i(t + dt)|/N_{Bas} < 10^{-4}$, where $i$ is the fixed-point iteration index, $N_{Bas}$ is the dimension of the matrices (the total number of basis-functions), and $|\cdots|$ represents the Euclidean norm. Upon convergence, the time propagation continues with the same time step, $dt$, unless convergence is achieved in the very first fixed-point iteration, in which case the propagation proceeds with a doubled time-step. In case were the iterations fail to converge within 5 cycles, $dt$ is halved and the dynamics is rolled-back to the previous time-step.

An important measurable that this time propagation scheme provides is the temporal dependence of the current flowing through the active molecule, which resides in the EM section. To evaluate this quantity, one may invoke the equation of motion (Eq. (11)) to isolate the dynamics of the EM single-particle density matrix block. Care, however, should be taken regarding which representation is to be used for this purpose. In the site representation, the overlap matrix mixes contributions of the $L$, $R$, and $EM$ blocks, thus preventing a straight-forward separation of their contributions to the current. In the state representation, within a TDDFT treatment, an equation of motion for $\widetilde{\widetilde{\mathcal{P}}}$ is lacking. Note that Eq. (11) describes the dynamics of $\dot{\widetilde{\widetilde{\mathcal{P}}}}$ and not of $\dot{\widetilde{\widetilde{\mathcal{P}}}}$ which, in turn, involves the unknown explicit dynamics of the transformation matrix $U$ (see SI sec. 1 for further details). These issues are absent in the block diagonal representation, which we use to perform the particle current calculation (see SI sec. 4):

$$J(t) = \frac{1}{\hbar}\Im\{tr_{EM}[\widetilde{\mathcal{P}}_{EM,L}(t)\widetilde{V}_{L,EM}(t) - \widetilde{\mathcal{P}}_{EM,R}(t)\widetilde{V}_{R,EM}(t)]\}, \tag{13}$$

where $tr[.]$ is the trace operation, $\Im\{.\}$ represents the imaginary part.

For steady-state current evaluations, the right-hand-side of Eq. (11) is nullified, $\dot{\widetilde{\widetilde{\mathcal{P}}}} = 0$, resulting in a Sylvester-type equation of the form:[64,87,89]

$$\left[-\frac{i}{\hbar}\widetilde{\widetilde{\mathcal{H}}}_{KS} - \frac{\Gamma}{2}(L + R)\right]\widetilde{\widetilde{\mathcal{P}}}^{sts} + \widetilde{\widetilde{\mathcal{P}}}^{sts}\left[\frac{i}{\hbar}\widetilde{\widetilde{\mathcal{H}}}_{KS} - \frac{\Gamma}{2}(L + R)\right] = -\Gamma\widetilde{\widetilde{\mathcal{P}}}^0. \tag{14}$$

Here, $\widetilde{\widetilde{\mathcal{P}}}^{sts}$ is the steady-state density matrix in the state representation,

$$L = \begin{pmatrix} I_L & 0 & 0 \\ 0 & 0 & 0 \\ 0 & 0 & 0 \end{pmatrix} \text{ and } R = \begin{pmatrix} 0 & 0 & 0 \\ 0 & 0 & 0 \\ 0 & 0 & I_R \end{pmatrix}, \tag{15}$$

are the projection matrices onto the left and the right lead states, respectively, and

$$\widetilde{\widetilde{\mathcal{P}}}^0 = \begin{pmatrix} \widetilde{\widetilde{\mathcal{P}}}_L^0 & 0 & 0 \\ 0 & 0 & 0 \\ 0 & 0 & \widetilde{\widetilde{\mathcal{P}}}_R^0 \end{pmatrix}. \tag{16}$$

Eq. (14) is solved iteratively using the following scheme. Given the density matrix at iteration $i$, $\widetilde{\widetilde{\mathcal{P}}}_i^{sts}$, the KS Hamiltonian is constructed, and the Sylvester equation is solved to give $\widetilde{\widetilde{\mathcal{P}}}_{i+1}^{sts}$. A damping scheme admixing these two density matrices, with weights $a_{mix}$ and $(1 - a_{mix})$, respectively, is then used to construct the next step KS Hamiltonian matrix, $\widetilde{\widetilde{\mathcal{H}}}_{KS}$. At each iteration step, $\widetilde{\widetilde{\mathcal{P}}}_i^{sts}$ is transformed to the block diagonal basis, $\widetilde{\mathcal{P}}_i^{sts}$, and the particle current is calculated using Eq. (13) The process is repeated until the steady-state current is converged, such that its RMS value during $N = 20$ consecutive iterations is smaller than a preset value (chosen as $10^{-10}$ a.u. for the hydrogen chain calculations).

The entire simulation scheme is implemented in Python,[111] which makes recurrent calls at each time-step to the Gaussian suite of programs[112] to evaluate the KS Hamiltonian (and overlap) matrix elements in the atomic orbital basis. We note that the current proof-of-concept implementation employs spin-compensated electron densities.

## 4. Results

The developed methodology is first benchmarked using a simple linear hydrogen chain molecular junction model system. The real-space model consists of two 180 hydrogen atom leads, bridged by a 20-atom EM section, out of which the central two atoms serve as the scattering molecular region. A uniform inter-atomic separation of 0.988 Å is used throughout the chain except for the distance between the two central atoms and the adjacent extended molecule sections, which is purposely stretched to 1.4 Å. This serves to weaken the coupling between the central hydrogen molecule and the hydrogen chain leads. The Perdew-Burke-Ernzerhof (PBE) [113,114] generalized gradient exchange-correlation density functional approximation is used along with the atomic centered Gaussian type STO-3G basis-set[115] for the two lead models, and the double-$\zeta$ 6-31G$^{**}$ basis set[116–118] is used for the extended molecule section (see SI sec. 5 for further details). The driving rate is chosen as $\hbar\Gamma = 0.61$ eV to yield a reasonably smooth density of states (DOS) at the lead sections (see SI sec. 2).

Figure 3 presents the time dependent total current flowing through the EM section, calculated using Eq. (13) for several bias voltages. To improve numerical stability, the simulation starts from the ground state density of the system and the bias voltage is turned on gradually using a hyperbolic tangent switching function (see SI sec. 6). We confirm that during the propagation, all state occupations remain bound to [0:1], namely the positive definiteness condition and Pauli's exclusion principle are both obeyed (see inset of Fig. Figure *3*). Steady-state values (represented by crosses) are obtained by solving Eq. (14), using the Sylvester equation solver implemented in the SciPy[119] package, starting from a density matrix (and its corresponding Kohn-Sham Hamiltonian) from the plateau region of the dynamic calculation. The excellent correspondence between the steady-state currents obtained using the dynamical and the Sylvester calculations indicates that the DLvN EOM indeed reached a stable stationary state.

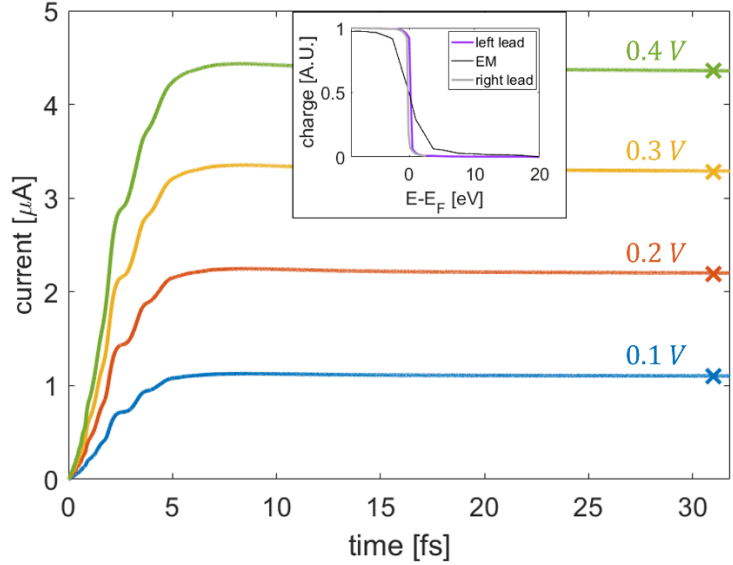

Figure 3: Time-dependent current calculated at various bias voltages for a 380 atoms hydrogen chain with $N_M = 2$, $N_{EM} = 20$, and $N_{L/R} = 180$. Bias voltages of 0.1 (blue), 0.2 (orange), 0.3 (yellow), and 0.4 (green) $V$ are considered with reservoir electronic temperatures of $T_L = T_R = 315.7\ K$ and a driving rate of $\hbar\Gamma = 0.61$ eV. The colored × marks designate the corresponding Sylvester steady-state currents calculated via Eq. (14) with an admixture weight of $a_{mix} = 0.99$. Inset: left lead (purple), right lead (grey), and extended molecule (black) steady-state occupations obtained at $t = 25$ fs under a bias voltage of 0.2 $V$.

Going beyond the total current flowing through the system, our approach allows also to analyze the spatially resolved current density within the chain:

$$\boldsymbol{j}(\mathbf{r},t) = \frac{e}{2m_e} Tr\left[\boldsymbol{\mathcal{P}}(t) \cdot \left(\boldsymbol{A}(\mathbf{r}) - \boldsymbol{A}^{\mathrm{T}}(\mathbf{r})\right)\right], \qquad (17)$$

where $\boldsymbol{A}(\mathbf{r})$ is a matrix defined in the atomic orbital basis as $\boldsymbol{A}(\mathbf{r})_{\mu\nu} = \frac{\hbar}{i}\phi_\mu(\mathbf{r})\nabla\phi_\nu(\mathbf{r})$ (see SI sec. 4). Figure 4 illustrates the spatially resolved steady-state current density for the hydrogen chain junction, whose total current is shown in Figure 3, under a bias voltage of 0.3 V obtained using Eq. (17). The axial (z) component of the steady-state current density is mostly uniform along the EM section with some expected variations near the nuclei positions (see sec. 7 of the SI for a plot of the current density integrated over the xy plane along the EM section). In the seamline between the weakly coupled molecule and the lead sections of the EM region (vertical black dashed lines) the current density reduces, indicating that it spreads over a larger cross section. The driven lead sections exhibit a less uniform current density map. This can be attributed to the fact that the uniform driving rate, $\Gamma$, applied in the state-representation of the junction translates to a spatially varying driving rate in real-space, with a larger value near the lead/EM interface region, where electrons are absorbed or injected. As a consequence, near the interface region of the sink lead, the direction of the current is reversed (see streamlines on the right side of Figure 4). This artificial behavior in the (unphysical) driven lead regions, however, has minor influence on the EM currents and negligible effect on the current flowing through the molecule itself. Far away from the interface, the current in the leads decays indicating that the lead approaches the equilibrium state of the implicit bath to which it is coupled.

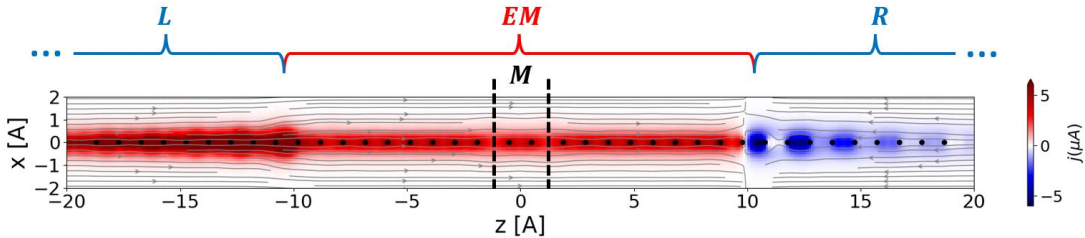

Figure 4: Spatially resolved steady-state current map of the hydrogen chain molecular junction model under a bias voltage of $V = 0.3\ V$. The heat map represents the axial (z) current component, the streamlines represent the projection of the local current density vector on the x-z plane, and the black dots represent the position of the hydrogen atoms in the vicinity of the extended molecule section. The borderline between the extended molecule section and the source and sink driven leads are located at $\pm 10.3$ Å, and the seamlines between the weakly coupled molecule and the left and right segments of the EM are located at $\pm 1.2$ Å, as marked by the vertical black dashed lines.

Following the benchmark hydrogen chain calculations, we now turn to discuss a more realistic model junction consisting of two graphene nanoislands bridged by a benzene molecule (see Figure 1b, junction model coordinates can be found in SI section 9). The geometry of the system was optimized via the LAMMPS[120] software using the reactive empirical bond-order potential (REBO).[121,122] The energy minimization was performed using the Fast Inertial Relaxation Engine (FIRE) algorithm [123] with a force tolerance of $10^{-6}$ eV/Å. The vertical coordinates of all carbon atoms were fixed to keep the structure planar, so as to mimic a substrate supported junction model. The EM section is chosen to include the benzene unit and its two adjacent CH$_2$ groups. The total current was calculated using Eq. (13) with the PBE[113,114] functional and the STO-3G and 6-31G** atomic centered Gaussian type basis-sets for the lead and EM sections, respectively.[115] A driving rate of $\hbar\Gamma = 1.09$ eV was used to yield a smooth lead section DOS (see SI sec. 6). Figure 5 shows the current dynamics flowing through the EM section for several bias voltages calculated using the DLvN EOM, as well as the corresponding steady-state currents obtained from the solution of the Sylvester equation (14). The good agreement between the DLvN and Sylvester steady-state currents indicates that the DLvN EOM has reached a true stationary state of the system.

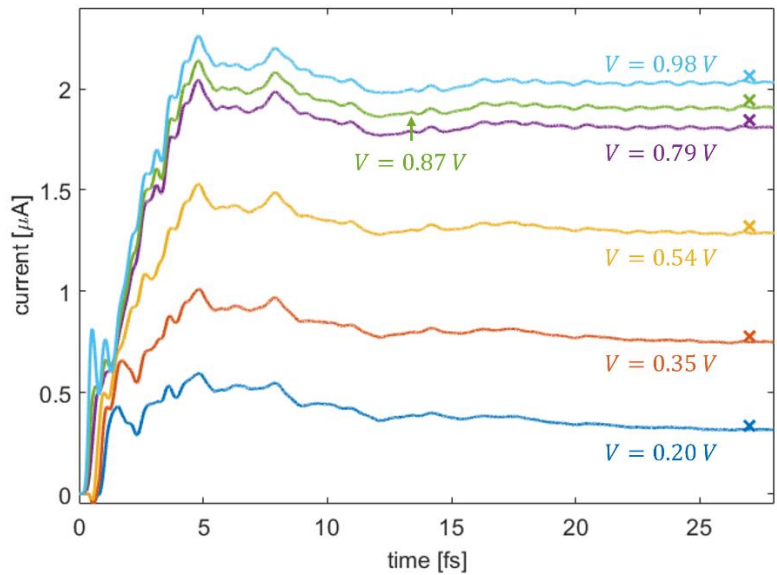

Figure 5: Time-dependent current calculated at various bias voltages for the GNR shown in Fig. 1b. Different bias voltages are considered (all values are marked in the figure) with reservoir electronic temperatures of $T_L = T_R = 315.7$ K and a driving rate of $\hbar\Gamma = 1.09$ eV. The colored × marks designate the corresponding Sylvester steady-state currents calculated via Eq. (14) with an admixture weight of $a_{mix} = 0.999$.

## 5. Conclusions

The simple hydrogen chain and graphitic junction examples presented above demonstrate the potential of the first-principles driven Liouville von Neumann methodology introduced herein for describing electron dynamics and thermodynamics in open quantum systems driven far from equilibrium. Using the DLvN approach to impose open boundary conditions on finite systems described by time-dependent density functional theory opens the door for studying key dynamical phenomena related to the fields of molecular electronics and spintronics, thermodynamics, and quantum technology. The continuous improvement of the underlying TD-DFT approximations ensures the increased accuracy and reliability of predictions made using the developed methodology. Future generalizations towards spin-uncompensated and non-collinear descriptions, as well as coupled electron-nuclear dynamics, will extend the applicability of the DLvN approach to describe magnetization dynamics and decoherence under external time-dependent stimuli.

## Acknowledgement

A.O. gratefully acknowledges the support of the Adams Fellowship Program of the Israel Academy of Sciences and Humanities. J.E.P. acknowledges support from the Office of Basic Energy Sciences, US Department of Energy (Grant No. DE-SC0005027). O.H. is grateful for the generous financial support of the Ministry of Science and Technology of Israel under project number 3-16244, the Center for Nanoscience and Nanotechnology of Tel Aviv University, and the Heineman Chair in Physical Chemistry. A.N. acknowledges support of the U.S. National Science Foundation under Grant No. CHE1953701. The authors thank Dr. Xiang Gao for help with geometry optimization of the graphitic junction model.

# Electron Dynamics in Open Quantum Systems: The Driven Liouville-von Neuman Methodology within Time Dependent Density Functional Theory

## Supporting Information

Annabelle Oz,[1,2] Abraham Nitzan,[1,2,3] Oded Hod,[1,2] Juan E. Peralta[4]

[1] *Department of Physical Chemistry, School of Chemistry, The Raymond and Beverly Sackler Faculty of Exact Sciences, Tel Aviv University, Tel Aviv, IL 6997801*

[2] *The Sackler Center for Computational Molecular and Materials Science, Tel Aviv University, Tel Aviv, IL 6997801*

[3] *Department of Chemistry, University of Pennsylvania, Philadelphia, PA, USA 19103*

[4] *Department of Physics, Central Michigan University, Mount Pleasant, MI, USA 48859*

# Contents

# 1. The driven Liouville von Neumann equation of motion in the realm of time-dependent density functional theory

In this section, we provide a detailed formulation of the driven Liouville von Neumann (DLvN) equation of motion (EOM) within the framework of time-dependent density functional theory in an atom-centered non-orthogonal basis-set representation. We start from the standard time-dependent Kohn-Sham (KS) equation written for the individual KS orbitals $|\phi_n\rangle$ as follows (atomic units are used throughout):

$$|\dot{\phi}_n\rangle = -i\mathcal{H}_{KS}|\phi_n\rangle. \tag{S1}$$

Here, $\mathcal{H}_{KS}$ is the KS Hamiltonian operator and $i = \sqrt{-1}$. Next, we span the KS orbitals within a localized basis-set representation $\{|\chi_\mu\rangle\}$:

$$|\phi_n\rangle = \sum_\mu c_\mu^n |\chi_\mu\rangle, \tag{S2}$$

where $c_\mu^n$ is the $\mu^{\text{th}}$ expansion coefficient of KS orbital $|\phi_n\rangle$. Plugging Eq. (S2) into Eq. (S1) and assuming that the basis orbitals are constant in time we obtain:

$$\sum_\mu \dot{c}_\mu^n |\chi_\mu\rangle = -i\mathcal{H}_{KS} \sum_\mu c_\mu^n |\chi_\mu\rangle. \tag{S3}$$

Multiplying Eq. (S3) by $\langle\chi_\nu|$ we obtain:

$$\sum_\mu \dot{c}_\mu^n \langle\chi_\nu|\chi_\mu\rangle = -i \sum_\mu c_\mu^n \langle\chi_\nu|\mathcal{H}_{KS}|\chi_\mu\rangle. \tag{S4}$$

Defining the overlap and KS Hamiltonian matrix elements as $S_{\nu\mu} \equiv \langle\chi_\nu|\chi_\mu\rangle$ and $\mathcal{H}_{KS_{\nu\mu}} \equiv \langle\chi_\nu|\mathcal{H}_{KS}|\chi_\mu\rangle$, respectively, Eq. (S4) becomes:

$$\sum_\mu S_{\nu\mu} \dot{c}_\mu^n = -i \sum_\mu \mathcal{H}_{KS_{\nu\mu}} c_\mu^n. \tag{S5}$$

Since this equation is valid for all values of the indices $\nu$ and $n$ it can be written in matrix form as:

$$\boldsymbol{S}\dot{\boldsymbol{C}} = -i\boldsymbol{\mathcal{H}}_{KS}\boldsymbol{C}. \tag{S6}$$

Multiplying by the inverse of the overlap matrix, $\boldsymbol{S}^{-1}$, on the left we obtain:

$$\dot{\boldsymbol{C}} = -i\boldsymbol{S}^{-1}\boldsymbol{\mathcal{H}}_{KS}\boldsymbol{C}. \tag{S7}$$

Accordingly, one can write the EOM for the complex transpose coefficient matrix as follows:

$$\dot{\boldsymbol{C}}^\dagger = [-i\boldsymbol{S}^{-1}\boldsymbol{\mathcal{H}}_{KS}\boldsymbol{C}]^\dagger = i\boldsymbol{C}^\dagger \boldsymbol{\mathcal{H}}_{KS}^\dagger (\boldsymbol{S}^{-1})^\dagger = i\boldsymbol{C}^\dagger \boldsymbol{\mathcal{H}}_{KS} (\boldsymbol{S}^\dagger)^{-1} = i\boldsymbol{C}^\dagger \boldsymbol{\mathcal{H}}_{KS}\boldsymbol{S}^{-1}, \tag{S8}$$

where we used the relation $(\boldsymbol{S}^{-1})^\dagger = (\boldsymbol{S}^\dagger)^{-1}$ $(\boldsymbol{I} = \boldsymbol{I}^\dagger = (\boldsymbol{S}\boldsymbol{S}^{-1})^\dagger = (\boldsymbol{S}^{-1})^\dagger \boldsymbol{S}^\dagger \Rightarrow (\boldsymbol{S}^{-1})^\dagger = (\boldsymbol{S}^\dagger)^{-1})$ and the fact that the overlap and Kohn-Sham matrices are Hermitian, such that $\boldsymbol{S}^\dagger = \boldsymbol{S}$ and $\boldsymbol{\mathcal{H}}_{KS} = \boldsymbol{\mathcal{H}}_{KS}^\dagger$. The latter relation stems from the fact that the density matrix, upon which $\boldsymbol{\mathcal{H}}_{KS}$ depends, is Hermitian by construction (see Eq. (S9) below) and so are all the operators within $\boldsymbol{\mathcal{H}}_{KS}$ (kinetic energy, Hartree, exchange, correlation, and external potential).

We can now define the single-particle density matrix in the localized basis-set representation as:

$$\mathcal{P} = \mathcal{C}n\mathcal{C}^\dagger, \tag{S9}$$

where $n$ is a diagonal matrix holding the occupation numbers of the different single-particle states on its diagonal. The time evolution of the density matrix is obtained by its time derivative:

$$\dot{\mathcal{P}} = \dot{\mathcal{C}}n\mathcal{C}^\dagger + \mathcal{C}n\dot{\mathcal{C}}^\dagger + \mathcal{C}\dot{n}\mathcal{C}^\dagger. \tag{S10}$$

Here, the first two terms on the right-hand-side correspond to pure orbital dynamics, whereas the third term represents the dynamics of the orbital occupations. Inserting Eqs. (S7) and (S8) into Eq. (S10) we obtain:

$$\dot{\mathcal{P}} = -i\mathcal{S}^{-1}\mathcal{H}_{KS}\mathcal{C}n\mathcal{C}^\dagger + i\mathcal{C}n\mathcal{C}^\dagger\mathcal{H}_{KS}\mathcal{S}^{-1} + \mathcal{C}\dot{n}\mathcal{C}^\dagger = -i\mathcal{S}^{-1}\mathcal{H}_{KS}\mathcal{P} + i\mathcal{P}\mathcal{H}_{KS}\mathcal{S}^{-1} + \mathcal{C}\dot{n}\mathcal{C}^\dagger. \tag{S11}$$

In microcanonical and canonical time-domain time-dependent density functional theory simulations, the overall number of particles in the system is conserved. In these cases, a customary ansatz is to propagate only the occupied subspace thus setting $\dot{n} = 0$, assuming that the occupied KS orbital populations do not vary with time and that the virtual orbitals remain unpopulated. The entire dynamics is thus overloaded on the occupied molecular orbital manifold via the corresponding expansion coefficients. This resembles choosing the Schrödinger representation (propagating the wave functions) instead of its Heisenberg counterpart (propagating the number operator). For open systems, however, one can no longer assume that $\dot{n} = 0$ and an explicit equation of motion should be provided to describe its dynamics. Within the DLvN approach the following EOM governs this dynamics:

$$\mathcal{C}\dot{n}\mathcal{C}^\dagger = -i\mathcal{S}^{-1}\mathcal{H}_{AH}\mathcal{P} + i\mathcal{P}\mathcal{H}_{AH}^\dagger\mathcal{S}^{-1} = -i\mathcal{S}^{-1}\mathcal{H}_{AH}\mathcal{P} - i\mathcal{P}\mathcal{H}_{AH}\mathcal{S}^{-1} = -i[\mathcal{S}^{-1}\mathcal{H}_{AH}\mathcal{P} + \mathcal{P}\mathcal{H}_{AH}\mathcal{S}^{-1}]. \tag{S12}$$

Here, $\mathcal{H}_{AH} = -\mathcal{H}_{AH}^\dagger$ is an anti-Hermitian matrix that, in principle, can assume the most general form of $\mathcal{H}_{AH} = \mathcal{H}_{AH}^{re} - i\mathcal{H}_{AH}^{im}$, where $\mathcal{H}_{AH}^{re}$ is a real anti-symmetric matrix such that $(\mathcal{H}_{AH}^{re})^T = -\mathcal{H}_{AH}^{re}$ and $\mathcal{H}_{AH}^{im}$ is a real symmetric matrix obeying $(\mathcal{H}_{AH}^{im})^T = \mathcal{H}_{AH}^{im}$. To rationalize this choice, we now plug Eq. (S12) in Eq. (S11) to obtain:

$$\dot{\mathcal{P}} = -i\mathcal{S}^{-1}\mathcal{H}_{KS}\mathcal{P} + i\mathcal{P}\mathcal{H}_{KS}\mathcal{S}^{-1} - i\mathcal{S}^{-1}\mathcal{H}_{AH}\mathcal{P} - i\mathcal{P}\mathcal{H}_{AH}\mathcal{S}^{-1} = -i\mathcal{S}^{-1}(\mathcal{H}_{KS} + \mathcal{H}_{AH})\mathcal{P} + i\mathcal{P}(\mathcal{H}_{KS} - \mathcal{H}_{AH})\mathcal{S}^{-1} = -i\mathcal{S}^{-1}(\mathcal{H}_{KS} + \mathcal{H}_{AH})\mathcal{P} + i\mathcal{P}(\mathcal{H}_{KS} + \mathcal{H}_{AH})^\dagger\mathcal{S}^{-1}. \tag{S13}$$

This equation assumes the form of a Liouville–von Neumann equation for a microcanonical (or canonical) system but with a general Hamiltonian matrix, $\mathcal{H}_{KS} + \mathcal{H}_{AH}$, that is neither Hermitian nor anti-Hermitian. The latter can be viewed as a dressed Hamiltonian, where we identify $\mathcal{H}_{AH}$ as a self-energy like term representing the effects of the coupling of the system to an implicit bath. Note, however, that $\mathcal{H}_{AH}$ is energy independent and hence should be viewed as an approximation of the self-energy within the wide band limit.

To obtain the explicit expression of $\mathcal{H}_{AH}$ within the DLvN EOM we divide the system into three sections comprising of the left lead, the (extended-)molecule, and the right lead. $\mathcal{H}_{AH}$ then serves to mimic the effect of coupling of the lead sections to implicit Fermionic baths, characterized by equilibrium Fermi-Dirac distributions with given chemical potentials and electronic temperatures. To this end, we first neglect $\mathcal{H}_{AH}^{re}$, which is equivalent to neglecting the real-part of the implicit baths' self-energies that induce lead level shifts due to the lead/implicit-bath couplings. This approximation becomes valid for sufficiently large lead models, with a relatively uniform and dense manifold of states, such that the level shifts become small with respect to the inter-level spacing. The remaining imaginary part, $\mathcal{H}_{AH}^{im}$, marked for brevity as $\mathbf{\Gamma}$, introduces a finite lifetime (broadening) to the various lead levels due to their coupling to the implicit single-particle states of the reservoir. Hence, within the DLvN approach, the dressed KS Hamiltonian acquires the form:

$$\mathcal{H}_{KS} \rightarrow \mathcal{H}_{KS} - i\mathbf{\Gamma}. \tag{S14}$$

In this form, $\mathbf{\Gamma}$ can be identified as an imaginary absorbing potential added to the lead sections of the original KS system serving to absorb outgoing electrons near the system boundaries (thus preventing their back-reflection into the system). Naturally, in order to avoid complete electronic depletion of the system, complementary emitting potentials should also be introduced in order to inject thermalized electrons into the system, as shown below.

Using the dressed Hamiltonian form of Eq. (S14) in Eq. (S13) we obtain:

$$\dot{\mathcal{P}} = -i\mathcal{S}^{-1}(\mathcal{H}_{KS} - i\mathbf{\Gamma})\mathcal{P} + i\mathcal{P}(\mathcal{H}_{KS} - i\mathbf{\Gamma})^{\dagger}\mathcal{S}^{-1} = -i\mathcal{S}^{-1}\mathcal{H}_{KS}\mathcal{P} + i\mathcal{P}\mathcal{H}_{KS}^{\dagger}\mathcal{S}^{-1} - \mathcal{S}^{-1}\mathbf{\Gamma}\mathcal{P} - \mathcal{P}\mathbf{\Gamma}^{\dagger}\mathcal{S}^{-1} = -i\mathcal{S}^{-1}\mathcal{H}_{KS}\mathcal{P} + i\mathcal{P}\mathcal{H}_{KS}\mathcal{S}^{-1} - \mathcal{S}^{-1}\mathbf{\Gamma}\mathcal{P} - \mathcal{P}\mathbf{\Gamma}^{\dagger}\mathcal{S}^{-1}. \tag{S15}$$

We can now multiply Eq. (S15) by $\mathcal{S}$ from left and from right to obtain:

$$\mathcal{S}\dot{\mathcal{P}}\mathcal{S} = -i\mathcal{H}_{KS}\mathcal{P}\mathcal{S} + i\mathcal{S}\mathcal{P}\mathcal{H}_{KS} - \mathbf{\Gamma}\mathcal{P}\mathcal{S} - \mathcal{S}\mathcal{P}\mathbf{\Gamma}^{\dagger}. \tag{S16}$$

Next, we introduce a block diagonalization transformation, $\mathbf{U}_b$, into Eq. (S16) to nullify the off-diagonal overlap matrix blocks as follows:[1]

$$\left[\left(U_b^{\dagger}\right)^{-1}U_b^{\dagger}\right]\mathcal{S}[U_b U_b^{-1}]\dot{\mathcal{P}}\left[\left(U_b^{\dagger}\right)^{-1}U_b^{\dagger}\right]\mathcal{S}[U_b U_b^{-1}] =$$

$$= -i\left[\left(U_b^{\dagger}\right)^{-1}U_b^{\dagger}\right]\mathcal{H}_{KS}[U_b U_b^{-1}]\mathcal{P}\left[\left(U_b^{\dagger}\right)^{-1}U_b^{\dagger}\right]\mathcal{S}[U_b U_b^{-1}] +$$

$$+i\left[\left(U_b^{\dagger}\right)^{-1}U_b^{\dagger}\right]\mathcal{S}[U_b U_b^{-1}]\mathcal{P}\left[\left(U_b^{\dagger}\right)^{-1}U_b^{\dagger}\right]\mathcal{H}_{KS}[U_b U_b^{-1}] - \mathbf{\Gamma}\mathcal{P}\left[\left(U_b^{\dagger}\right)^{-1}U_b^{\dagger}\right]\mathcal{S}[U_b U_b^{-1}] -$$

$$\left[\left(U_b^{\dagger}\right)^{-1}U_b^{\dagger}\right]\mathcal{S}[U_b U_b^{-1}]\mathcal{P}\mathbf{\Gamma}^{\dagger}, \tag{S17}$$

where

$$U_b \equiv \begin{pmatrix} I_L & -S_L^{-1}S_{L,EM} & 0 \\ 0 & I_{EM} & 0 \\ 0 & -S_R^{-1}S_{R,EM} & I_R \end{pmatrix} ; \quad U_b^{-1} = \begin{pmatrix} I_L & S_L^{-1}S_{L,EM} & 0 \\ 0 & I_{EM} & 0 \\ 0 & S_R^{-1}S_{R,EM} & I_R \end{pmatrix}. \tag{S18}$$

Defining

$$\begin{cases} \widetilde{\mathcal{S}} \equiv U_b^\dagger \mathcal{S} U_b \\ \widetilde{\mathcal{H}}_{KS} \equiv U_b^\dagger \mathcal{H}_{KS} U_b \end{cases} \tag{S19}$$

Eq. (S17) can be rewritten as follows:

$$(U_b^\dagger)^{-1}\widetilde{\mathcal{S}}U_b^{-1}\dot{\mathcal{P}}(U_b^\dagger)^{-1}\widetilde{\mathcal{S}}U_b^{-1} = -i(U_b^\dagger)^{-1}\widetilde{\mathcal{H}}_{KS}U_b^{-1}\mathcal{P}(U_b^\dagger)^{-1}\widetilde{\mathcal{S}}U_b^{-1} +$$
$$i(U_b^\dagger)^{-1}\widetilde{\mathcal{S}}U_b^{-1}\mathcal{P}(U_b^\dagger)^{-1}\widetilde{\mathcal{H}}_{KS}U_b^{-1} - \Gamma\mathcal{P}(U_b^\dagger)^{-1}\widetilde{\mathcal{S}}U_b^{-1} - (U_b^\dagger)^{-1}\widetilde{\mathcal{S}}U_b^{-1}\mathcal{P}\Gamma^\dagger. \tag{S20}$$

Next, we multiply Eq. (S20) by $U_b^\dagger$ on the left and by $U_b$ on the right, to obtain:

$$\widetilde{\mathcal{S}}U_b^{-1}\dot{\mathcal{P}}(U_b^\dagger)^{-1}\widetilde{\mathcal{S}} =$$
$$= -i\widetilde{\mathcal{H}}_{KS}U_b^{-1}\mathcal{P}(U_b^\dagger)^{-1}\widetilde{\mathcal{S}} + i\widetilde{\mathcal{S}}U_b^{-1}\mathcal{P}(U_b^\dagger)^{-1}\widetilde{\mathcal{H}}_{KS} - U_b^\dagger\Gamma\mathcal{P}(U_b^\dagger)^{-1}\widetilde{\mathcal{S}} - \widetilde{\mathcal{S}}U_b^{-1}\mathcal{P}\Gamma^\dagger U_b \tag{S21}$$

Introducing $U_b U_b^{-1}$ and $(U_b^\dagger)^{-1}U_b^\dagger$ in the last two terms, respectively, yields:

$$\widetilde{\mathcal{S}}U_b^{-1}\dot{\mathcal{P}}(U_b^\dagger)^{-1}\widetilde{\mathcal{S}} = -i\widetilde{\mathcal{H}}_{KS}U_b^{-1}\mathcal{P}(U_b^\dagger)^{-1}\widetilde{\mathcal{S}} + i\widetilde{\mathcal{S}}U_b^{-1}\mathcal{P}(U_b^\dagger)^{-1}\widetilde{\mathcal{H}}_{KS} - U_b^\dagger\Gamma U_b U_b^{-1}\mathcal{P}(U_b^\dagger)^{-1}\widetilde{\mathcal{S}} -$$
$$\widetilde{\mathcal{S}}U_b^{-1}\mathcal{P}(U_b^\dagger)^{-1}U_b^\dagger\Gamma^\dagger U_b. \tag{S22}$$

Next, we define:

$$\begin{cases} \widetilde{\mathcal{P}} \equiv U_b^{-1}\mathcal{P}(U_b^\dagger)^{-1} \\ \widetilde{\dot{\mathcal{P}}} \equiv U_b^{-1}\dot{\mathcal{P}}(U_b^\dagger)^{-1} \\ \widetilde{\Gamma} \equiv U_b^\dagger \Gamma U_b \end{cases} \tag{S23}$$

We note that since $U_b$ is a fixed transformation (time-independent within the fixed nuclei Born-Oppenheimer approximation) the relation $\dot{\widetilde{\mathcal{P}}} = \widetilde{\dot{\mathcal{P}}}$ holds. With these definitions we obtain:

$$\widetilde{\mathcal{S}}\dot{\widetilde{\mathcal{P}}}\widetilde{\mathcal{S}} = -i\widetilde{\mathcal{H}}_{KS}\widetilde{\mathcal{P}}\widetilde{\mathcal{S}} + i\widetilde{\mathcal{S}}\widetilde{\mathcal{P}}\widetilde{\mathcal{H}}_{KS} - \widetilde{\Gamma}\widetilde{\mathcal{P}}\widetilde{\mathcal{S}} - \widetilde{\mathcal{S}}\widetilde{\mathcal{P}}\widetilde{\Gamma}^\dagger, \tag{S24}$$

where we have used the fact that $\widetilde{\Gamma}^\dagger = \left(U_b^\dagger \Gamma U_b\right)^\dagger = U_b^\dagger \Gamma^\dagger U_b$.

Next, we introduce the site-to-state transformation:

$$U \equiv \begin{pmatrix} U_L & 0 & 0 \\ 0 & U_{EM} & 0 \\ 0 & 0 & U_R \end{pmatrix}, \tag{S25}$$

such that $\widetilde{\widetilde{H}}_{KS_i} = U_i^\dagger \widetilde{H}_{KS_i} U_i$ is diagonal and $U_i^\dagger \widetilde{S}_i U_i = I_i$ are unit submatrices of the appropriate dimensions. With this, Eq. (S24) can be rewritten as:

$$(U^\dagger)^{-1}U^\dagger \tilde{\mathbf{S}} U U^{-1} \tilde{\mathcal{P}} (U^\dagger)^{-1} U^\dagger \tilde{\mathbf{S}} U U^{-1} =$$
$$= -i(U^\dagger)^{-1} U^\dagger \widetilde{\mathcal{H}}_{KS} U U^{-1} \tilde{\mathcal{P}} (U^\dagger)^{-1} U^\dagger \tilde{\mathbf{S}} U U^{-1} + i(U^\dagger)^{-1} U^\dagger \tilde{\mathbf{S}} U U^{-1} \tilde{\mathcal{P}} (U^\dagger)^{-1} U^\dagger \widetilde{\mathcal{H}}_{KS} U U^{-1} -$$
$$\tilde{\mathbf{\Gamma}} U U^{-1} \tilde{\mathcal{P}} (U^\dagger)^{-1} U^\dagger \tilde{\mathbf{S}} U U^{-1} - (U^\dagger)^{-1} U^\dagger \tilde{\mathbf{S}} U U^{-1} \tilde{\mathcal{P}} (U^\dagger)^{-1} U^\dagger \tilde{\mathbf{\Gamma}}^\dagger. \tag{S26}$$

Using the relation $U^\dagger \tilde{\mathbf{S}} U = I$ we get:

$$(U^\dagger)^{-1} U^{-1} \tilde{\mathcal{P}} (U^\dagger)^{-1} U^{-1} =$$
$$= -i(U^\dagger)^{-1} U^\dagger \widetilde{\mathcal{H}}_{KS} U U^{-1} \tilde{\mathcal{P}} (U^\dagger)^{-1} U^{-1} + i(U^\dagger)^{-1} U^{-1} \tilde{\mathcal{P}} (U^\dagger)^{-1} U^\dagger \widetilde{\mathcal{H}}_{KS} U U^{-1} -$$
$$\tilde{\mathbf{\Gamma}} U U^{-1} \tilde{\mathcal{P}} (U^\dagger)^{-1} U^{-1} - (U^\dagger)^{-1} U^{-1} \tilde{\mathcal{P}} (U^\dagger)^{-1} U^\dagger \tilde{\mathbf{\Gamma}}^\dagger \tag{S27}$$

Next, we define:

$$\begin{cases} \widetilde{\widetilde{\mathcal{H}}}_{KS} \equiv U^\dagger \widetilde{\mathcal{H}}_{KS} U \\ \widetilde{\widetilde{\mathcal{P}}} \equiv U^{-1} \tilde{\mathcal{P}} (U^\dagger)^{-1} \\ \widetilde{\widetilde{\dot{\mathcal{P}}}} \equiv U^{-1} \dot{\tilde{\mathcal{P}}} (U^\dagger)^{-1} \end{cases}, \tag{S28}$$

to obtain:

$$(U^\dagger)^{-1} \widetilde{\widetilde{\dot{\mathcal{P}}}} U^{-1} = -i(U^\dagger)^{-1} \widetilde{\widetilde{\mathcal{H}}}_{KS} \widetilde{\widetilde{\mathcal{P}}} U^{-1} + i(U^\dagger)^{-1} \widetilde{\widetilde{\mathcal{P}}} \widetilde{\widetilde{\mathcal{H}}}_{KS} U^{-1} - \tilde{\mathbf{\Gamma}} U \widetilde{\widetilde{\mathcal{P}}} U^{-1} - (U^\dagger)^{-1} \widetilde{\widetilde{\mathcal{P}}} U^\dagger \tilde{\mathbf{\Gamma}}^\dagger. \tag{S29}$$

Multiplying by $U^\dagger$ on the left and $U$ on the right we arrive at:

$$\widetilde{\widetilde{\dot{\mathcal{P}}}} = -i\widetilde{\widetilde{\mathcal{H}}}_{KS} \widetilde{\widetilde{\mathcal{P}}} + i\widetilde{\widetilde{\mathcal{P}}} \widetilde{\widetilde{\mathcal{H}}}_{KS} - U^\dagger \tilde{\mathbf{\Gamma}} U \widetilde{\widetilde{\mathcal{P}}} - \widetilde{\widetilde{\mathcal{P}}} U^\dagger \tilde{\mathbf{\Gamma}}^\dagger U. \tag{S30}$$

Defining:

$$\widetilde{\widetilde{\mathbf{\Gamma}}} \equiv U^\dagger \tilde{\mathbf{\Gamma}} U \tag{S31}$$

We finally obtain:

$$\widetilde{\widetilde{\dot{\mathcal{P}}}} = -i\left[\widetilde{\widetilde{\mathcal{H}}}_{KS}, \widetilde{\widetilde{\mathcal{P}}}\right] - \widetilde{\widetilde{\mathbf{\Gamma}}} \widetilde{\widetilde{\mathcal{P}}} - \widetilde{\widetilde{\mathcal{P}}} \widetilde{\widetilde{\mathbf{\Gamma}}}^\dagger. \tag{S32}$$

where $\widetilde{\widetilde{\mathbf{\Gamma}}}^\dagger = \left(U^\dagger \tilde{\mathbf{\Gamma}} U\right)^\dagger = U^\dagger \tilde{\mathbf{\Gamma}}^\dagger U$. In its simplest form $\widetilde{\widetilde{\mathbf{\Gamma}}}$ is written as:

$$\widetilde{\widetilde{\mathbf{\Gamma}}} = \widetilde{\widetilde{\mathbf{\Gamma}}}^\dagger = \gamma \begin{pmatrix} I_L & 0 & 0 \\ 0 & 0 & 0 \\ 0 & 0 & I_R \end{pmatrix}, \tag{S33}$$

which represents uniform broadening of all left and right lead levels. Hence, the last two terms in Eq. (S32) can be written as:

$$-\gamma \begin{pmatrix} I_L & 0 & 0 \\ 0 & 0 & 0 \\ 0 & 0 & I_R \end{pmatrix} \widetilde{\widetilde{\mathcal{P}}} - \gamma \widetilde{\widetilde{\mathcal{P}}} \begin{pmatrix} I_L & 0 & 0 \\ 0 & 0 & 0 \\ 0 & 0 & I_R \end{pmatrix} =$$

$$= -\gamma \begin{pmatrix} I_L & 0 & 0 \\ 0 & 0 & 0 \\ 0 & 0 & I_R \end{pmatrix} \begin{pmatrix} \widetilde{\widetilde{\mathcal{P}}}_L & \widetilde{\widetilde{\mathcal{P}}}_{L,EM} & \widetilde{\widetilde{\mathcal{P}}}_{LR} \\ \widetilde{\widetilde{\mathcal{P}}}_{EM,L} & \widetilde{\widetilde{\mathcal{P}}}_{EM} & \widetilde{\widetilde{\mathcal{P}}}_{EM,R} \\ \widetilde{\widetilde{\mathcal{P}}}_{RL} & \widetilde{\widetilde{\mathcal{P}}}_{R,EM} & \widetilde{\widetilde{\mathcal{P}}}_R \end{pmatrix} - \gamma \begin{pmatrix} \widetilde{\widetilde{\mathcal{P}}}_L & \widetilde{\widetilde{\mathcal{P}}}_{L,EM} & \widetilde{\widetilde{\mathcal{P}}}_{LR} \\ \widetilde{\widetilde{\mathcal{P}}}_{EM,L} & \widetilde{\widetilde{\mathcal{P}}}_{EM} & \widetilde{\widetilde{\mathcal{P}}}_{EM,R} \\ \widetilde{\widetilde{\mathcal{P}}}_{RL} & \widetilde{\widetilde{\mathcal{P}}}_{R,EM} & \widetilde{\widetilde{\mathcal{P}}}_R \end{pmatrix} \begin{pmatrix} I_L & 0 & 0 \\ 0 & 0 & 0 \\ 0 & 0 & I_R \end{pmatrix} =$$

$$= -\gamma \begin{pmatrix} \widetilde{\widetilde{\mathcal{P}}}_L & \widetilde{\widetilde{\mathcal{P}}}_{L,EM} & \widetilde{\widetilde{\mathcal{P}}}_{LR} \\ \mathbf{0} & \mathbf{0} & \mathbf{0} \\ \widetilde{\widetilde{\mathcal{P}}}_{RL} & \widetilde{\widetilde{\mathcal{P}}}_{R,EM} & \widetilde{\widetilde{\mathcal{P}}}_R \end{pmatrix} - \gamma \begin{pmatrix} \widetilde{\widetilde{\mathcal{P}}}_L & \mathbf{0} & \widetilde{\widetilde{\mathcal{P}}}_{LR} \\ \widetilde{\widetilde{\mathcal{P}}}_{EM,L} & \mathbf{0} & \widetilde{\widetilde{\mathcal{P}}}_{EM,R} \\ \widetilde{\widetilde{\mathcal{P}}}_{RL} & \mathbf{0} & \widetilde{\widetilde{\mathcal{P}}}_R \end{pmatrix} = -\gamma \begin{pmatrix} 2\widetilde{\widetilde{\mathcal{P}}}_L & \widetilde{\widetilde{\mathcal{P}}}_{L,EM} & 2\widetilde{\widetilde{\mathcal{P}}}_{LR} \\ \widetilde{\widetilde{\mathcal{P}}}_{EM,L} & \mathbf{0} & \widetilde{\widetilde{\mathcal{P}}}_{EM,R} \\ 2\widetilde{\widetilde{\mathcal{P}}}_{RL} & \widetilde{\widetilde{\mathcal{P}}}_{R,EM} & 2\widetilde{\widetilde{\mathcal{P}}}_R \end{pmatrix} =$$

$$= -2\gamma \begin{pmatrix} \widetilde{\widetilde{\mathcal{P}}}_L & \frac{1}{2}\widetilde{\widetilde{\mathcal{P}}}_{L,EM} & \widetilde{\widetilde{\mathcal{P}}}_{LR} \\ \frac{1}{2}\widetilde{\widetilde{\mathcal{P}}}_{EM,L} & \mathbf{0} & \frac{1}{2}\widetilde{\widetilde{\mathcal{P}}}_{EM,R} \\ \widetilde{\widetilde{\mathcal{P}}}_{RL} & \frac{1}{2}\widetilde{\widetilde{\mathcal{P}}}_{R,EM} & \widetilde{\widetilde{\mathcal{P}}}_R \end{pmatrix} \tag{S34}$$

The source term is then obtained by considering electrons that travel in the implicit reservoir toward the left and right leads with equilibrium distributions $\widetilde{\widetilde{\mathcal{P}}}_L^0$ and $\widetilde{\widetilde{\mathcal{P}}}_R^0$. Upon reaching the reservoir/lead interface they are adsorbed at a rate of $2\gamma$ and are injected into the system at the same rate. This can be described by the following term, which drives the system at the lead sections towards the equilibrium state of leads that are coupled to the corresponding external implicit reservoirs and decoupled from the extended molecule section:

$$\gamma \begin{pmatrix} I_L & \mathbf{0} & \mathbf{0} \\ \mathbf{0} & \mathbf{0} & \mathbf{0} \\ \mathbf{0} & \mathbf{0} & I_R \end{pmatrix} \begin{pmatrix} \widetilde{\widetilde{\mathcal{P}}}_L^0 & \mathbf{0} & \mathbf{0} \\ \mathbf{0} & \widetilde{\widetilde{\mathcal{P}}}_{EM}^0 & \mathbf{0} \\ \mathbf{0} & \mathbf{0} & \widetilde{\widetilde{\mathcal{P}}}_R^0 \end{pmatrix} + \gamma \begin{pmatrix} \widetilde{\widetilde{\mathcal{P}}}_L^0 & \mathbf{0} & \mathbf{0} \\ \mathbf{0} & \widetilde{\widetilde{\mathcal{P}}}_{EM}^0 & \mathbf{0} \\ \mathbf{0} & \mathbf{0} & \widetilde{\widetilde{\mathcal{P}}}_R^0 \end{pmatrix} \begin{pmatrix} I_L & \mathbf{0} & \mathbf{0} \\ \mathbf{0} & \mathbf{0} & \mathbf{0} \\ \mathbf{0} & \mathbf{0} & I_R \end{pmatrix} = \gamma \begin{pmatrix} \widetilde{\widetilde{\mathcal{P}}}_L^0 & \mathbf{0} & \mathbf{0} \\ \mathbf{0} & \mathbf{0} & \mathbf{0} \\ \mathbf{0} & \mathbf{0} & \widetilde{\widetilde{\mathcal{P}}}_R^0 \end{pmatrix} +$$

$$\gamma \begin{pmatrix} \widetilde{\widetilde{\mathcal{P}}}_L^0 & \mathbf{0} & \mathbf{0} \\ \mathbf{0} & \mathbf{0} & \mathbf{0} \\ \mathbf{0} & \mathbf{0} & \widetilde{\widetilde{\mathcal{P}}}_R^0 \end{pmatrix} = 2\gamma \begin{pmatrix} \widetilde{\widetilde{\mathcal{P}}}_L^0 & \mathbf{0} & \mathbf{0} \\ \mathbf{0} & \mathbf{0} & \mathbf{0} \\ \mathbf{0} & \mathbf{0} & \widetilde{\widetilde{\mathcal{P}}}_R^0 \end{pmatrix}. \tag{S35}$$

Inserting the expressions of Eqs. (S34) and (S35) into Eq. (S32) and defining $\Gamma \equiv 2\gamma$ we obtain:

$$\dot{\widetilde{\widetilde{\mathcal{P}}}} = -i\left[\widetilde{\widetilde{\mathcal{H}}}_{KS}, \widetilde{\widetilde{\mathcal{P}}}\right] - \Gamma \begin{pmatrix} \widetilde{\widetilde{\mathcal{P}}}_L - \widetilde{\widetilde{\mathcal{P}}}_L^0 & \frac{1}{2}\widetilde{\widetilde{\mathcal{P}}}_{L,EM} & \widetilde{\widetilde{\mathcal{P}}}_{LR} \\ \frac{1}{2}\widetilde{\widetilde{\mathcal{P}}}_{EM,L} & \mathbf{0} & \frac{1}{2}\widetilde{\widetilde{\mathcal{P}}}_{EM,R} \\ \widetilde{\widetilde{\mathcal{P}}}_{RL} & \frac{1}{2}\widetilde{\widetilde{\mathcal{P}}}_{R,EM} & \widetilde{\widetilde{\mathcal{P}}}_R - \widetilde{\widetilde{\mathcal{P}}}_R^0 \end{pmatrix}. \tag{S36}$$

Note that within the realm of TDDFT, $\dot{\widetilde{\widetilde{\mathcal{P}}}}$, which is the state representation of $\dot{\mathcal{P}}$, is not the time derivative of $\widetilde{\widetilde{\mathcal{P}}}$, namely $\dot{\widetilde{\widetilde{\mathcal{P}}}} \neq \dot{\widetilde{\widetilde{\mathcal{P}}}}$. This results from the fact that the KS Hamiltonian matrix has implicit time-dependence via its dependence on the density matrix and hence the $U$ transformation matrix varies with time, as well. Since $\widetilde{\widetilde{\mathcal{P}}} \equiv U^{-1}\widetilde{\mathcal{P}}(U^\dagger)^{-1}$ (see Eq. (S28)), its time derivative, $\dot{\widetilde{\widetilde{\mathcal{P}}}}$, should include the time derivative of $U$. Lacking an explicit equation of motion for $U$, we are thus forced to perform the propagation step in the site representation. To this end, we use Eqs. (S19), (S23), and (S28) within Eq. (S36) as follows:

$$U^{-1}U_b^{-1}\dot{\mathcal{P}}(U_b^\dagger)^{-1}(U^\dagger)^{-1} =$$

$$= -i\left[U^\dagger U_b^\dagger \mathcal{H}_{KS} U_b U, U^{-1}U_b^{-1}\mathcal{P}(U_b^\dagger)^{-1}(U^\dagger)^{-1}\right] - \Gamma \begin{pmatrix} \widetilde{\widetilde{\mathcal{P}}}_L - \widetilde{\widetilde{\mathcal{P}}}_L^0 & \frac{1}{2}\widetilde{\widetilde{\mathcal{P}}}_{L,EM} & \widetilde{\widetilde{\mathcal{P}}}_{LR} \\ \frac{1}{2}\widetilde{\widetilde{\mathcal{P}}}_{EM,L} & 0 & \frac{1}{2}\widetilde{\widetilde{\mathcal{P}}}_{EM,R} \\ \widetilde{\widetilde{\mathcal{P}}}_{RL} & \frac{1}{2}\widetilde{\widetilde{\mathcal{P}}}_{R,EM} & \widetilde{\widetilde{\mathcal{P}}}_R - \widetilde{\widetilde{\mathcal{P}}}_R^0 \end{pmatrix}$$

$$= -i\left[U^\dagger U_b^\dagger \mathcal{H}_{KS} U_b U U^{-1} U_b^{-1}\mathcal{P}(U_b^\dagger)^{-1}(U^\dagger)^{-1} - U^{-1}U_b^{-1}\mathcal{P}(U_b^\dagger)^{-1}(U^\dagger)^{-1} U^\dagger U_b^\dagger \mathcal{H}_{KS} U_b U\right]$$

$$-\Gamma \begin{pmatrix} \widetilde{\widetilde{\mathcal{P}}}_L - \widetilde{\widetilde{\mathcal{P}}}_L^0 & \frac{1}{2}\widetilde{\widetilde{\mathcal{P}}}_{L,EM} & \widetilde{\widetilde{\mathcal{P}}}_{LR} \\ \frac{1}{2}\widetilde{\widetilde{\mathcal{P}}}_{EM,L} & 0 & \frac{1}{2}\widetilde{\widetilde{\mathcal{P}}}_{EM,R} \\ \widetilde{\widetilde{\mathcal{P}}}_{RL} & \frac{1}{2}\widetilde{\widetilde{\mathcal{P}}}_{R,EM} & \widetilde{\widetilde{\mathcal{P}}}_R - \widetilde{\widetilde{\mathcal{P}}}_R^0 \end{pmatrix} =$$

$$= -iU^\dagger U_b^\dagger \mathcal{H}_{KS} \mathcal{P}(U_b^\dagger)^{-1}(U^\dagger)^{-1} + iU^{-1}U_b^{-1}\mathcal{P}\mathcal{H}_{KS} U_b U - \Gamma \begin{pmatrix} \widetilde{\widetilde{\mathcal{P}}}_L - \widetilde{\widetilde{\mathcal{P}}}_L^0 & \frac{1}{2}\widetilde{\widetilde{\mathcal{P}}}_{L,EM} & \widetilde{\widetilde{\mathcal{P}}}_{LR} \\ \frac{1}{2}\widetilde{\widetilde{\mathcal{P}}}_{EM,L} & 0 & \frac{1}{2}\widetilde{\widetilde{\mathcal{P}}}_{EM,R} \\ \widetilde{\widetilde{\mathcal{P}}}_{RL} & \frac{1}{2}\widetilde{\widetilde{\mathcal{P}}}_{R,EM} & \widetilde{\widetilde{\mathcal{P}}}_R - \widetilde{\widetilde{\mathcal{P}}}_R^0 \end{pmatrix}. \quad (S37)$$

Multiplying by $U_b U$ on the left and $U^\dagger U_b^\dagger$ on the right we obtain:

$$\dot{\mathcal{P}} = -iU_b U U^\dagger U_b^\dagger \mathcal{H}_{KS} \mathcal{P} + i\mathcal{P}\mathcal{H}_{KS} U_b U U^\dagger U_b^\dagger - \Gamma U_b U \begin{pmatrix} \widetilde{\widetilde{\mathcal{P}}}_L - \widetilde{\widetilde{\mathcal{P}}}_L^0 & \frac{1}{2}\widetilde{\widetilde{\mathcal{P}}}_{L,EM} & \widetilde{\widetilde{\mathcal{P}}}_{LR} \\ \frac{1}{2}\widetilde{\widetilde{\mathcal{P}}}_{EM,L} & 0 & \frac{1}{2}\widetilde{\widetilde{\mathcal{P}}}_{EM,R} \\ \widetilde{\widetilde{\mathcal{P}}}_{RL} & \frac{1}{2}\widetilde{\widetilde{\mathcal{P}}}_{R,EM} & \widetilde{\widetilde{\mathcal{P}}}_R - \widetilde{\widetilde{\mathcal{P}}}_R^0 \end{pmatrix} U^\dagger U_b^\dagger. \quad (S38)$$

Since, by construction, the transformation $U$ obeys the relation $U^\dagger \widetilde{S} U = I$, we may write $\widetilde{S} = (U^\dagger)^{-1}U^{-1} = (UU^\dagger)^{-1}$, such that $UU^\dagger = \widetilde{S}^{-1}$. Similarly, from Eq. (S19) we have $\widetilde{S} \equiv U_b^\dagger S U_b$. We may therefore write $\widetilde{S}^{-1} = (U_b^\dagger S U_b)^{-1} = U_b^{-1} S^{-1}(U_b^\dagger)^{-1}$. Solving for $S^{-1}$ we obtain $S^{-1} = U_b \widetilde{S}^{-1} U_b^\dagger$. Hence, we obtain $U_b U U^\dagger U_b^\dagger = U_b \widetilde{S}^{-1} U_b^\dagger = S^{-1}$ such that:

$$\dot{\mathcal{P}} = -iS^{-1}\mathcal{H}_{KS}\mathcal{P} + i\mathcal{P}\mathcal{H}_{KS}S^{-1} - \Gamma U_b U \begin{pmatrix} \widetilde{\widetilde{\mathcal{P}}}_L - \widetilde{\widetilde{\mathcal{P}}}_L^0 & \frac{1}{2}\widetilde{\widetilde{\mathcal{P}}}_{L,EM} & \widetilde{\widetilde{\mathcal{P}}}_{LR} \\ \frac{1}{2}\widetilde{\widetilde{\mathcal{P}}}_{EM,L} & 0 & \frac{1}{2}\widetilde{\widetilde{\mathcal{P}}}_{EM,R} \\ \widetilde{\widetilde{\mathcal{P}}}_{RL} & \frac{1}{2}\widetilde{\widetilde{\mathcal{P}}}_{R,EM} & \widetilde{\widetilde{\mathcal{P}}}_R - \widetilde{\widetilde{\mathcal{P}}}_R^0 \end{pmatrix} U^\dagger U_b^\dagger. \quad (S39)$$

This is the DLvN EOM—our working equation—given in the site representation, which we use for the time propagation.

To further simplify this expression and avoid the full matrix transformations appearing in the driving term of Eq. (S39) we may separate it into the sink and source terms and treat them separately. Using

Eqs. (S18), (S25), and (S34) the sink term can be written as:

$$\mathcal{L}_{sink} = U_b U \begin{pmatrix} \widetilde{\widetilde{\mathcal{P}}}_L & \frac{1}{2}\widetilde{\widetilde{\mathcal{P}}}_{L,EM} & \widetilde{\widetilde{\mathcal{P}}}_{LR} \\ \frac{1}{2}\widetilde{\widetilde{\mathcal{P}}}_{EM,L} & 0 & \frac{1}{2}\widetilde{\widetilde{\mathcal{P}}}_{EM,R} \\ \widetilde{\widetilde{\mathcal{P}}}_{RL} & \frac{1}{2}\widetilde{\widetilde{\mathcal{P}}}_{R,EM} & \widetilde{\widetilde{\mathcal{P}}}_R \end{pmatrix} U^\dagger U_b^\dagger =$$

$$= \frac{1}{2} U_b U \left[ \begin{pmatrix} I_L & 0 & 0 \\ 0 & 0 & 0 \\ 0 & 0 & I_R \end{pmatrix} \widetilde{\widetilde{\mathcal{P}}} + \widetilde{\widetilde{\mathcal{P}}} \begin{pmatrix} I_L & 0 & 0 \\ 0 & 0 & 0 \\ 0 & 0 & I_R \end{pmatrix} \right] U^\dagger U_b^\dagger \tag{S40}$$

Using Eqs. (S23) and (S28) we now obtain:

$$\mathcal{L}_{sink} = \frac{1}{2} U_b U \left[ \begin{pmatrix} I_L & 0 & 0 \\ 0 & 0 & 0 \\ 0 & 0 & I_R \end{pmatrix} U^{-1} U_b^{-1} \mathcal{P} \left( U_b^\dagger \right)^{-1} (U^\dagger)^{-1} \right.$$

$$\left. + U^{-1} U_b^{-1} \mathcal{P} \left( U_b^\dagger \right)^{-1} (U^\dagger)^{-1} \begin{pmatrix} I_L & 0 & 0 \\ 0 & 0 & 0 \\ 0 & 0 & I_R \end{pmatrix} \right] U^\dagger U_b^\dagger =$$

$$= \frac{1}{2} U_b U \begin{pmatrix} I_L & 0 & 0 \\ 0 & 0 & 0 \\ 0 & 0 & I_R \end{pmatrix} U^{-1} U_b^{-1} \mathcal{P} + \frac{1}{2} \mathcal{P} \left( U_b^\dagger \right)^{-1} (U^\dagger)^{-1} \begin{pmatrix} I_L & 0 & 0 \\ 0 & 0 & 0 \\ 0 & 0 & I_R \end{pmatrix} U^\dagger U_b^\dagger \tag{S41}$$

We may now use the relation $\begin{pmatrix} I_L & 0 & 0 \\ 0 & 0 & 0 \\ 0 & 0 & I_R \end{pmatrix} = I - \begin{pmatrix} 0 & 0 & 0 \\ 0 & I_{EM} & 0 \\ 0 & 0 & 0 \end{pmatrix}$ to write:

$$\mathcal{L}_{sink} = \frac{1}{2} U_b U \left[ I - \begin{pmatrix} 0 & 0 & 0 \\ 0 & I_{EM} & 0 \\ 0 & 0 & 0 \end{pmatrix} \right] U^{-1} U_b^{-1} \mathcal{P} + \frac{1}{2} \mathcal{P} \left( U_b^\dagger \right)^{-1} (U^\dagger)^{-1} \left[ I - \begin{pmatrix} 0 & 0 & 0 \\ 0 & I_{EM} & 0 \\ 0 & 0 & 0 \end{pmatrix} \right] U^\dagger U_b^\dagger =$$

$$= \mathcal{P} - \frac{1}{2} U_b U \begin{pmatrix} 0 & 0 & 0 \\ 0 & I_{EM} & 0 \\ 0 & 0 & 0 \end{pmatrix} U^{-1} U_b^{-1} \mathcal{P} - \frac{1}{2} \mathcal{P} \left( U_b^\dagger \right)^{-1} (U^\dagger)^{-1} \begin{pmatrix} 0 & 0 & 0 \\ 0 & I_{EM} & 0 \\ 0 & 0 & 0 \end{pmatrix} U^\dagger U_b^\dagger \tag{S42}$$

Using Eq. (S25) for the transformation matrix $U$ we can write:

$$U \begin{pmatrix} 0 & 0 & 0 \\ 0 & I_{EM} & 0 \\ 0 & 0 & 0 \end{pmatrix} U^{-1} = \begin{pmatrix} U_L & 0 & 0 \\ 0 & U_{EM} & 0 \\ 0 & 0 & U_R \end{pmatrix} \begin{pmatrix} 0 & 0 & 0 \\ 0 & I_{EM} & 0 \\ 0 & 0 & 0 \end{pmatrix} \begin{pmatrix} U_L^{-1} & 0 & 0 \\ 0 & U_{EM}^{-1} & 0 \\ 0 & 0 & U_R^{-1} \end{pmatrix} =$$

$$\begin{pmatrix} U_L & 0 & 0 \\ 0 & U_{EM} & 0 \\ 0 & 0 & U_R \end{pmatrix} \begin{pmatrix} 0 & 0 & 0 \\ 0 & U_{EM}^{-1} & 0 \\ 0 & 0 & 0 \end{pmatrix} = \begin{pmatrix} 0 & 0 & 0 \\ 0 & U_{EM} U_{EM}^{-1} & 0 \\ 0 & 0 & 0 \end{pmatrix} = \begin{pmatrix} 0 & 0 & 0 \\ 0 & I_{EM} & 0 \\ 0 & 0 & 0 \end{pmatrix} \tag{S43}$$

Therefore, we have

$$U_b U \begin{pmatrix} 0 & 0 & 0 \\ 0 & I_{EM} & 0 \\ 0 & 0 & 0 \end{pmatrix} U^{-1} U_b^{-1} = U_b \begin{pmatrix} 0 & 0 & 0 \\ 0 & I_{EM} & 0 \\ 0 & 0 & 0 \end{pmatrix} U_b^{-1} =$$

$$\begin{pmatrix} I_L & -S_L^{-1}S_{L,EM} & 0 \\ 0 & I_{EM} & 0 \\ 0 & -S_R^{-1}S_{R,EM} & I_R \end{pmatrix} \begin{pmatrix} 0 & 0 & 0 \\ 0 & I_{EM} & 0 \\ 0 & 0 & 0 \end{pmatrix} \begin{pmatrix} I_L & S_L^{-1}S_{L,EM} & 0 \\ 0 & I_{EM} & 0 \\ 0 & S_R^{-1}S_{R,EM} & I_R \end{pmatrix} =$$

$$\begin{pmatrix} I_L & -S_L^{-1}S_{L,EM} & 0 \\ 0 & I_{EM} & 0 \\ 0 & -S_R^{-1}S_{R,EM} & I_R \end{pmatrix} \begin{pmatrix} 0 & 0 & 0 \\ 0 & I_{EM} & 0 \\ 0 & 0 & 0 \end{pmatrix} = \begin{pmatrix} 0 & -S_L^{-1}S_{L,EM} & 0 \\ 0 & I_{EM} & 0 \\ 0 & -S_R^{-1}S_{R,EM} & 0 \end{pmatrix}, \tag{S44}$$

where we have used Eq. (S18) for the expressions of $U_b$ and $U_b^{-1}$. Similarly,

$$(U^\dagger)^{-1} \begin{pmatrix} 0 & 0 & 0 \\ 0 & I_{EM} & 0 \\ 0 & 0 & 0 \end{pmatrix} U^\dagger = \begin{pmatrix} (U_L^\dagger)^{-1} & 0 & 0 \\ 0 & (U_{EM}^\dagger)^{-1} & 0 \\ 0 & 0 & (U_R^\dagger)^{-1} \end{pmatrix} \begin{pmatrix} 0 & 0 & 0 \\ 0 & I_{EM} & 0 \\ 0 & 0 & 0 \end{pmatrix} \begin{pmatrix} U_L^\dagger & 0 & 0 \\ 0 & U_{EM}^\dagger & 0 \\ 0 & 0 & U_R^\dagger \end{pmatrix}$$

$$= \begin{pmatrix} {U_L^\dagger}^{-1} & 0 & 0 \\ 0 & {U_{EM}^\dagger}^{-1} & 0 \\ 0 & 0 & {U_R^\dagger}^{-1} \end{pmatrix} \begin{pmatrix} 0 & 0 & 0 \\ 0 & U_{EM}^\dagger & 0 \\ 0 & 0 & 0 \end{pmatrix} = \begin{pmatrix} 0 & 0 & 0 \\ 0 & (U_{EM}^\dagger)^{-1} U_{EM}^\dagger & 0 \\ 0 & 0 & 0 \end{pmatrix} = \begin{pmatrix} 0 & 0 & 0 \\ 0 & I_{EM} & 0 \\ 0 & 0 & 0 \end{pmatrix}, \tag{S45}$$

and therefore:

$$(U_b^\dagger)^{-1}(U^\dagger)^{-1} \begin{pmatrix} 0 & 0 & 0 \\ 0 & I_{EM} & 0 \\ 0 & 0 & 0 \end{pmatrix} U^\dagger U_b^\dagger =$$

$$= \begin{pmatrix} I_L & 0 & 0 \\ S_{EM,L}S_L^{-1} & I_{EM} & S_{EM,R}S_R^{-1} \\ 0 & 0 & I_R \end{pmatrix} \begin{pmatrix} 0 & 0 & 0 \\ 0 & I_{EM} & 0 \\ 0 & 0 & 0 \end{pmatrix} \begin{pmatrix} I_L & 0 & 0 \\ -S_{EM,L}S_L^{-1} & I_{EM} & -S_{EM,R}S_R^{-1} \\ 0 & 0 & I_R \end{pmatrix} =$$

$$= \begin{pmatrix} I_L & 0 & 0 \\ S_{EM,L}S_L^{-1} & I_{EM} & S_{EM,R}S_R^{-1} \\ 0 & 0 & I_R \end{pmatrix} \begin{pmatrix} 0 & 0 & 0 \\ -S_{EM,L}S_L^{-1} & I_{EM} & -S_{EM,R}S_R^{-1} \\ 0 & 0 & 0 \end{pmatrix} =$$

$$\begin{pmatrix} 0 & 0 & 0 \\ -S_{EM,L}S_L^{-1} & I_{EM} & -S_{EM,R}S_R^{-1} \\ 0 & 0 & 0 \end{pmatrix}. \tag{S46}$$

Inserting Eqs. (S44) and (S46) in Eq. (S42) yields:

$$\mathcal{L}_{sink} = \mathcal{P} - \frac{1}{2} \begin{pmatrix} 0 & -S_L^{-1}S_{L,EM} & 0 \\ 0 & I_{EM} & 0 \\ 0 & -S_R^{-1}S_{R,EM} & 0 \end{pmatrix} \begin{pmatrix} \mathcal{P}_L & \mathcal{P}_{L,EM} & \mathcal{P}_{LR} \\ \mathcal{P}_{EM,L} & \mathcal{P}_{EM} & \mathcal{P}_{EM,R} \\ \mathcal{P}_{RL} & \mathcal{P}_{R,EM} & \mathcal{P}_R \end{pmatrix}$$

$$- \frac{1}{2} \begin{pmatrix} \mathcal{P}_L & \mathcal{P}_{L,EM} & \mathcal{P}_{LR} \\ \mathcal{P}_{EM,L} & \mathcal{P}_{EM} & \mathcal{P}_{EM,R} \\ \mathcal{P}_{RL} & \mathcal{P}_{R,EM} & \mathcal{P}_R \end{pmatrix} \begin{pmatrix} 0 & 0 & 0 \\ -S_{EM,L}S_L^{-1} & I_{EM} & -S_{EM,R}S_R^{-1} \\ 0 & 0 & 0 \end{pmatrix} =$$

$$= \mathcal{P} - \frac{1}{2}\begin{pmatrix} -S_L^{-1}S_{L,EM}\mathcal{P}_{EM,L} & -S_L^{-1}S_{L,EM}\mathcal{P}_{EM} & -S_L^{-1}S_{L,EM}\mathcal{P}_{EM,R} \\ \mathcal{P}_{EM,L} & \mathcal{P}_{EM} & \mathcal{P}_{EM,R} \\ -S_R^{-1}S_{R,EM}\mathcal{P}_{EM,L} & -S_R^{-1}S_{R,EM}\mathcal{P}_{EM} & -S_R^{-1}S_{R,EM}\mathcal{P}_{EM,R} \end{pmatrix} -$$

$$\frac{1}{2}\begin{pmatrix} -\mathcal{P}_{L,EM}S_{EM,L}S_L^{-1} & \mathcal{P}_{L,EM} & -\mathcal{P}_{L,EM}S_{EM,R}S_R^{-1} \\ -\mathcal{P}_{EM}S_{EM,L}S_L^{-1} & \mathcal{P}_{EM} & -\mathcal{P}_{EM}S_{EM,R}S_R^{-1} \\ -\mathcal{P}_{R,EM}S_{EM,L}S_L^{-1} & \mathcal{P}_{R,EM} & -\mathcal{P}_{R,EM}S_{EM,R}S_R^{-1} \end{pmatrix} \quad (S47)$$

$$= \mathcal{P} - \frac{1}{2}\begin{pmatrix} -S_L^{-1}S_{L,EM}\mathcal{P}_{EM,L} - \mathcal{P}_{L,EM}S_{EM,L}S_L^{-1} & \mathcal{P}_{L,EM} - S_L^{-1}S_{L,EM}\mathcal{P}_{EM} & -S_L^{-1}S_{L,EM}\mathcal{P}_{EM,R} - \mathcal{P}_{L,EM}S_{EM,R}S_R^{-1} \\ \mathcal{P}_{EM,L} - \mathcal{P}_{EM}S_{EM,L}S_L^{-1} & 2\mathcal{P}_{EM} & \mathcal{P}_{EM,R} - \mathcal{P}_{EM}S_{EM,R}S_R^{-1} \\ -S_R^{-1}S_{R,EM}\mathcal{P}_{EM,L} - \mathcal{P}_{R,EM}S_{EM,L}S_L^{-1} & \mathcal{P}_{R,EM} - S_R^{-1}S_{R,EM}\mathcal{P}_{EM} & -S_R^{-1}S_{R,EM}\mathcal{P}_{EM,R} - \mathcal{P}_{R,EM}S_{EM,R}S_R^{-1} \end{pmatrix},$$

or equivalently:

$$\mathcal{L}_{sink} =$$

$$\begin{pmatrix} \mathcal{P}_L + \frac{S_L^{-1}S_{L,EM}\mathcal{P}_{EM,L} + \mathcal{P}_{L,EM}S_{EM,L}S_L^{-1}}{2} & \frac{\mathcal{P}_{L,EM} + S_L^{-1}S_{L,EM}\mathcal{P}_{EM}}{2} & \mathcal{P}_{LR} + \frac{S_L^{-1}S_{L,EM}\mathcal{P}_{EM,R} + \mathcal{P}_{L,EM}S_{EM,R}S_R^{-1}}{2} \\ \frac{\mathcal{P}_{EM,L} + \mathcal{P}_{EM}S_{EM,L}S_L^{-1}}{2} & 0 & \frac{\mathcal{P}_{EM,R} + \mathcal{P}_{EM}S_{EM,R}S_R^{-1}}{2} \\ \mathcal{P}_{RL} + \frac{S_R^{-1}S_{R,EM}\mathcal{P}_{EM,L} + \mathcal{P}_{R,EM}S_{EM,L}S_L^{-1}}{2} & \frac{\mathcal{P}_{R,EM} + S_R^{-1}S_{R,EM}\mathcal{P}_{EM}}{2} & \mathcal{P}_R + \frac{S_R^{-1}S_{R,EM}\mathcal{P}_{EM,R} + \mathcal{P}_{R,EM}S_{EM,R}S_R^{-1}}{2} \end{pmatrix}. \quad (S48)$$

Considering next the source term using Eq. (S25) we have:

$$\mathcal{L}_{source} = U_b U \begin{pmatrix} \widetilde{\widetilde{\mathcal{P}}}_L^0 & 0 & 0 \\ 0 & 0 & 0 \\ 0 & 0 & \widetilde{\widetilde{\mathcal{P}}}_R^0 \end{pmatrix} U^\dagger U_b^\dagger =$$

$$= U_b \begin{pmatrix} U_L & 0 & 0 \\ 0 & U_{EM} & 0 \\ 0 & 0 & U_R \end{pmatrix} \begin{pmatrix} \widetilde{\widetilde{\mathcal{P}}}_L^0 & 0 & 0 \\ 0 & 0 & 0 \\ 0 & 0 & \widetilde{\widetilde{\mathcal{P}}}_R^0 \end{pmatrix} \begin{pmatrix} U_L^\dagger & 0 & 0 \\ 0 & U_{EM}^\dagger & 0 \\ 0 & 0 & U_R^\dagger \end{pmatrix} U_b^\dagger =$$

$$= U_b \begin{pmatrix} U_L & 0 & 0 \\ 0 & U_{EM} & 0 \\ 0 & 0 & U_R \end{pmatrix} \begin{pmatrix} \widetilde{\widetilde{\mathcal{P}}}_L^0 U_L^\dagger & 0 & 0 \\ 0 & 0 & 0 \\ 0 & 0 & \widetilde{\widetilde{\mathcal{P}}}_R^0 U_R^\dagger \end{pmatrix} U_b^\dagger = U_b \begin{pmatrix} U_L \widetilde{\widetilde{\mathcal{P}}}_L^0 U_L^\dagger & 0 & 0 \\ 0 & 0 & 0 \\ 0 & 0 & U_R \widetilde{\widetilde{\mathcal{P}}}_R^0 U_R^\dagger \end{pmatrix} U_b^\dagger. \quad (S49)$$

Using Eqs. (S18) and (S49) we may now write:

$$\mathcal{L}_{source} = \begin{pmatrix} I_L & -S_L^{-1}S_{L,EM} & 0 \\ 0 & I_{EM} & 0 \\ 0 & -S_R^{-1}S_{R,EM} & I_R \end{pmatrix} \begin{pmatrix} U_L \widetilde{\widetilde{\mathcal{P}}}_L^0 U_L^\dagger & 0 & 0 \\ 0 & 0 & 0 \\ 0 & 0 & U_R \widetilde{\widetilde{\mathcal{P}}}_R^0 U_R^\dagger \end{pmatrix} \begin{pmatrix} I_L & 0 & 0 \\ -S_{EM,L}S_L^{-1} & I_{EM} & -S_{EM,R}S_R^{-1} \\ 0 & 0 & I_R \end{pmatrix} =$$

$$= \begin{pmatrix} I_L & -S_L^{-1}S_{L,EM} & 0 \\ 0 & I_{EM} & 0 \\ 0 & -S_R^{-1}S_{R,EM} & I_R \end{pmatrix} \begin{pmatrix} U_L \widetilde{\widetilde{\mathcal{P}}}_L^0 U_L^\dagger & 0 & 0 \\ 0 & 0 & 0 \\ 0 & 0 & U_R \widetilde{\widetilde{\mathcal{P}}}_R^0 U_R^\dagger \end{pmatrix} = \begin{pmatrix} U_L \widetilde{\widetilde{\mathcal{P}}}_L^0 U_L^\dagger & 0 & 0 \\ 0 & 0 & 0 \\ 0 & 0 & U_R \widetilde{\widetilde{\mathcal{P}}}_R^0 U_R^\dagger \end{pmatrix}. \quad (S50)$$

Since $\widetilde{\widetilde{\mathcal{P}}}_{L/R}^0$ are diagonal matrices, we obtain the following simplified expression for their transformed matrix elements:

$$\left(\boldsymbol{U}_{L/R}\widetilde{\widetilde{\boldsymbol{\mathcal{P}}}}^0_{L/R}\boldsymbol{U}^\dagger_{L/R}\right)_{ij} = \sum_k \sum_l (\boldsymbol{U}_{L/R})_{ik} \left(\widetilde{\widetilde{\boldsymbol{\mathcal{P}}}}^0_{L/R}\right)_{kl} (\boldsymbol{U}^\dagger_{L/R})_{lj} = \sum_k \sum_l (\boldsymbol{U}_{L/R})_{ik} f(\varepsilon^k_{L/R}, \mu_{L/R}) \delta_{kl} (\boldsymbol{U}^\dagger_{L/R})_{lj}$$

$$= \Sigma_k (\boldsymbol{U}_{L/R})_{ik} (\boldsymbol{U}^\dagger_{L/R})_{kj} f(\varepsilon^k_{L/R}, \mu_{L/R}) = \Sigma_k (\boldsymbol{U}_{L/R})_{ik} (\boldsymbol{U}_{L/R})^*_{jk} f(\varepsilon^k_{L/R}, \mu_{L/R}), \tag{S51}$$

where $f(\varepsilon, \mu) = \left[e^{(\varepsilon-\mu)/k_B T} + 1\right]^{-1}$ is the Fermi Dirac distribution, $k_B$ is Boltmann's constant, $T$ is the electronic temperature, $\varepsilon^k_{L/R}$ is the $k^{th}$ eigenvalue of the $L/R$ lead, and $\mu_{L/R}$ is the corresponding chemical potential. We note that in the present implementation we resort to Eq. (S39) for the propagation without considering the above simplifications. In practice, the propagation is performed as follows:

1. Construct a junction model with predefined lead and extended molecule sections.
2. Perform a ground state calculation to obtain the overlap matrix, $\boldsymbol{\mathcal{S}}$, and the initial $\boldsymbol{\mathcal{H}}_{KS}$ and $\boldsymbol{\mathcal{P}}$ matrices in the site representation.
3. Build the matrix transformation $\boldsymbol{U}_b$ (Eq. (S18)).
4. Transform $\boldsymbol{\mathcal{H}}_{KS} \to \widetilde{\boldsymbol{\mathcal{H}}}_{KS}$ (Eq. (S19)).
5. Calculate $\boldsymbol{U}_{L/R}$ and $\boldsymbol{\varepsilon}_{L/R}$ by solving the generalized eigenstate equations and transform $\widetilde{\boldsymbol{\mathcal{H}}}_{KS} \to \widetilde{\widetilde{\boldsymbol{\mathcal{H}}}}_{KS}$ (Eq. (S28)).
6. Construct the $\widetilde{\widetilde{\boldsymbol{\mathcal{P}}}}^0_L$ and $\widetilde{\widetilde{\boldsymbol{\mathcal{P}}}}^0_R$ blocks using the left and right lead model eigenstates, $\boldsymbol{\varepsilon}_{L/R}$, obtained in step 5 above.
7. Propagate $\boldsymbol{\mathcal{P}}$ (Eq. (S39)).
8. Construct the new $\boldsymbol{\mathcal{H}}_{KS}$ from the new $\boldsymbol{\mathcal{P}}$.
9. If the time has not exceeded the maximal time, return to step 4.

## 2. Driving rate sensitivity test

While the driving rate, $\Gamma$, appearing in Eq. (11) of the main text can, in principle, be determined from the self-energy of the semi-infinite lead models,[2] in the current implementation we use it as a free parameter. To determine the value to be used in the dynamical simulations, we broaden the discrete energy levels of the finite lead models with Lorentzian functions of different widths and adopt the Lorentzian width parameter that provides a density of states that represents well that of a semi-infinite system (not too narrow to provide a discrete spectrum and not too wide to artificially wash out the electronic structure features of the lead) as our $\Gamma$ value for the time-dependent calculations.[3–5] Figure S1 compares the density of states of the hydrogen chain studied in Fig. 3 of the main text for several Lorentzian broadening widths, from which we select the value of $\hbar\Gamma = 0.61$ eV for the dynamical calculations performed for the hydrogen chain junction in the main text. To verify that our results are relatively insensitive to this choice, we present in Figure S2: the steady-state current as a function of $\Gamma$ showing that above a value of ~0.3 eV the steady-state current weakly depends on $\Gamma$ within the relevant range, set by the density of states analysis discussed above.

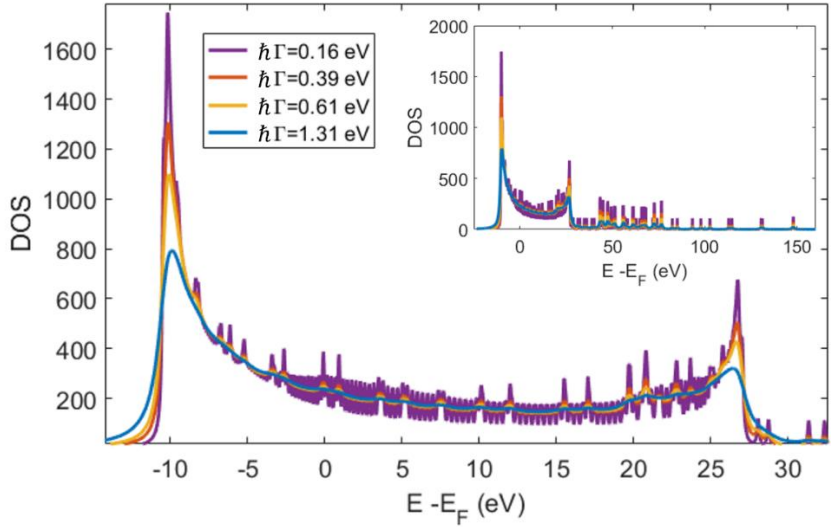

Figure S1: The artificially broadened density of states of the hydrogen chain junction model considered in Fig. 3 of the main text for four Lorentzian widths. The inset shows the full broadened spectra.

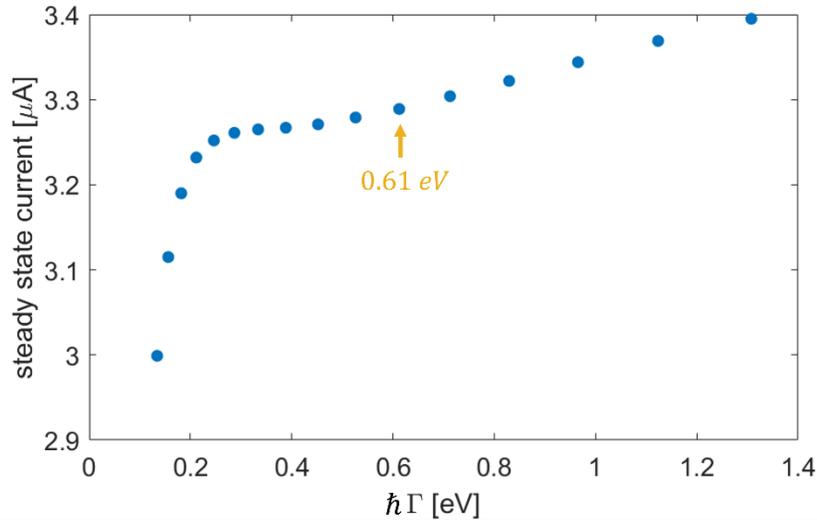

Figure S2: Steady-state current of the hydrogen chain junction model considered in Fig. 3 of the main text calculated at a bias voltage of 0.2 V with different values of the driving rate $\Gamma$. The value used in the main text is marked by the yellow arrow.

The broadened lead density of states for the graphitic junction model depicted in Fig. 1b of the main text appears in Fig. S5. The adopted value of $\hbar\Gamma = 1.09$ eV provides adequate broadening of the energy levels to results in a density of states that satisfactorily represents that of the infinite nanoribbon lead electronic structure (Fig. S6).

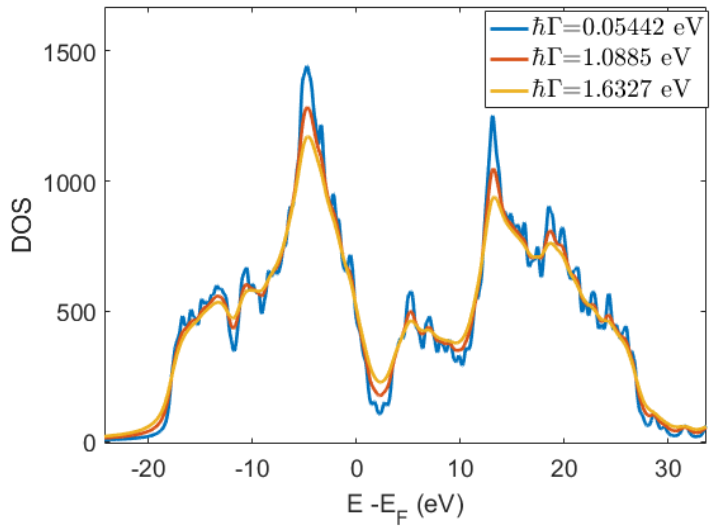

Figure S3: The artificially broadened density of states of the graphitic nano-ribbon junction model shown in Fig. 1b of the main text for three Lorentzian widths.

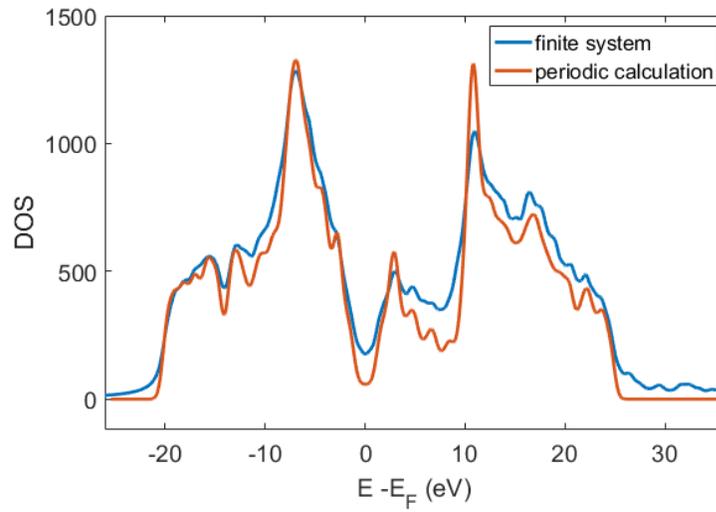

Figure S4: The artificially broadened density of states ($\hbar\Gamma = 1.09\ eV$) of the graphitic nano-ribbon junction model shown in Fig. 1b of the main text compared to that obtained using explicit periodic boundary conditions calculations.

# 3. Propagators

The driven Liouville von Neumann equation of motion for the single-particle density matrix (Eq. (S39) above) could, in principle, be propagated using one of many available propagation schemes.[6–10] However, in this particular case, care must be taken, since the propagation involves a non-unitary time evolution, where the number of electrons is not constant. Hence, we have implemented and tested the numerical stability of several propagation schemes that can handle non-unitary propagation, including the fourth-order Runge-Kutta, Heun, and Ralston methods using a solver based on Butcher tableaux, as well as the implicit Euler, trapezoid, and midpoint methods with adaptive time steps[11]. For the latter three methods, the implicit equations are solved utilizing a simple fixed-point algorithm that automatically adapts the time step according to the number of iterations needed to satisfy the implicit equations in the fixed-point loop. More details regarding the algorithm are provided in the main text.

Figure S5 shows the time-dependent current obtained using the various propagation schemes considered for the hydrogen chain of Fig. 3 of the main text under a bias voltage of 0.3 V and using a driving rate of $\hbar\Gamma = 0.61\ eV$. Among all non-unitary propagation methods considered, the currents dynamics obtained using the Heun and Ralston methods (not shown in the figure) diverged within 0.2 fs with the chosen parameters, whereas the implicit Euler demonstrated a superior stability for propagating the DLvN EOM. Therefore, this method was adopted to perform the calculations presented in the main text.

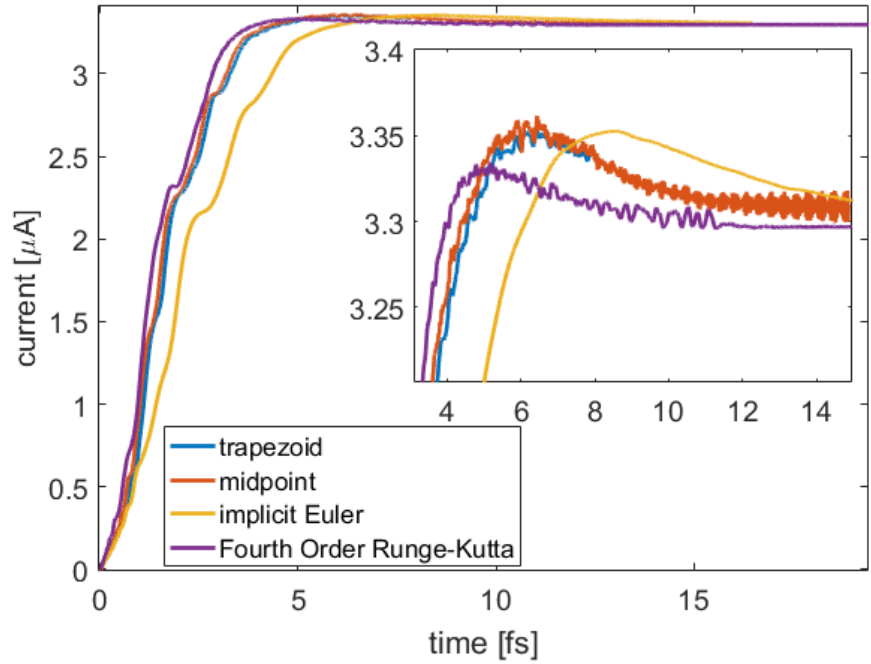

Figure S5: Comparison of the performance of different propagation schemes including the fourth-order Runge-Kutta (using a solver based on Butcher tableaux), implicit Euler, trapezoid, and the midpoint propagators. The model system used for this comparison is the hydrogen chain studied in Fig. 3 in the main text, under a bias voltage of 0.3 V and a driving rate of $\hbar \Gamma = 0.61$ eV.

## 4. Total current calculation

Within the DLvN scheme the construction of the Kohn-Sham Hamiltonian and the propagation are performed in the site (atomic basis) representation (Eq. (S39)), whereas the boundary conditions are applied in the state representation. For calculating the electronic current flowing through the extended molecule section, however, one must transform to the block-diagonal representation. The reason for this is that in the site representation the off-diagonal overlap blocks mix the EM and driven lead section bases, making it unfeasible to calculate the pure EM current contribution. While in the state representation this problem is remedied, one should recall that in this representation we lack an equation for $\dot{\tilde{\tilde{\mathcal{P}}}}$, having instead an equation for $\dot{\tilde{\mathcal{P}}}$. This, in turn, prohibits the actual current calculation. In the block-diagonal basis, these two problems are eliminated as, on the one hand the off-diagonal overlap blocks vanish and on the other hand there is an explicit equation of motion for $\dot{\tilde{\mathcal{P}}}$, as demonstrated below.

We are interested in calculating the instantaneous particle current flowing between the $L$ and $R$ driven lead sections through the $EM$ region. This can be obtained from the expression for the time derivative of the particle number in this region, $\dot{N}_{EM}$. To this end, the relation $N = tr(\mathcal{P}S)$ can be used, where $N$ is the total number of electrons, $S$ is the overlap matrix, and $\mathcal{P}$ is the single particle density matrix. Note that this expression holds also for the case of fractional occupations encountered in our out-of-equilibrium calculations (see SI section 8. Total number of electrons for the case of fractional occupations). Care should be taken, however, when taking the partial trace of this expression to obtain the number of electrons in the $EM$ section, as the density matrix is complex Hermitian. In this case we have,

$$[tr_{EM}(\mathcal{P}S)]^* = \left[\sum_{\nu \in EM}\sum_{\lambda} \mathcal{P}_{\nu\lambda} S_{\lambda\nu}\right]^* = \sum_{\nu \in EM}\sum_{\lambda} \mathcal{P}^*_{\nu\lambda} S^*_{\lambda\nu} = \sum_{\nu \in EM}\sum_{\lambda} \mathcal{P}_{\lambda\nu} S_{\lambda\nu} = \sum_{\nu \in EM}\sum_{\lambda} \mathcal{P}_{\lambda\nu} S_{\nu\lambda} =$$
$$= \sum_{\nu \in EM}\sum_{\lambda} S_{\nu\lambda}\mathcal{P}_{\lambda\nu} = tr_{EM}(S\mathcal{P}), \tag{S52}$$

and

$$[tr_{EM}(S\mathcal{P})]^* = \left[\sum_{\nu \in EM}\sum_{\lambda} S_{\nu\lambda} \mathcal{P}_{\lambda\nu}\right]^* = \sum_{\nu \in EM}\sum_{\lambda} S^*_{\nu\lambda} \mathcal{P}^*_{\lambda\nu} = \sum_{\nu \in EM}\sum_{\lambda} S_{\nu\lambda} \mathcal{P}_{\nu\lambda} = \sum_{\nu \in EM}\sum_{\lambda} S_{\lambda\nu} \mathcal{P}_{\nu\lambda} =$$
$$= \sum_{\nu \in EM}\sum_{\lambda} \mathcal{P}_{\nu\lambda} S_{\lambda\nu} = tr_{EM}(\mathcal{P}S), \tag{S53}$$

where we used the fact that $\mathcal{P}$ is Hermitian and $S$ is real and symmetric. We therefore see that $tr(\mathcal{P}S)$ is not necessarily real and therefore cannot represent the particle number in the $EM$ section. This can be remedied by using the Löwdin symmetric version of the trace formula:

$$N = tr\left(S^{\frac{1}{2}}\mathcal{P}S^{\frac{1}{2}}\right), \tag{S54}$$

whose partial trace is real:

$$\left[tr_{EM}\left(S^{\frac{1}{2}}\mathcal{P}S^{\frac{1}{2}}\right)\right]^* = \left[\sum_{\mu\in EM}\sum_{\nu}\sum_{\lambda} S^{\frac{1}{2}}_{\mu\nu}\mathcal{P}_{\nu\lambda}S^{\frac{1}{2}}_{\lambda\mu}\right]^* = \sum_{\mu\in EM}\sum_{\nu}\sum_{\lambda} S^{\frac{1}{2}*}_{\mu\nu}\mathcal{P}^*_{\nu\lambda}S^{\frac{1}{2}*}_{\lambda\mu} =$$

$$= \sum_{\mu\in EM}\sum_{\nu}\sum_{\lambda} S^{\frac{1}{2}}_{\mu\nu}\mathcal{P}_{\nu\lambda}S^{\frac{1}{2}}_{\lambda\mu} = \sum_{\mu\in EM}\sum_{\nu}\sum_{\lambda} S^{\frac{1}{2}}_{\lambda\mu}\mathcal{P}_{\nu\lambda}S^{\frac{1}{2}}_{\mu\nu} = tr_{EM}\left(S^{\frac{1}{2}}\mathcal{P}S^{\frac{1}{2}}\right). \tag{S55}$$

The full trace obeys the cyclic property so we can write:

$$N = tr\left(S^{\frac{1}{2}}\mathcal{P}S^{\frac{1}{2}}\right) = tr(\mathcal{P}S) = tr(S\mathcal{P}) = \frac{1}{2}[tr(\mathcal{P}S) + tr(S\mathcal{P})]. \tag{S56}$$

The reason for introducing the last term in Eq. (S56) is that per Eqs. (S52) and (S53) its partial trace over the $EM$ section is real. This expression can be transformed to the block diagonal representation using the transformations of Eqs. (S18) and (S19) in SI section 1:

$$N(t) = \frac{1}{2}[tr(\mathcal{P}S) + tr(S\mathcal{P})] =$$

$$= \frac{1}{2}\left[tr\left(U_b\widetilde{\mathcal{P}}(t)U_b^\dagger(U_b^\dagger)^{-1}\widetilde{S}U_b^{-1}\right) + tr\left((U_b^\dagger)^{-1}\widetilde{S}U_b^{-1}U_b\widetilde{\mathcal{P}}(t)U_b^\dagger\right)\right] = \tag{S57}$$

$$= \frac{1}{2}\left[tr(U_b\widetilde{\mathcal{P}}(t)\widetilde{S}U_b^{-1}) + tr\left((U_b^\dagger)^{-1}\widetilde{S}\widetilde{\mathcal{P}}(t)U_b^\dagger\right)\right] = \frac{1}{2}\left[tr(\widetilde{\mathcal{P}}(t)\widetilde{S}U_b^{-1}U_b) + tr\left(\widetilde{S}\widetilde{\mathcal{P}}(t)U_b^\dagger(U_b^\dagger)^{-1}\right)\right] =$$

$$= \frac{1}{2}\left[tr(\widetilde{\mathcal{P}}(t)\widetilde{S}) + tr\left(\widetilde{S}\widetilde{\mathcal{P}}(t)\right)\right]$$

Since the block diagonalization transformation only rotates the $EM$ basis to make it diagonal to the $L$ and $R$ bases without modifying the latter (while readjusting the $L/EM$ and $R/EM$ Kohn-Sham Hamiltonian coupling blocks), the partial sums over the $L$, $EM$, and $R$ indices retain their spatial interpretation as belonging to the corresponding system sections. We can, therefore, write the full trace as the sum of partial traces over the separate system sections:

$$N(t) = \frac{1}{2}\left[tr(\widetilde{\mathcal{P}}(t)\widetilde{S}) + tr\left(\widetilde{S}\widetilde{\mathcal{P}}(t)\right)\right] = \tag{S58}$$

$$= \frac{1}{2}\left\{\left[tr_L(\widetilde{\mathcal{P}}(t)\widetilde{S}) + tr_L\left(\widetilde{S}\widetilde{\mathcal{P}}(t)\right)\right] + \left[tr_{EM}(\widetilde{\mathcal{P}}(t)\widetilde{S}) + tr_{EM}\left(\widetilde{S}\widetilde{\mathcal{P}}(t)\right)\right] + \left[tr_R(\widetilde{\mathcal{P}}(t)\widetilde{S}) + tr_R\left(\widetilde{S}\widetilde{\mathcal{P}}(t)\right)\right]\right\},$$

where we identify:

$$N_{\alpha=L,EM,R}(t) = \frac{1}{2}\left[tr_\alpha(\widetilde{\mathcal{P}}(t)\widetilde{S}) + tr_\alpha\left(\widetilde{S}\widetilde{\mathcal{P}}(t)\right)\right] = \frac{1}{2}tr_\alpha[\widetilde{\mathcal{P}}(t)\widetilde{S} + \widetilde{S}\widetilde{\mathcal{P}}(t)] \tag{S59}$$

as the instantaneous number of electrons in the different sections.

Since $\widetilde{S}$ is time independent (for fixed nuclei positions) we may express the temporal change in the number of particles in the *EM* section as:

$$\dot{N}_{EM} = \frac{1}{2} tr_{EM}\left(\dot{\tilde{\mathcal{P}}}\tilde{\mathcal{S}} + \tilde{\mathcal{S}}\dot{\tilde{\mathcal{P}}}\right). \tag{S60}$$

As explained above, to obtain an expression for the average total current flowing through the *EM* section we now need to write the DLvN EOM for $\tilde{\mathcal{P}}$ in the block diagonal basis. This can be achieved by transforming the DLvN EOM from the state representation, where the boundary conditions are readily applied, to the block diagonal basis (see Eq. 11 of the main text). The DLvN EOM in the state representation is given by:

$$\dot{\tilde{\tilde{\mathcal{P}}}} = -i\left[\tilde{\tilde{\mathcal{H}}}_{KS}, \tilde{\tilde{\mathcal{P}}}\right] - \Gamma\begin{pmatrix} \tilde{\tilde{\mathcal{P}}}_L - \tilde{\tilde{\mathcal{P}}}_L^0 & \frac{1}{2}\tilde{\tilde{\mathcal{P}}}_{L,EM} & \tilde{\tilde{\mathcal{P}}}_{LR} \\ \frac{1}{2}\tilde{\tilde{\mathcal{P}}}_{EM,L} & 0 & \frac{1}{2}\tilde{\tilde{\mathcal{P}}}_{EM,R} \\ \tilde{\tilde{\mathcal{P}}}_{RL} & \frac{1}{2}\tilde{\tilde{\mathcal{P}}}_{R,EM} & \tilde{\tilde{\mathcal{P}}}_R - \tilde{\tilde{\mathcal{P}}}_R^0 \end{pmatrix}. \tag{S61}$$

Using the back transformation from the state- to the block diagonal representation (the inverse transformation of Eq. (S28) of SI section 1), and the fact that $U_b$ is time-independent (see Eq. (S18) of SI section 1) so that $\dot{\tilde{\mathcal{P}}} = \frac{d}{dt}\left[U_b^{-1}\mathcal{P}(U_b^\dagger)^{-1}\right] = U_b^{-1}\dot{\mathcal{P}}(U_b^\dagger)^{-1} = \tilde{\mathcal{P}}$, we may write:

$$\dot{\tilde{\mathcal{P}}} = \tilde{\mathcal{P}} = U\dot{\tilde{\tilde{\mathcal{P}}}}U^\dagger = -iU\left[\tilde{\tilde{\mathcal{H}}}_{KS}, \tilde{\tilde{\mathcal{P}}}\right]U^\dagger - \Gamma U\begin{pmatrix} \tilde{\tilde{\mathcal{P}}}_L - \tilde{\tilde{\mathcal{P}}}_L^0 & \frac{1}{2}\tilde{\tilde{\mathcal{P}}}_{L,EM} & \tilde{\tilde{\mathcal{P}}}_{LR} \\ \frac{1}{2}\tilde{\tilde{\mathcal{P}}}_{EM,L} & 0 & \frac{1}{2}\tilde{\tilde{\mathcal{P}}}_{EM,R} \\ \tilde{\tilde{\mathcal{P}}}_{RL} & \frac{1}{2}\tilde{\tilde{\mathcal{P}}}_{R,EM} & \tilde{\tilde{\mathcal{P}}}_R - \tilde{\tilde{\mathcal{P}}}_R^0 \end{pmatrix}U^\dagger. \tag{S62}$$

We shall first transform the driving term on the right-hand side from the state- to the block diagonal representation. To this end, we rewrite it in the following form:

$$-\Gamma U\begin{pmatrix} \tilde{\tilde{\mathcal{P}}}_L - \tilde{\tilde{\mathcal{P}}}_L^0 & \frac{1}{2}\tilde{\tilde{\mathcal{P}}}_{L,EM} & \tilde{\tilde{\mathcal{P}}}_{LR} \\ \frac{1}{2}\tilde{\tilde{\mathcal{P}}}_{EM,L} & 0 & \frac{1}{2}\tilde{\tilde{\mathcal{P}}}_{EM,R} \\ \tilde{\tilde{\mathcal{P}}}_{RL} & \frac{1}{2}\tilde{\tilde{\mathcal{P}}}_{R,EM} & \tilde{\tilde{\mathcal{P}}}_R - \tilde{\tilde{\mathcal{P}}}_R^0 \end{pmatrix}U^\dagger = -\frac{\Gamma}{2}U\begin{pmatrix} I_L & 0 & 0 \\ 0 & 0 & 0 \\ 0 & 0 & I_R \end{pmatrix}\tilde{\tilde{\mathcal{P}}}U^\dagger - \frac{\Gamma}{2}U\tilde{\tilde{\mathcal{P}}}\begin{pmatrix} I_L & 0 & 0 \\ 0 & 0 & 0 \\ 0 & 0 & I_R \end{pmatrix}U^\dagger +$$

$$+\frac{\Gamma}{2}U\begin{pmatrix} I_L & 0 & 0 \\ 0 & 0 & 0 \\ 0 & 0 & I_R \end{pmatrix}\begin{pmatrix} \tilde{\tilde{\mathcal{P}}}_L^0 & 0 & 0 \\ 0 & \tilde{\tilde{\mathcal{P}}}_{EM}^0 & 0 \\ 0 & 0 & \tilde{\tilde{\mathcal{P}}}_R^0 \end{pmatrix}U^\dagger + \frac{\Gamma}{2}U\begin{pmatrix} \tilde{\tilde{\mathcal{P}}}_L^0 & 0 & 0 \\ 0 & \tilde{\tilde{\mathcal{P}}}_{EM}^0 & 0 \\ 0 & 0 & \tilde{\tilde{\mathcal{P}}}_R^0 \end{pmatrix}\begin{pmatrix} I_L & 0 & 0 \\ 0 & 0 & 0 \\ 0 & 0 & I_R \end{pmatrix}U^\dagger. \tag{S63}$$

The first term on the right-hand-side in Eq. (S63) reads:

$$U\begin{pmatrix} I_L & 0 & 0 \\ 0 & 0 & 0 \\ 0 & 0 & I_R \end{pmatrix}\tilde{\tilde{\mathcal{P}}}U^\dagger = U\begin{pmatrix} I_L & 0 & 0 \\ 0 & 0 & 0 \\ 0 & 0 & I_R \end{pmatrix}U^{-1}U\tilde{\tilde{\mathcal{P}}}U^\dagger =$$

$$= \begin{pmatrix} U_L & 0 & 0 \\ 0 & U_{EM} & 0 \\ 0 & 0 & U_R \end{pmatrix} \begin{pmatrix} I_L & 0 & 0 \\ 0 & 0 & 0 \\ 0 & 0 & I_R \end{pmatrix} \begin{pmatrix} U_L^{-1} & 0 & 0 \\ 0 & U_{EM}^{-1} & 0 \\ 0 & 0 & U_R^{-1} \end{pmatrix} \widetilde{\mathcal{P}} = \quad (S64)$$

$$= \begin{pmatrix} U_L & 0 & 0 \\ 0 & U_{EM} & 0 \\ 0 & 0 & U_R \end{pmatrix} \begin{pmatrix} U_L^{-1} & 0 & 0 \\ 0 & 0 & 0 \\ 0 & 0 & U_R^{-1} \end{pmatrix} \widetilde{\mathcal{P}} = \begin{pmatrix} I_L & 0 & 0 \\ 0 & 0 & 0 \\ 0 & 0 & I_R \end{pmatrix} \widetilde{\mathcal{P}} =$$

$$= \begin{pmatrix} I_L & 0 & 0 \\ 0 & 0 & 0 \\ 0 & 0 & I_R \end{pmatrix} \begin{pmatrix} \widetilde{\mathcal{P}}_L & \widetilde{\mathcal{P}}_{L,EM} & \widetilde{\mathcal{P}}_{L,R} \\ \widetilde{\mathcal{P}}_{EM,L} & \widetilde{\mathcal{P}}_{EM} & \widetilde{\mathcal{P}}_{EM,R} \\ \widetilde{\mathcal{P}}_{R,L} & \widetilde{\mathcal{P}}_{R,EM} & \widetilde{\mathcal{P}}_R \end{pmatrix} = \begin{pmatrix} \widetilde{\mathcal{P}}_L & \widetilde{\mathcal{P}}_{L,EM} & \widetilde{\mathcal{P}}_{L,R} \\ 0 & 0 & 0 \\ \widetilde{\mathcal{P}}_{R,L} & \widetilde{\mathcal{P}}_{R,EM} & \widetilde{\mathcal{P}}_R \end{pmatrix}.$$

Similarly, the second term on the right-hand-side reads:

$$U\widetilde{\mathcal{P}} \begin{pmatrix} I_L & 0 & 0 \\ 0 & 0 & 0 \\ 0 & 0 & I_R \end{pmatrix} U^\dagger = U\widetilde{\mathcal{P}} U^\dagger (U^\dagger)^{-1} \begin{pmatrix} I_L & 0 & 0 \\ 0 & 0 & 0 \\ 0 & 0 & I_R \end{pmatrix} U^\dagger =$$

$$= \widetilde{\mathcal{P}} \begin{pmatrix} (U_L^\dagger)^{-1} & 0 & 0 \\ 0 & (U_{EM}^\dagger)^{-1} & 0 \\ 0 & 0 & (U_R^\dagger)^{-1} \end{pmatrix} \begin{pmatrix} I_L & 0 & 0 \\ 0 & 0 & 0 \\ 0 & 0 & I_R \end{pmatrix} \begin{pmatrix} U_L^\dagger & 0 & 0 \\ 0 & U_{EM}^\dagger & 0 \\ 0 & 0 & U_R^\dagger \end{pmatrix} = \quad (S65)$$

$$= \widetilde{\mathcal{P}} \begin{pmatrix} (U_L^\dagger)^{-1} & 0 & 0 \\ 0 & (U_{EM}^\dagger)^{-1} & 0 \\ 0 & 0 & (U_R^\dagger)^{-1} \end{pmatrix} \begin{pmatrix} U_L^\dagger & 0 & 0 \\ 0 & 0 & 0 \\ 0 & 0 & U_R^\dagger \end{pmatrix} = \widetilde{\mathcal{P}} \begin{pmatrix} I_L & 0 & 0 \\ 0 & 0 & 0 \\ 0 & 0 & I_R \end{pmatrix} =$$

$$= \begin{pmatrix} \widetilde{\mathcal{P}}_L & \widetilde{\mathcal{P}}_{L,EM} & \widetilde{\mathcal{P}}_{L,R} \\ \widetilde{\mathcal{P}}_{EM,L} & \widetilde{\mathcal{P}}_{EM} & \widetilde{\mathcal{P}}_{EM,R} \\ \widetilde{\mathcal{P}}_{R,L} & \widetilde{\mathcal{P}}_{R,EM} & \widetilde{\mathcal{P}}_R \end{pmatrix} \begin{pmatrix} I_L & 0 & 0 \\ 0 & 0 & 0 \\ 0 & 0 & I_R \end{pmatrix} = \begin{pmatrix} \widetilde{\mathcal{P}}_L & 0 & \widetilde{\mathcal{P}}_{L,R} \\ \widetilde{\mathcal{P}}_{EM,L} & 0 & \widetilde{\mathcal{P}}_{EM,R} \\ \widetilde{\mathcal{P}}_{R,L} & 0 & \widetilde{\mathcal{P}}_R \end{pmatrix}.$$

The third term gives:

$$U \begin{pmatrix} I_L & 0 & 0 \\ 0 & 0 & 0 \\ 0 & 0 & I_R \end{pmatrix} \begin{pmatrix} \widetilde{\widetilde{\mathcal{P}}}_L^0 & 0 & 0 \\ 0 & \widetilde{\widetilde{\mathcal{P}}}_{EM}^0 & 0 \\ 0 & 0 & \widetilde{\widetilde{\mathcal{P}}}_R^0 \end{pmatrix} U^\dagger = \begin{pmatrix} U_L & 0 & 0 \\ 0 & U_{EM} & 0 \\ 0 & 0 & U_R \end{pmatrix} \begin{pmatrix} \widetilde{\widetilde{\mathcal{P}}}_L^0 & 0 & 0 \\ 0 & 0 & 0 \\ 0 & 0 & \widetilde{\widetilde{\mathcal{P}}}_R^0 \end{pmatrix} \begin{pmatrix} U_L^\dagger & 0 & 0 \\ 0 & U_{EM}^\dagger & 0 \\ 0 & 0 & U_R^\dagger \end{pmatrix} =$$

$$= \begin{pmatrix} U_L & 0 & 0 \\ 0 & U_{EM} & 0 \\ 0 & 0 & U_R \end{pmatrix} \begin{pmatrix} \widetilde{\widetilde{\mathcal{P}}}_L^0 U_L^\dagger & 0 & 0 \\ 0 & 0 & 0 \\ 0 & 0 & \widetilde{\widetilde{\mathcal{P}}}_R^0 U_R^\dagger \end{pmatrix} = \begin{pmatrix} U_L \widetilde{\widetilde{\mathcal{P}}}_L^0 U_L^\dagger & 0 & 0 \\ 0 & 0 & 0 \\ 0 & 0 & U_R \widetilde{\widetilde{\mathcal{P}}}_R^0 U_R^\dagger \end{pmatrix} \equiv \begin{pmatrix} \widetilde{\mathcal{P}}_L^0 & 0 & 0 \\ 0 & 0 & 0 \\ 0 & 0 & \widetilde{\mathcal{P}}_R^0 \end{pmatrix}, \quad (S66)$$

and the fourth term gives:

$$U \begin{pmatrix} \widetilde{\widetilde{\mathcal{P}}}_L^0 & 0 & 0 \\ 0 & \widetilde{\widetilde{\mathcal{P}}}_{EM}^0 & 0 \\ 0 & 0 & \widetilde{\widetilde{\mathcal{P}}}_R^0 \end{pmatrix} \begin{pmatrix} I_L & 0 & 0 \\ 0 & 0 & 0 \\ 0 & 0 & I_R \end{pmatrix} U^\dagger = \begin{pmatrix} U_L & 0 & 0 \\ 0 & U_{EM} & 0 \\ 0 & 0 & U_R \end{pmatrix} \begin{pmatrix} \widetilde{\widetilde{\mathcal{P}}}_L^0 & 0 & 0 \\ 0 & 0 & 0 \\ 0 & 0 & \widetilde{\widetilde{\mathcal{P}}}_R^0 \end{pmatrix} \begin{pmatrix} U_L^\dagger & 0 & 0 \\ 0 & U_{EM}^\dagger & 0 \\ 0 & 0 & U_R^\dagger \end{pmatrix} =$$

$$= \begin{pmatrix} U_L & 0 & 0 \\ 0 & U_{EM} & 0 \\ 0 & 0 & U_R \end{pmatrix} \begin{pmatrix} \widetilde{\widetilde{\mathcal{P}}}_L^0 U_L^\dagger & 0 & 0 \\ 0 & 0 & 0 \\ 0 & 0 & \widetilde{\widetilde{\mathcal{P}}}_R^0 U_R^\dagger \end{pmatrix} = \begin{pmatrix} U_L \widetilde{\widetilde{\mathcal{P}}}_L^0 U_L^\dagger & 0 & 0 \\ 0 & 0 & 0 \\ 0 & 0 & U_R \widetilde{\widetilde{\mathcal{P}}}_R^0 U_R^\dagger \end{pmatrix} \equiv \begin{pmatrix} \widetilde{\mathcal{P}}_L^0 & 0 & 0 \\ 0 & 0 & 0 \\ 0 & 0 & \widetilde{\mathcal{P}}_R^0 \end{pmatrix}. \quad \text{(S67)}$$

Summing all four contributions to the driving term we thus obtain:

$$-\Gamma U \begin{pmatrix} \widetilde{\widetilde{\mathcal{P}}}_L - \widetilde{\widetilde{\mathcal{P}}}_L^0 & \frac{1}{2}\widetilde{\widetilde{\mathcal{P}}}_{L,EM} & \widetilde{\widetilde{\mathcal{P}}}_{LR} \\ \frac{1}{2}\widetilde{\widetilde{\mathcal{P}}}_{EM,L} & 0 & \frac{1}{2}\widetilde{\widetilde{\mathcal{P}}}_{EM,R} \\ \widetilde{\widetilde{\mathcal{P}}}_{RL} & \frac{1}{2}\widetilde{\widetilde{\mathcal{P}}}_{R,EM} & \widetilde{\widetilde{\mathcal{P}}}_R - \widetilde{\widetilde{\mathcal{P}}}_R^0 \end{pmatrix} U^\dagger =$$

$$= -\frac{\Gamma}{2}\begin{pmatrix} \widetilde{\mathcal{P}}_L & \widetilde{\mathcal{P}}_{L,EM} & \widetilde{\mathcal{P}}_{L,R} \\ 0 & 0 & 0 \\ \widetilde{\mathcal{P}}_{R,L} & \widetilde{\mathcal{P}}_{R,EM} & \widetilde{\mathcal{P}}_R \end{pmatrix} - \frac{\Gamma}{2}\begin{pmatrix} \widetilde{\mathcal{P}}_L & 0 & \widetilde{\mathcal{P}}_{L,R} \\ \widetilde{\mathcal{P}}_{EM,L} & 0 & \widetilde{\mathcal{P}}_{EM,R} \\ \widetilde{\mathcal{P}}_{R,L} & 0 & \widetilde{\mathcal{P}}_R \end{pmatrix} + \frac{\Gamma}{2}\begin{pmatrix} \widetilde{\mathcal{P}}_L^0 & 0 & 0 \\ 0 & 0 & 0 \\ 0 & 0 & \widetilde{\mathcal{P}}_R^0 \end{pmatrix} + \frac{\Gamma}{2}\begin{pmatrix} \widetilde{\mathcal{P}}_L^0 & 0 & 0 \\ 0 & 0 & 0 \\ 0 & 0 & \widetilde{\mathcal{P}}_R^0 \end{pmatrix} =$$

$$= -\Gamma \begin{pmatrix} \widetilde{\mathcal{P}}_L - \widetilde{\mathcal{P}}_L^0 & \frac{1}{2}\widetilde{\mathcal{P}}_{L,EM} & \widetilde{\mathcal{P}}_{L,R} \\ \frac{1}{2}\widetilde{\mathcal{P}}_{EM,L} & 0 & \frac{1}{2}\widetilde{\mathcal{P}}_{EM,R} \\ \widetilde{\mathcal{P}}_{R,L} & \frac{1}{2}\widetilde{\mathcal{P}}_{R,EM} & \widetilde{\mathcal{P}}_R - \widetilde{\mathcal{P}}_R^0 \end{pmatrix}. \quad \text{(S68)}$$

We now turn to treat the commutator term in Eq. (S62):

$$U\left[\widetilde{\widetilde{\mathcal{H}}}_{KS}, \widetilde{\widetilde{\mathcal{P}}}\right] U^\dagger = U\widetilde{\widetilde{\mathcal{H}}}_{KS}\widetilde{\widetilde{\mathcal{P}}}U^\dagger - U\widetilde{\widetilde{\mathcal{P}}}\widetilde{\widetilde{\mathcal{H}}}_{KS}U^\dagger =$$

$$= UU^\dagger \widetilde{\mathcal{H}}_{KS} UU^{-1} \widetilde{\mathcal{P}}(U^\dagger)^{-1} U^\dagger - UU^{-1}\widetilde{\mathcal{P}}(U^\dagger)^{-1} U^\dagger \widetilde{\mathcal{H}}_{KS} UU^\dagger = UU^\dagger \widetilde{\mathcal{H}}_{KS} \widetilde{\mathcal{P}} - \widetilde{\mathcal{P}}\widetilde{\mathcal{H}}_{KS} UU^\dagger. \quad \text{(S69)}$$

Since $U$ obeys the relation $U^\dagger \widetilde{S} U = I$, we may write $\widetilde{S} = (U^\dagger)^{-1} U^{-1} = (UU^\dagger)^{-1}$, such that $UU^\dagger = \widetilde{S}^{-1}$. Therefore, we have:

$$U\left[\widetilde{\widetilde{\mathcal{H}}}_{KS}, \widetilde{\widetilde{\mathcal{P}}}\right] U^\dagger = \widetilde{S}^{-1}\widetilde{\mathcal{H}}_{KS}\widetilde{\mathcal{P}} - \widetilde{\mathcal{P}}\widetilde{\mathcal{H}}_{KS}\widetilde{S}^{-1}. \quad \text{(S70)}$$

Collecting the terms of $\dot{\widetilde{\mathcal{P}}}$ (Eqs. (S62), (S68), and (S70)) we therefore obtain:

$$\dot{\widetilde{\mathcal{P}}} = -i\left(\widetilde{S}^{-1}\widetilde{\mathcal{H}}_{KS}\widetilde{\mathcal{P}} - \widetilde{\mathcal{P}}\widetilde{\mathcal{H}}_{KS}\widetilde{S}^{-1}\right) - \Gamma \begin{pmatrix} \widetilde{\mathcal{P}}_L - \widetilde{\mathcal{P}}_L^0 & \frac{1}{2}\widetilde{\mathcal{P}}_{L,EM} & \widetilde{\mathcal{P}}_{L,R} \\ \frac{1}{2}\widetilde{\mathcal{P}}_{EM,L} & 0 & \frac{1}{2}\widetilde{\mathcal{P}}_{EM,R} \\ \widetilde{\mathcal{P}}_{R,L} & \frac{1}{2}\widetilde{\mathcal{P}}_{R,EM} & \widetilde{\mathcal{P}}_R - \widetilde{\mathcal{P}}_R^0 \end{pmatrix}. \quad \text{(S71)}$$

Substituting this in Eq. (S60) for the time derivative of the particle number in the *EM* section yields:

$$\dot{N}_{EM} = \frac{1}{2}tr_{EM}\left(\dot{\widetilde{\mathcal{P}}}\widetilde{S} + \widetilde{S}\dot{\widetilde{\mathcal{P}}}\right) =$$

$$= -\frac{i}{2}tr_{EM}\left[\left(\widetilde{S}^{-1}\widetilde{\mathcal{H}}_{KS}\widetilde{\mathcal{P}} - \widetilde{\mathcal{P}}\widetilde{\mathcal{H}}_{KS}\widetilde{S}^{-1}\right)\widetilde{S} + \widetilde{S}\left(\widetilde{S}^{-1}\widetilde{\mathcal{H}}_{KS}\widetilde{\mathcal{P}} - \widetilde{\mathcal{P}}\widetilde{\mathcal{H}}_{KS}\widetilde{S}^{-1}\right)\right] - \quad \text{(S72)}$$

$$-\frac{\Gamma}{2} tr_{EM} \left[ \begin{pmatrix} \widetilde{\mathcal{P}}_L - \widetilde{\mathcal{P}}_L^0 & \frac{1}{2}\widetilde{\mathcal{P}}_{L,EM} & \widetilde{\mathcal{P}}_{L,R} \\ \frac{1}{2}\widetilde{\mathcal{P}}_{EM,L} & 0 & \frac{1}{2}\widetilde{\mathcal{P}}_{EM,R} \\ \widetilde{\mathcal{P}}_{R,L} & \frac{1}{2}\widetilde{\mathcal{P}}_{R,EM} & \widetilde{\mathcal{P}}_R - \widetilde{\mathcal{P}}_R^0 \end{pmatrix} \widetilde{S} + \widetilde{S} \begin{pmatrix} \widetilde{\mathcal{P}}_L - \widetilde{\mathcal{P}}_L^0 & \frac{1}{2}\widetilde{\mathcal{P}}_{L,EM} & \widetilde{\mathcal{P}}_{L,R} \\ \frac{1}{2}\widetilde{\mathcal{P}}_{EM,L} & 0 & \frac{1}{2}\widetilde{\mathcal{P}}_{EM,R} \\ \widetilde{\mathcal{P}}_{R,L} & \frac{1}{2}\widetilde{\mathcal{P}}_{R,EM} & \widetilde{\mathcal{P}}_R - \widetilde{\mathcal{P}}_R^0 \end{pmatrix} \right].$$

The overlap matrix in the block diagonal basis assumes the following form:

$$\widetilde{S} \equiv U_b^\dagger S U_b =$$

$$= \begin{pmatrix} I_L & 0 & 0 \\ -S_{EM,L}S_L^{-1} & I_{EM} & -S_{EM,R}S_R^{-1} \\ 0 & 0 & I_R \end{pmatrix} \begin{pmatrix} S_L & S_{L,EM} & 0 \\ S_{EM,L} & S_{EM} & S_{EM,R} \\ 0 & S_{R,EM} & S_R \end{pmatrix} \begin{pmatrix} I_L & -S_L^{-1}S_{L,EM} & 0 \\ 0 & I_{EM} & 0 \\ 0 & -S_R^{-1}S_{R,EM} & I_R \end{pmatrix} =$$

$$= \begin{pmatrix} I_L & 0 & 0 \\ -S_{EM,L}S_L^{-1} & I_{EM} & -S_{EM,R}S_R^{-1} \\ 0 & 0 & I_R \end{pmatrix} \begin{pmatrix} S_L & 0 & 0 \\ S_{EM,L} & -S_{EM,L}S_L^{-1}S_{L,EM} + S_{EM} - S_{EM,R}S_R^{-1}S_{R,EM} & S_{EM,R} \\ 0 & 0 & S_R \end{pmatrix} =$$

$$= \begin{pmatrix} S_L & 0 & 0 \\ 0 & -S_{EM,L}S_L^{-1}S_{L,EM} + S_{EM} - S_{EM,R}S_R^{-1}S_{R,EM} & 0 \\ 0 & 0 & S_R \end{pmatrix} = \quad (S73)$$

$$= \begin{pmatrix} S_L & 0 & 0 \\ 0 & \widetilde{S}_{EM} & 0 \\ 0 & 0 & S_R \end{pmatrix},$$

where we have defined $\widetilde{S}_{EM} \equiv S_{EM} - S_{EM,L}S_L^{-1}S_{L,EM} - S_{EM,R}S_R^{-1}S_{R,EM}$. We can now use this to evaluate the different terms appearing in Eq. (S72). Starting from the driving term contributions we have:

$$\begin{pmatrix} \widetilde{\mathcal{P}}_L - \widetilde{\mathcal{P}}_L^0 & \frac{1}{2}\widetilde{\mathcal{P}}_{L,EM} & \widetilde{\mathcal{P}}_{L,R} \\ \frac{1}{2}\widetilde{\mathcal{P}}_{EM,L} & 0 & \frac{1}{2}\widetilde{\mathcal{P}}_{EM,R} \\ \widetilde{\mathcal{P}}_{R,L} & \frac{1}{2}\widetilde{\mathcal{P}}_{R,EM} & \widetilde{\mathcal{P}}_R - \widetilde{\mathcal{P}}_R^0 \end{pmatrix} \widetilde{S} = \begin{pmatrix} \widetilde{\mathcal{P}}_L - \widetilde{\mathcal{P}}_L^0 & \frac{1}{2}\widetilde{\mathcal{P}}_{L,EM} & \widetilde{\mathcal{P}}_{L,R} \\ \frac{1}{2}\widetilde{\mathcal{P}}_{EM,L} & 0 & \frac{1}{2}\widetilde{\mathcal{P}}_{EM,R} \\ \widetilde{\mathcal{P}}_{R,L} & \frac{1}{2}\widetilde{\mathcal{P}}_{R,EM} & \widetilde{\mathcal{P}}_R - \widetilde{\mathcal{P}}_R^0 \end{pmatrix} \begin{pmatrix} S_L & 0 & 0 \\ 0 & \widetilde{S}_{EM} & 0 \\ 0 & 0 & S_R \end{pmatrix} =$$

$$= \begin{pmatrix} (\widetilde{\mathcal{P}}_L - \widetilde{\mathcal{P}}_L^0)S_L & \frac{1}{2}\widetilde{\mathcal{P}}_{L,EM}\widetilde{S}_{EM} & \widetilde{\mathcal{P}}_{L,R}S_R \\ \frac{1}{2}\widetilde{\mathcal{P}}_{EM,L}S_L & 0 & \frac{1}{2}\widetilde{\mathcal{P}}_{EM,R}S_R \\ \widetilde{\mathcal{P}}_{R,L}S_L & \frac{1}{2}\widetilde{\mathcal{P}}_{R,EM}\widetilde{S}_{EM} & (\widetilde{\mathcal{P}}_R - \widetilde{\mathcal{P}}_R^0)S_R \end{pmatrix}, \quad (S74)$$

and

$$\tilde{S}\begin{pmatrix} \tilde{\mathcal{P}}_L - \tilde{\mathcal{P}}_L^0 & \frac{1}{2}\tilde{\mathcal{P}}_{L,EM} & \tilde{\mathcal{P}}_{L,R} \\ \frac{1}{2}\tilde{\mathcal{P}}_{EM,L} & 0 & \frac{1}{2}\tilde{\mathcal{P}}_{EM,R} \\ \tilde{\mathcal{P}}_{R,L} & \frac{1}{2}\tilde{\mathcal{P}}_{R,EM} & \tilde{\mathcal{P}}_R - \tilde{\mathcal{P}}_R^0 \end{pmatrix} = \begin{pmatrix} S_L & 0 & 0 \\ 0 & \tilde{S}_{EM} & 0 \\ 0 & 0 & S_R \end{pmatrix} \begin{pmatrix} \tilde{\mathcal{P}}_L - \tilde{\mathcal{P}}_L^0 & \frac{1}{2}\tilde{\mathcal{P}}_{L,EM} & \tilde{\mathcal{P}}_{L,R} \\ \frac{1}{2}\tilde{\mathcal{P}}_{EM,L} & 0 & \frac{1}{2}\tilde{\mathcal{P}}_{EM,R} \\ \tilde{\mathcal{P}}_{R,L} & \frac{1}{2}\tilde{\mathcal{P}}_{R,EM} & \tilde{\mathcal{P}}_R - \tilde{\mathcal{P}}_R^0 \end{pmatrix} =$$

$$= \begin{pmatrix} S_L(\tilde{\mathcal{P}}_L - \tilde{\mathcal{P}}_L^0) & \frac{1}{2}\tilde{S}_L\tilde{\mathcal{P}}_{L,EM} & S_L\tilde{\mathcal{P}}_{L,R} \\ \frac{1}{2}S_{EM}\tilde{\mathcal{P}}_{EM,L} & 0 & \frac{1}{2}S_{EM}\tilde{\mathcal{P}}_{EM,R} \\ S_R\tilde{\mathcal{P}}_{R,L} & \frac{1}{2}\tilde{S}_R\tilde{\mathcal{P}}_{R,EM} & S_R(\tilde{\mathcal{P}}_R - \tilde{\mathcal{P}}_R^0) \end{pmatrix}. \quad (S75)$$

Altogether, the driving term contribution to $\dot{N}_{EM}$ in Eq. (S72) is given by (Eqs. (S74) and (S75)):

$$-\frac{\Gamma}{2}tr_{EM}\left[\begin{pmatrix} \tilde{\mathcal{P}}_L - \tilde{\mathcal{P}}_L^0 & \frac{1}{2}\tilde{\mathcal{P}}_{L,EM} & \tilde{\mathcal{P}}_{L,R} \\ \frac{1}{2}\tilde{\mathcal{P}}_{EM,L} & 0 & \frac{1}{2}\tilde{\mathcal{P}}_{EM,R} \\ \tilde{\mathcal{P}}_{R,L} & \frac{1}{2}\tilde{\mathcal{P}}_{R,EM} & \tilde{\mathcal{P}}_R - \tilde{\mathcal{P}}_R^0 \end{pmatrix}\tilde{S} + \tilde{S}\begin{pmatrix} \tilde{\mathcal{P}}_L - \tilde{\mathcal{P}}_L^0 & \frac{1}{2}\tilde{\mathcal{P}}_{L,EM} & \tilde{\mathcal{P}}_{L,R} \\ \frac{1}{2}\tilde{\mathcal{P}}_{EM,L} & 0 & \frac{1}{2}\tilde{\mathcal{P}}_{EM,R} \\ \tilde{\mathcal{P}}_{R,L} & \frac{1}{2}\tilde{\mathcal{P}}_{R,EM} & \tilde{\mathcal{P}}_R - \tilde{\mathcal{P}}_R^0 \end{pmatrix}\right] =$$

$$= -\frac{\Gamma}{2}tr_{EM}\left[\begin{pmatrix} (\tilde{\mathcal{P}}_L - \tilde{\mathcal{P}}_L^0)S_L + S_L(\tilde{\mathcal{P}}_L - \tilde{\mathcal{P}}_L^0) & \frac{1}{2}(\tilde{\mathcal{P}}_{L,EM}\tilde{S}_{EM} + \tilde{S}_L\tilde{\mathcal{P}}_{L,EM}) & \tilde{\mathcal{P}}_{L,R}S_R + S_L\tilde{\mathcal{P}}_{L,R} \\ \frac{1}{2}(\tilde{\mathcal{P}}_{EM,L}S_L + S_{EM}\tilde{\mathcal{P}}_{EM,L}) & 0 & \frac{1}{2}(\tilde{\mathcal{P}}_{EM,R}S_R + S_{EM}\tilde{\mathcal{P}}_{EM,R}) \\ \tilde{\mathcal{P}}_{R,L}S_L + S_R\tilde{\mathcal{P}}_{R,L} & \frac{1}{2}(\tilde{\mathcal{P}}_{R,EM}\tilde{S}_{EM} + \tilde{S}_R\tilde{\mathcal{P}}_{R,EM}) & (\tilde{\mathcal{P}}_R - \tilde{\mathcal{P}}_R^0)S_R + S_R(\tilde{\mathcal{P}}_R - \tilde{\mathcal{P}}_R^0) \end{pmatrix}\right]$$

$$= -\frac{\Gamma}{2}tr_{EM}\begin{pmatrix} [(\tilde{\mathcal{P}}_L - \tilde{\mathcal{P}}_L^0), S_L]_+ & \frac{1}{2}(\tilde{\mathcal{P}}_{L,EM}\tilde{S}_{EM} + \tilde{S}_L\tilde{\mathcal{P}}_{L,EM}) & \tilde{\mathcal{P}}_{L,R}S_R + S_L\tilde{\mathcal{P}}_{L,R} \\ \frac{1}{2}(\tilde{\mathcal{P}}_{EM,L}S_L + S_{EM}\tilde{\mathcal{P}}_{EM,L}) & 0 & \frac{1}{2}(\tilde{\mathcal{P}}_{EM,R}S_R + S_{EM}\tilde{\mathcal{P}}_{EM,R}) \\ \tilde{\mathcal{P}}_{R,L}S_L + S_R\tilde{\mathcal{P}}_{R,L} & \frac{1}{2}(\tilde{\mathcal{P}}_{R,EM}\tilde{S}_{EM} + \tilde{S}_R\tilde{\mathcal{P}}_{R,EM}) & [(\tilde{\mathcal{P}}_R - \tilde{\mathcal{P}}_R^0), S_R]_+ \end{pmatrix} \quad (S76)$$

$$= 0,$$

where $[A, B]_+ = AB + BA$ is the anticommutator. We therefore see that the driving term does not contribute to the expression of the total instantaneous current flowing through the *EM* section, as expected.

Finally, we evaluate the contribution of the commutator in Eq. (S72):

$$-\frac{i}{2}tr_{EM}\left[(\tilde{S}^{-1}\tilde{\mathcal{H}}_{KS}\tilde{\mathcal{P}} - \tilde{\mathcal{P}}\tilde{\mathcal{H}}_{KS}\tilde{S}^{-1})\tilde{S} + \tilde{S}(\tilde{S}^{-1}\tilde{\mathcal{H}}_{KS}\tilde{\mathcal{P}} - \tilde{\mathcal{P}}\tilde{\mathcal{H}}_{KS}\tilde{S}^{-1})\right] =$$

$$= -\frac{i}{2}tr_{EM}\left[\tilde{S}^{-1}\tilde{\mathcal{H}}_{KS}\tilde{\mathcal{P}}\tilde{S} - \tilde{S}\tilde{\mathcal{P}}\tilde{\mathcal{H}}_{KS}\tilde{S}^{-1} + \tilde{\mathcal{H}}_{KS}\tilde{\mathcal{P}} - \tilde{\mathcal{P}}\tilde{\mathcal{H}}_{KS}\right]. \quad (S77)$$

The first term reads:

$$\tilde{S}^{-1}\tilde{\mathcal{H}}_{KS}\tilde{\mathcal{P}}\tilde{S} =$$

$$= \begin{pmatrix} S_L^{-1} & 0 & 0 \\ 0 & \tilde{S}_{EM}^{-1} & 0 \\ 0 & 0 & S_R^{-1} \end{pmatrix} \begin{pmatrix} H_L & \tilde{V}_{L,EM} & 0 \\ \tilde{V}_{EM,L} & \tilde{H}_{EM} & \tilde{V}_{EM,R} \\ 0 & \tilde{V}_{R,EM} & H_R \end{pmatrix} \begin{pmatrix} \tilde{\mathcal{P}}_L & \tilde{\mathcal{P}}_{L,EM} & \tilde{\mathcal{P}}_{LR} \\ \tilde{\mathcal{P}}_{EM,L} & \tilde{\mathcal{P}}_{EM} & \tilde{\mathcal{P}}_{EM,R} \\ \tilde{\mathcal{P}}_{R,L} & \tilde{\mathcal{P}}_{R,EM} & \tilde{\mathcal{P}}_R \end{pmatrix} \begin{pmatrix} S_L & 0 & 0 \\ 0 & \tilde{S}_{EM} & 0 \\ 0 & 0 & S_R \end{pmatrix} =$$

$$
= \begin{pmatrix} S_L^{-1}H_L & S_L^{-1}\widetilde{V}_{L,EM} & 0 \\ \widetilde{S}_{EM}^{-1}\widetilde{V}_{EM,L} & \widetilde{S}_{EM}^{-1}\widetilde{H}_{EM} & \widetilde{S}_{EM}^{-1}\widetilde{V}_{EM,R} \\ 0 & S_R^{-1}\widetilde{V}_{R,EM} & S_R^{-1}H_R \end{pmatrix} \begin{pmatrix} \widetilde{\mathcal{P}}_L S_L & \widetilde{\mathcal{P}}_{L,EM}\widetilde{S}_{EM} & \widetilde{\mathcal{P}}_{LR}S_R \\ \widetilde{\mathcal{P}}_{EM,L}S_L & \widetilde{\mathcal{P}}_{EM}\widetilde{S}_{EM} & \widetilde{\mathcal{P}}_{EM,R}S_R \\ \widetilde{\mathcal{P}}_{R,L}S_L & \widetilde{\mathcal{P}}_{R,EM}\widetilde{S}_{EM} & \widetilde{\mathcal{P}}_R S_R \end{pmatrix} =
\tag{S78}
$$

$$
= \begin{pmatrix} S_L^{-1}H_L\widetilde{\mathcal{P}}_L S_L + S_L^{-1}\widetilde{V}_{L,EM}\widetilde{\mathcal{P}}_{EM,L}S_L & S_L^{-1}H_L\widetilde{\mathcal{P}}_{L,EM}\widetilde{S}_{EM} + S_L^{-1}\widetilde{V}_{L,EM}\widetilde{\mathcal{P}}_{EM}\widetilde{S}_{EM} & S_L^{-1}H_L\widetilde{\mathcal{P}}_{LR}S_R + S_L^{-1}\widetilde{V}_{L,EM}\widetilde{\mathcal{P}}_{EM,R}S_R \\ \widetilde{S}_{EM}^{-1}\widetilde{V}_{EM,L}\widetilde{\mathcal{P}}_L S_L + \widetilde{S}_{EM}^{-1}\widetilde{H}_{EM}\widetilde{\mathcal{P}}_{EM,L}S_L + \widetilde{S}_{EM}^{-1}\widetilde{V}_{EM,R}\widetilde{\mathcal{P}}_{R,L}S_L & \widetilde{S}_{EM}^{-1}\widetilde{V}_{EM,L}\widetilde{\mathcal{P}}_{L,EM}\widetilde{S}_{EM} + \widetilde{S}_{EM}^{-1}\widetilde{H}_{EM}\widetilde{\mathcal{P}}_{EM}\widetilde{S}_{EM} + \widetilde{S}_{EM}^{-1}\widetilde{V}_{EM,R}\widetilde{\mathcal{P}}_{R,EM}\widetilde{S}_{EM} & \widetilde{S}_{EM}^{-1}\widetilde{V}_{EM,L}\widetilde{\mathcal{P}}_{LR}S_R + \widetilde{S}_{EM}^{-1}\widetilde{H}_{EM}\widetilde{\mathcal{P}}_{EM,R}S_R + \widetilde{S}_{EM}^{-1}\widetilde{V}_{EM,R}\widetilde{\mathcal{P}}_R S_R \\ S_R^{-1}\widetilde{V}_{R,EM}\widetilde{\mathcal{P}}_{EM,L}S_L + S_R^{-1}H_R\widetilde{\mathcal{P}}_{R,L}S_L & S_R^{-1}\widetilde{V}_{R,EM}\widetilde{\mathcal{P}}_{EM}\widetilde{S}_{EM} + S_R^{-1}H_R\widetilde{\mathcal{P}}_{R,EM}\widetilde{S}_{EM} & S_R^{-1}\widetilde{V}_{R,EM}\widetilde{\mathcal{P}}_{EM,R}S_R + S_R^{-1}H_R\widetilde{\mathcal{P}}_R S_R \end{pmatrix},
$$

whose *EM* block is:

$$
\left(\widetilde{S}^{-1}\widetilde{\mathcal{H}}_{KS}\widetilde{\mathcal{P}}\widetilde{S}\right)_{EM} = \widetilde{S}_{EM}^{-1}\widetilde{V}_{EM,L}\widetilde{\mathcal{P}}_{L,EM}\widetilde{S}_{EM} + \widetilde{S}_{EM}^{-1}\widetilde{H}_{EM}\widetilde{\mathcal{P}}_{EM}\widetilde{S}_{EM} + \widetilde{S}_{EM}^{-1}\widetilde{V}_{EM,R}\widetilde{\mathcal{P}}_{R,EM}\widetilde{S}_{EM}.
\tag{S79}
$$

The second term reads:

$$
\widetilde{S}\widetilde{\mathcal{P}}\widetilde{\mathcal{H}}_{KS}\widetilde{S}^{-1} =
$$

$$
= \begin{pmatrix} S_L & 0 & 0 \\ 0 & \widetilde{S}_{EM} & 0 \\ 0 & 0 & S_R \end{pmatrix} \begin{pmatrix} \widetilde{\mathcal{P}}_L & \widetilde{\mathcal{P}}_{L,EM} & \widetilde{\mathcal{P}}_{LR} \\ \widetilde{\mathcal{P}}_{EM,L} & \widetilde{\mathcal{P}}_{EM} & \widetilde{\mathcal{P}}_{EM,R} \\ \widetilde{\mathcal{P}}_{R,L} & \widetilde{\mathcal{P}}_{R,EM} & \widetilde{\mathcal{P}}_R \end{pmatrix} \begin{pmatrix} H_L & \widetilde{V}_{L,EM} & 0 \\ \widetilde{V}_{EM,L} & \widetilde{H}_{EM} & \widetilde{V}_{EM,R} \\ 0 & \widetilde{V}_{R,EM} & H_R \end{pmatrix} \begin{pmatrix} S_L^{-1} & 0 & 0 \\ 0 & \widetilde{S}_{EM}^{-1} & 0 \\ 0 & 0 & S_R^{-1} \end{pmatrix} =
$$

$$
= \begin{pmatrix} S_L\widetilde{\mathcal{P}}_L & S_L\widetilde{\mathcal{P}}_{L,EM} & S_L\widetilde{\mathcal{P}}_{LR} \\ \widetilde{S}_{EM}\widetilde{\mathcal{P}}_{EM,L} & \widetilde{S}_{EM}\widetilde{\mathcal{P}}_{EM} & \widetilde{S}_{EM}\widetilde{\mathcal{P}}_{EM,R} \\ S_R\widetilde{\mathcal{P}}_{R,L} & S_R\widetilde{\mathcal{P}}_{R,EM} & S_R\widetilde{\mathcal{P}}_R \end{pmatrix} \begin{pmatrix} H_L S_L^{-1} & \widetilde{V}_{L,EM}\widetilde{S}_{EM}^{-1} & 0 \\ \widetilde{V}_{EM,L}S_L^{-1} & \widetilde{H}_{EM}\widetilde{S}_{EM}^{-1} & \widetilde{V}_{EM,R}S_R^{-1} \\ 0 & \widetilde{V}_{R,EM}\widetilde{S}_{EM}^{-1} & H_R S_R^{-1} \end{pmatrix} =
\tag{S80}
$$

$$
= \begin{pmatrix} S_L\widetilde{\mathcal{P}}_L H_L S_L^{-1} + S_L\widetilde{\mathcal{P}}_{L,EM}\widetilde{V}_{EM,L}S_L^{-1} & S_L\widetilde{\mathcal{P}}_L\widetilde{V}_{L,EM}\widetilde{S}_{EM}^{-1} + S_L\widetilde{\mathcal{P}}_{L,EM}\widetilde{H}_{EM}\widetilde{S}_{EM}^{-1} + S_L\widetilde{\mathcal{P}}_{LR}\widetilde{V}_{R,EM}\widetilde{S}_{EM}^{-1} & S_L\widetilde{\mathcal{P}}_{L,EM}\widetilde{V}_{EM,R}S_R^{-1} + S_L\widetilde{\mathcal{P}}_{LR}H_R S_R^{-1} \\ \widetilde{S}_{EM}\widetilde{\mathcal{P}}_{EM,L}H_L S_L^{-1} + \widetilde{S}_{EM}\widetilde{\mathcal{P}}_{EM}\widetilde{V}_{EM,L}S_L^{-1} & \widetilde{S}_{EM}\widetilde{\mathcal{P}}_{EM,L}\widetilde{V}_{L,EM}\widetilde{S}_{EM}^{-1} + \widetilde{S}_{EM}\widetilde{\mathcal{P}}_{EM}\widetilde{H}_{EM}\widetilde{S}_{EM}^{-1} + \widetilde{S}_{EM}\widetilde{\mathcal{P}}_{EM,R}\widetilde{V}_{R,EM}\widetilde{S}_{EM}^{-1} & \widetilde{S}_{EM}\widetilde{\mathcal{P}}_{EM}\widetilde{V}_{EM,R}S_R^{-1} + \widetilde{S}_{EM}\widetilde{\mathcal{P}}_{EM,R}H_R S_R^{-1} \\ S_R\widetilde{\mathcal{P}}_{R,L}H_L S_L^{-1} + S_R\widetilde{\mathcal{P}}_{R,EM}\widetilde{V}_{EM,L}S_L^{-1} & S_R\widetilde{\mathcal{P}}_{R,L}\widetilde{V}_{L,EM}\widetilde{S}_{EM}^{-1} + S_R\widetilde{\mathcal{P}}_{R,EM}\widetilde{H}_{EM}\widetilde{S}_{EM}^{-1} + S_R\widetilde{\mathcal{P}}_R\widetilde{V}_{R,EM}\widetilde{S}_{EM}^{-1} & S_R\widetilde{\mathcal{P}}_{R,EM}\widetilde{V}_{EM,R}S_R^{-1} + S_R\widetilde{\mathcal{P}}_R H_R S_R^{-1} \end{pmatrix},
$$

whose *EM* block is:

$$
\left(\widetilde{S}\widetilde{\mathcal{P}}\widetilde{\mathcal{H}}_{KS}\widetilde{S}^{-1}\right)_{EM} = \widetilde{S}_{EM}\widetilde{\mathcal{P}}_{EM,L}\widetilde{V}_{L,EM}\widetilde{S}_{EM}^{-1} + \widetilde{S}_{EM}\widetilde{\mathcal{P}}_{EM}\widetilde{H}_{EM}\widetilde{S}_{EM}^{-1} + \widetilde{S}_{EM}\widetilde{\mathcal{P}}_{EM,R}\widetilde{V}_{R,EM}\widetilde{S}_{EM}^{-1}.
\tag{S81}
$$

The third term reads:

$$
\widetilde{\mathcal{H}}_{KS}\widetilde{\mathcal{P}} = \begin{pmatrix} H_L & \widetilde{V}_{L,EM} & 0 \\ \widetilde{V}_{EM,L} & \widetilde{H}_{EM} & \widetilde{V}_{EM,R} \\ 0 & \widetilde{V}_{R,EM} & H_R \end{pmatrix} \begin{pmatrix} \widetilde{\mathcal{P}}_L & \widetilde{\mathcal{P}}_{L,EM} & \widetilde{\mathcal{P}}_{LR} \\ \widetilde{\mathcal{P}}_{EM,L} & \widetilde{\mathcal{P}}_{EM} & \widetilde{\mathcal{P}}_{EM,R} \\ \widetilde{\mathcal{P}}_{R,L} & \widetilde{\mathcal{P}}_{R,EM} & \widetilde{\mathcal{P}}_R \end{pmatrix} =
\tag{S82}
$$

$$
= \begin{pmatrix} H_L\widetilde{\mathcal{P}}_L + \widetilde{V}_{L,EM}\widetilde{\mathcal{P}}_{EM,L} & H_L\widetilde{\mathcal{P}}_{L,EM} + \widetilde{V}_{L,EM}\widetilde{\mathcal{P}}_{EM} & H_L\widetilde{\mathcal{P}}_{LR} + \widetilde{V}_{L,EM}\widetilde{\mathcal{P}}_{EM,R} \\ \widetilde{V}_{EM,L}\widetilde{\mathcal{P}}_L + \widetilde{H}_{EM}\widetilde{\mathcal{P}}_{EM,L} + \widetilde{V}_{EM,R}\widetilde{\mathcal{P}}_{R,L} & \widetilde{V}_{EM,L}\widetilde{\mathcal{P}}_{L,EM} + \widetilde{H}_{EM}\widetilde{\mathcal{P}}_{EM} + \widetilde{V}_{EM,R}\widetilde{\mathcal{P}}_{R,EM} & \widetilde{V}_{EM,L}\widetilde{\mathcal{P}}_{LR} + \widetilde{H}_{EM}\widetilde{\mathcal{P}}_{EM,R} + \widetilde{V}_{EM,R}\widetilde{\mathcal{P}}_R \\ \widetilde{V}_{R,EM}\widetilde{\mathcal{P}}_{EM,L} + H_R\widetilde{\mathcal{P}}_{R,L} & \widetilde{V}_{R,EM}\widetilde{\mathcal{P}}_{EM} + H_R\widetilde{\mathcal{P}}_{R,EM} & \widetilde{V}_{R,EM}\widetilde{\mathcal{P}}_{EM,R} + H_R\widetilde{\mathcal{P}}_R \end{pmatrix},
$$

whose *EM* block is:

$$
\left(\widetilde{\mathcal{H}}_{KS}\widetilde{\mathcal{P}}\right)_{EM} = \widetilde{V}_{EM,L}\widetilde{\mathcal{P}}_{L,EM} + \widetilde{H}_{EM}\widetilde{\mathcal{P}}_{EM} + \widetilde{V}_{EM,R}\widetilde{\mathcal{P}}_{R,EM}
\tag{S83}
$$

The fourth term reads:

$$
\widetilde{\mathcal{P}}\widetilde{\mathcal{H}}_{KS} = \begin{pmatrix} \widetilde{\mathcal{P}}_L & \widetilde{\mathcal{P}}_{L,EM} & \widetilde{\mathcal{P}}_{LR} \\ \widetilde{\mathcal{P}}_{EM,L} & \widetilde{\mathcal{P}}_{EM} & \widetilde{\mathcal{P}}_{EM,R} \\ \widetilde{\mathcal{P}}_{R,L} & \widetilde{\mathcal{P}}_{R,EM} & \widetilde{\mathcal{P}}_R \end{pmatrix} \begin{pmatrix} H_L & \widetilde{V}_{L,EM} & 0 \\ \widetilde{V}_{EM,L} & \widetilde{H}_{EM} & \widetilde{V}_{EM,R} \\ 0 & \widetilde{V}_{R,EM} & H_R \end{pmatrix} =
\tag{S84}
$$

$$= \begin{pmatrix} \widetilde{\mathcal{P}}_L H_L + \widetilde{\mathcal{P}}_{L,EM} \widetilde{V}_{EM,L} & \widetilde{\mathcal{P}}_L \widetilde{V}_{L,EM} + \widetilde{\mathcal{P}}_{L,EM} \widetilde{H}_{EM} + \widetilde{\mathcal{P}}_{LR} \widetilde{V}_{R,EM} & \widetilde{\mathcal{P}}_{L,EM} \widetilde{V}_{EM,R} + \widetilde{\mathcal{P}}_{LR} H_R \\ \widetilde{\mathcal{P}}_{EM,L} H_L + \widetilde{\mathcal{P}}_{EM} \widetilde{V}_{EM,L} & \widetilde{\mathcal{P}}_{EM,L} \widetilde{V}_{L,EM} + \widetilde{\mathcal{P}}_{EM} \widetilde{H}_{EM} + \widetilde{\mathcal{P}}_{EM,R} \widetilde{V}_{R,EM} & \widetilde{\mathcal{P}}_{EM} \widetilde{V}_{EM,R} + \widetilde{\mathcal{P}}_{EM,R} H_R \\ \widetilde{\mathcal{P}}_{R,L} H_L + \widetilde{\mathcal{P}}_{R,EM} \widetilde{V}_{EM,L} & \widetilde{\mathcal{P}}_{R,L} \widetilde{V}_{L,EM} + \widetilde{\mathcal{P}}_{R,EM} \widetilde{H}_{EM} + \widetilde{\mathcal{P}}_R \widetilde{V}_{R,EM} & \widetilde{\mathcal{P}}_{R,EM} \widetilde{V}_{EM,R} + \widetilde{\mathcal{P}}_R H_R \end{pmatrix},$$

whose *EM* block is:

$$\left(\widetilde{\mathcal{P}}\widetilde{\mathcal{H}}_{KS}\right)_{EM} = \widetilde{\mathcal{P}}_{EM,L} \widetilde{V}_{L,EM} + \widetilde{\mathcal{P}}_{EM} \widetilde{H}_{EM} + \widetilde{\mathcal{P}}_{EM,R} \widetilde{V}_{R,EM} \tag{S85}$$

Collecting all terms in Eqs. (S79), (S81), (S83), (S85) we may write the *EM* block of the commutator term in Eq. (S77) as follows:

$$-\frac{i}{2} tr_{EM} \left[ \widetilde{S}^{-1} \widetilde{\mathcal{H}}_{KS} \widetilde{\mathcal{P}} \widetilde{S} - \widetilde{S} \widetilde{\mathcal{P}} \widetilde{\mathcal{H}}_{KS} \widetilde{S}^{-1} + \widetilde{\mathcal{H}}_{KS} \widetilde{\mathcal{P}} - \widetilde{\mathcal{P}} \widetilde{\mathcal{H}}_{KS} \right] =$$

$$= -\frac{i}{2} tr_{EM} [\widetilde{S}_{EM}^{-1} \widetilde{V}_{EM,L} \widetilde{\mathcal{P}}_{L,EM} \widetilde{S}_{EM} + \widetilde{S}_{EM}^{-1} \widetilde{H}_{EM} \widetilde{\mathcal{P}}_{EM} \widetilde{S}_{EM} + \widetilde{S}_{EM}^{-1} \widetilde{V}_{EM,R} \widetilde{\mathcal{P}}_{R,EM} \widetilde{S}_{EM} - \tag{S86}$$

$$-\widetilde{S}_{EM} \widetilde{\mathcal{P}}_{EM,L} \widetilde{V}_{L,EM} \widetilde{S}_{EM}^{-1} - \widetilde{S}_{EM} \widetilde{\mathcal{P}}_{EM} \widetilde{H}_{EM} \widetilde{S}_{EM}^{-1} - \widetilde{S}_{EM} \widetilde{\mathcal{P}}_{EM,R} \widetilde{V}_{R,EM} \widetilde{S}_{EM}^{-1} +$$

$$+ \widetilde{V}_{EM,L} \widetilde{\mathcal{P}}_{L,EM} + \widetilde{H}_{EM} \widetilde{\mathcal{P}}_{EM} + \widetilde{V}_{EM,R} \widetilde{\mathcal{P}}_{R,EM} - \widetilde{\mathcal{P}}_{EM,L} \widetilde{V}_{L,EM} - \widetilde{\mathcal{P}}_{EM} \widetilde{H}_{EM} - \widetilde{\mathcal{P}}_{EM,R} \widetilde{V}_{R,EM}] =$$

This can be reordered as follows:

$$= -\frac{i}{2} tr_{EM} \left( \widetilde{S}_{EM}^{-1} \widetilde{H}_{EM} \widetilde{\mathcal{P}}_{EM} \widetilde{S}_{EM} - \widetilde{S}_{EM} \widetilde{\mathcal{P}}_{EM} \widetilde{H}_{EM} \widetilde{S}_{EM}^{-1} + \widetilde{H}_{EM} \widetilde{\mathcal{P}}_{EM} - \widetilde{\mathcal{P}}_{EM} \widetilde{H}_{EM} \right) -$$

$$-\frac{i}{2} tr_{EM} \left( \widetilde{S}_{EM}^{-1} \widetilde{V}_{EM,L} \widetilde{\mathcal{P}}_{L,EM} \widetilde{S}_{EM} - \widetilde{S}_{EM} \widetilde{\mathcal{P}}_{EM,L} \widetilde{V}_{L,EM} \widetilde{S}_{EM}^{-1} + \widetilde{V}_{EM,L} \widetilde{\mathcal{P}}_{L,EM} - \widetilde{\mathcal{P}}_{EM,L} \widetilde{V}_{L,EM} \right) - \tag{S87}$$

$$-\frac{i}{2} tr_{EM} \left( \widetilde{S}_{EM}^{-1} \widetilde{V}_{EM,R} \widetilde{\mathcal{P}}_{R,EM} \widetilde{S}_{EM} - \widetilde{S}_{EM} \widetilde{\mathcal{P}}_{EM,R} \widetilde{V}_{R,EM} \widetilde{S}_{EM}^{-1} + \widetilde{V}_{EM,R} \widetilde{\mathcal{P}}_{R,EM} - \widetilde{\mathcal{P}}_{EM,R} \widetilde{V}_{R,EM} \right).$$

In the first row of Eq. (S87), the partial *EM* trace obeys the cyclic property, since all the matrices involved are square matrices of dimension *EM*. Therefore, the contribution of this term vanishes:

$$tr_{EM} \left( \widetilde{S}_{EM}^{-1} \widetilde{H}_{EM} \widetilde{\mathcal{P}}_{EM} \widetilde{S}_{EM} - \widetilde{S}_{EM} \widetilde{\mathcal{P}}_{EM} \widetilde{H}_{EM} \widetilde{S}_{EM}^{-1} + \widetilde{H}_{EM} \widetilde{\mathcal{P}}_{EM} - \widetilde{\mathcal{P}}_{EM} \widetilde{H}_{EM} \right) =$$

$$= tr_{EM} \left( \widetilde{H}_{EM} \widetilde{\mathcal{P}}_{EM} \widetilde{S}_{EM} \widetilde{S}_{EM}^{-1} - \widetilde{\mathcal{P}}_{EM} \widetilde{H}_{EM} \widetilde{S}_{EM}^{-1} \widetilde{S}_{EM} + \widetilde{H}_{EM} \widetilde{\mathcal{P}}_{EM} - \widetilde{\mathcal{P}}_{EM} \widetilde{H}_{EM} \right) = \tag{S88}$$

$$= tr_{EM} \left( \widetilde{H}_{EM} \widetilde{\mathcal{P}}_{EM} - \widetilde{\mathcal{P}}_{EM} \widetilde{H}_{EM} + \widetilde{H}_{EM} \widetilde{\mathcal{P}}_{EM} - \widetilde{\mathcal{P}}_{EM} \widetilde{H}_{EM} \right) =$$

$$= 2 tr_{EM} \left( \widetilde{H}_{EM} \widetilde{\mathcal{P}}_{EM} - \widetilde{\mathcal{P}}_{EM} \widetilde{H}_{EM} \right) = 2 tr_{EM} \left( \widetilde{\mathcal{P}}_{EM} \widetilde{H}_{EM} - \widetilde{\mathcal{P}}_{EM} \widetilde{H}_{EM} \right) = 0.$$

Looking next into the first term on the second row of Eq. (S87) we can write:

$$tr_{EM} \left( \widetilde{S}_{EM}^{-1} \widetilde{V}_{EM,L} \widetilde{\mathcal{P}}_{L,EM} \widetilde{S}_{EM} \right) = \sum_{i \in EM} \sum_{j \in EM} \sum_{k \in L} \sum_{l \in EM} \left( \widetilde{S}_{EM}^{-1} \right)_{ij} \left( \widetilde{V}_{EM,L} \right)_{jk} \left( \widetilde{\mathcal{P}}_{L,EM} \right)_{kl} \left( \widetilde{S}_{EM} \right)_{li} =$$

$$= \sum_{j \in EM} \sum_{k \in L} \sum_{l \in EM} \sum_{i \in EM} \left( \widetilde{V}_{EM,L} \right)_{jk} \left( \widetilde{\mathcal{P}}_{L,EM} \right)_{kl} \left( \widetilde{S}_{EM} \right)_{li} \left( \widetilde{S}_{EM}^{-1} \right)_{ij} = \tag{S89}$$

$$= tr_{EM} \left( \widetilde{V}_{EM,L} \widetilde{\mathcal{P}}_{L,EM} \widetilde{S}_{EM} \widetilde{S}_{EM}^{-1} \right) = tr_{EM} \left( \widetilde{V}_{EM,L} \widetilde{\mathcal{P}}_{L,EM} \right),$$

where in the second row we switched the summation order and changed the orders of the summed elements. Similarly, for the second term on the second row of Eq. (S87) we have:

$$tr_{EM}(\tilde{S}_{EM}\tilde{\mathcal{P}}_{EM,L}\tilde{V}_{L,EM}\tilde{S}_{EM}^{-1}) = tr_{EM}(\tilde{\mathcal{P}}_{EM,L}\tilde{V}_{L,EM}), \tag{S90}$$

and for the two first terms in the third row:

$$tr_{EM}(\tilde{S}_{EM}^{-1}\tilde{V}_{EM,R}\tilde{\mathcal{P}}_{R,EM}\tilde{S}_{EM} - \tilde{S}_{EM}\tilde{\mathcal{P}}_{EM,R}\tilde{V}_{R,EM}\tilde{S}_{EM}^{-1}) = tr_{EM}(\tilde{V}_{EM,R}\tilde{\mathcal{P}}_{R,EM} - \tilde{\mathcal{P}}_{EM,R}\tilde{V}_{R,EM}). \tag{S91}$$

Collecting all terms in Eqs. (S87)-(S91) we have:

$$\dot{N}_{EM} = -i \cdot tr_{EM}(\tilde{V}_{EM,L}\tilde{\mathcal{P}}_{L,EM} - \tilde{\mathcal{P}}_{EM,L}\tilde{V}_{L,EM}) - i \cdot tr_{EM}(\tilde{V}_{EM,R}\tilde{\mathcal{P}}_{R,EM} - \tilde{\mathcal{P}}_{EM,R}\tilde{V}_{R,EM}). \tag{S92}$$

Using the fact that the density matrix and KS Hamiltonian matrix are Hermitian we can further write:

$$[tr_{EM}(\tilde{V}_{EM,L}\tilde{\mathcal{P}}_{L,EM})]^* = \left[\sum_{i \in EM}\sum_{j \in L}(\tilde{V}_{EM,L})_{ij}(\tilde{\mathcal{P}}_{L,EM})_{ji}\right]^* = \sum_{i \in EM}\sum_{j \in L}(\tilde{V}_{EM,L})_{ij}^*(\tilde{\mathcal{P}}_{L,EM})_{ji}^* =$$

$$= \sum_{i \in EM}\sum_{j \in L}(\tilde{V}_{L,EM})_{ji}(\tilde{\mathcal{P}}_{EM,L})_{ij} = \sum_{i \in EM}\sum_{j \in L}(\tilde{\mathcal{P}}_{EM,L})_{ij}(\tilde{V}_{L,EM})_{ji} = tr_{EM}(\tilde{\mathcal{P}}_{EM,L}\tilde{V}_{L,EM}). \tag{S93}$$

Similarly, we can write $[tr_{EM}(\tilde{V}_{EM,R}\tilde{\mathcal{P}}_{R,EM})]^* = tr_{EM}(\tilde{\mathcal{P}}_{EM,R}\tilde{V}_{R,EM})$, so that:

$$\dot{N}_{EM} = -i \cdot tr_{EM}(\tilde{V}_{EM,L}\tilde{\mathcal{P}}_{L,EM} - \tilde{\mathcal{P}}_{EM,L}\tilde{V}_{L,EM}) - i \cdot tr_{EM}(\tilde{V}_{EM,R}\tilde{\mathcal{P}}_{R,EM} - \tilde{\mathcal{P}}_{EM,R}\tilde{V}_{R,EM}) =$$

$$= -i \cdot tr_{EM}(\tilde{\mathcal{P}}_{EM,L}^*\tilde{V}_{L,EM}^* - \tilde{\mathcal{P}}_{EM,L}\tilde{V}_{L,EM}) - i \cdot tr_{EM}(\tilde{\mathcal{P}}_{EM,R}^*\tilde{V}_{R,EM}^* - \tilde{\mathcal{P}}_{EM,R}\tilde{V}_{R,EM}) =$$

$$= -i \cdot tr_{EM}[2i \cdot Im(\tilde{\mathcal{P}}_{EM,L}\tilde{V}_{L,EM})] - i \cdot tr_{EM}[2i \cdot Im(\tilde{\mathcal{P}}_{EM,R}\tilde{V}_{R,EM})] = \tag{S94}$$

$$= 2 \cdot tr_{EM}[Im(\tilde{\mathcal{P}}_{EM,L}\tilde{V}_{L,EM})] + 2 \cdot tr_{EM}[Im(\tilde{\mathcal{P}}_{EM,R}\tilde{V}_{R,EM})] =$$

$$= 2 \cdot Im[tr_{EM}(\tilde{\mathcal{P}}_{EM,L}\tilde{V}_{L,EM})] + 2 \cdot Im[tr_{EM}(\tilde{\mathcal{P}}_{EM,R}\tilde{V}_{R,EM})].$$

We can thus identify the first term in the last line of Eq. (S94) as the current flowing from the $L$ driven lead into the $EM$ section and the second term as the current flowing from the $R$ driven lead into the $EM$ section:

$$J_{L \to EM} = 2 \cdot Im[tr_{EM}(\tilde{\mathcal{P}}_{EM,L}\tilde{V}_{L,EM})], \tag{S95}$$

and

$$J_{R \to EM} = 2 \cdot Im[tr_{EM}(\tilde{\mathcal{P}}_{EM,R}\tilde{V}_{R,EM})]. \tag{S96}$$

Accordingly, the instantaneous average total current flowing through the $EM$ section at time $t$ is:

$$J(t) = 0.5(J_{L \to EM}(t) + J_{EM \to R}(t)) = Im\{tr_{EM}[\tilde{\mathcal{P}}_{EM,L}(t)\tilde{V}_{L,EM}(t)]\} - Im\{tr_{EM}[\tilde{\mathcal{P}}_{EM,R}(t)\tilde{V}_{R,EM}(t)]\}$$

$$= Im\{tr_{EM}[\tilde{\mathcal{P}}_{EM,L}(t)\tilde{V}_{L,EM}(t) - \tilde{\mathcal{P}}_{EM,R}(t)\tilde{V}_{R,EM}(t)]\}, \tag{S97}$$

which is the final expression that we use for the evaluation of the current in the block-diagonal representation.

# 5. Effects of the basis sets and size of leads on the calculated current

In all hydrogen chain benchmark calculations presented in the main text we used 180 atom lead models in combination with the STO-3G atomic centered basis-set to represent the Kohn-Sham orbitals and the PBE functional approximation. To evaluate the sensitivity of our results towards these choices we present in Figure S6a the steady-state current versus the applied bias voltage for lead models of 120, 180, and 240 hydrogen atoms calculated at the PBE/6-31G** level of theory with $\hbar\Gamma = 0.92, 0.61,$ and $0.46$ eV, respectively. Note that driving rate is varied according to the lead model dimensions to preserve the density of states plot of the lead. The results show that our choice of 180 atom lead model is well converged with the lead model size, with the largest difference between the current calculated using the 180 and 240 hydrogen chain lead models being ~15%. Figure S6b presents a basis-set sensitivity analysis for the molecular section. Using 180-hydrogen atom chain lead models with $\hbar\Gamma = 0.61\ eV$, we compare the steady-state current evaluated with three choices of basis-sets for the extended molecule region. The results indicate that our choice of 6-31G** basis-set provides currents converged up to 0.3% in the voltage range studied.

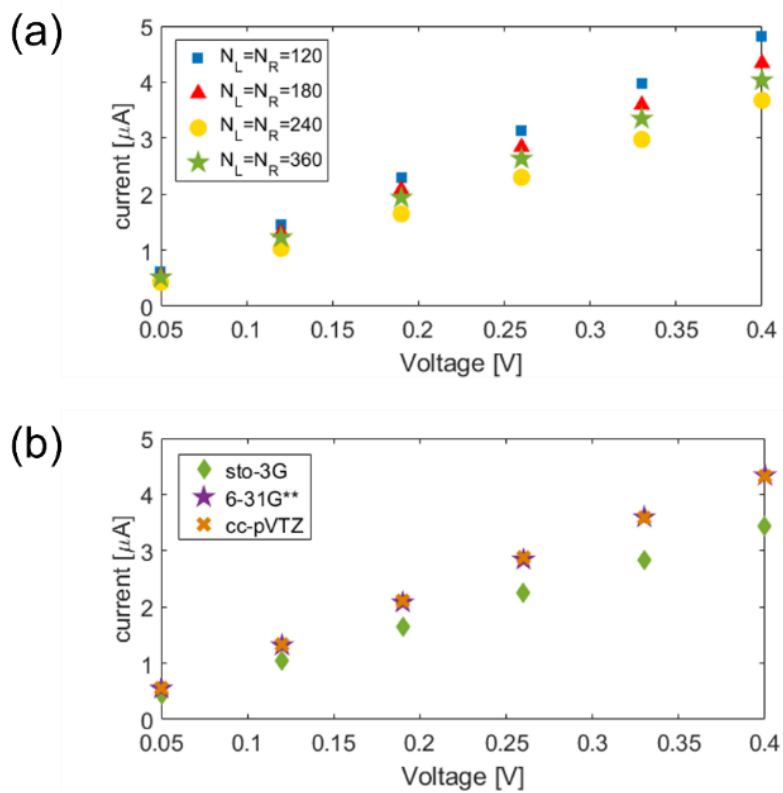

Figure S6: Analysis of the steady-state current sensitivity to the lead model size and the choice of lead basis-set. (a) Steady-state current vs. bias voltage calculated at the PBE/6-31G** level of theory for the *EM* section and PBE/STO-3G for the leads of hydrogen chain junction models with lead sections of 120 (blue squares), 180 (red triangles), 240 (yellow circles), and 360 (green stars) atoms (using $\hbar\Gamma = 0.92, 0.61, 0.46,$ and $0.31$ eV, respectively), an EM section of 20 hydrogen atoms and a weakly couple $H_2$ molecule. The interatomic distances are set as in Fig. 3 of the main text. (b) Same as (a) calculated with the 180 hydrogen atom lead models at the PBE/STO-3G level, and a 20 atom *EM* section described at the PBE/STO-3G (green diamonds), PBE/6-31G** (purple stars), and PBE/cc-pVTZ (orange crosses) levels.

# 6. Driving rate switching function

To increase numerical stability of our graphene nanoribbon junction transport calculations, the driving rate, $\Gamma$, was gradually increased from zero to its full value, $\Gamma_f$.[12] To that end, we chose a hyperbolic tangent switching function of the form:

$$\Gamma(t) = \frac{1}{2}\left(\tanh\left(\frac{t-t_0}{w}\right) + 1\right) \cdot \Gamma_f, \tag{S98}$$

where $t$ is the time, $t_0 = 0.05$ fs, and $w = 0.36$ fs.

Figure S7 compares the time-dependent current calculated at $V = 0.35$ V for this junction in the cases where $\Gamma$ is switched on abruptly (orange line) and gradually (blue line). When the driving rate $\hbar\Gamma$ is abruptly switched on to $\Gamma_f = 1.09$ eV, the current shows larger oscillations with respect to the gradual switch-on case. Since this initial transient dynamics is unphysical in our simulations starting from a somewhat arbitrary initial density matrix, it is numerically advantageous to switch on the driving rate gradually.

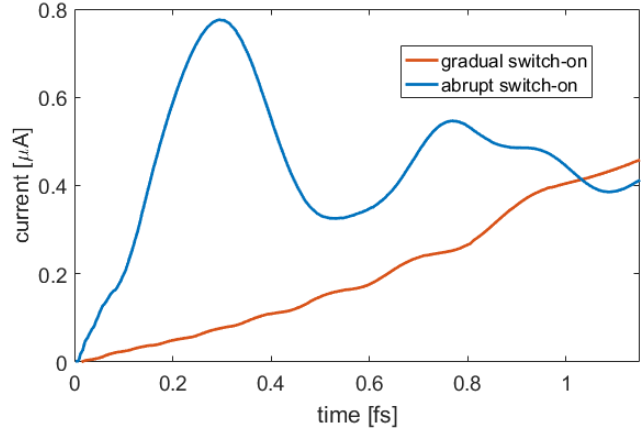

Figure S7: Comparison of the effect of gradual (orange) versus abrupt (blue) switching-on of the driving rate on the current dynamics through the graphene nano-ribbon junction model shown in Fig. 1b of the main text, under a bias voltage of $V = 0.35$ V.

## 7. Integrated current density along the extended molecule section

As a further consistency check of our current calculations, we compare in Figure S8 the steady-state current of the hydrogen chain model (Fig. 3 of the main text) evaluated from the partial trace of the partitioned single-particle density matrix (Eq. 13 of the main text) to that obtained by cross-sectional spatial integration of the current density (Eq. 18 of the main text). In the figure we plot the latter evaluated at several axial positions along the chain. As expected, the steady-state current is spatially uniform, increasing/decreasing only near the edges of the EM section, where the source/sink terms are applied. Furthermore, the integrated current matches well the one obtained via Eq. 13 of the main text (dashed orange line), further validating our methodology.

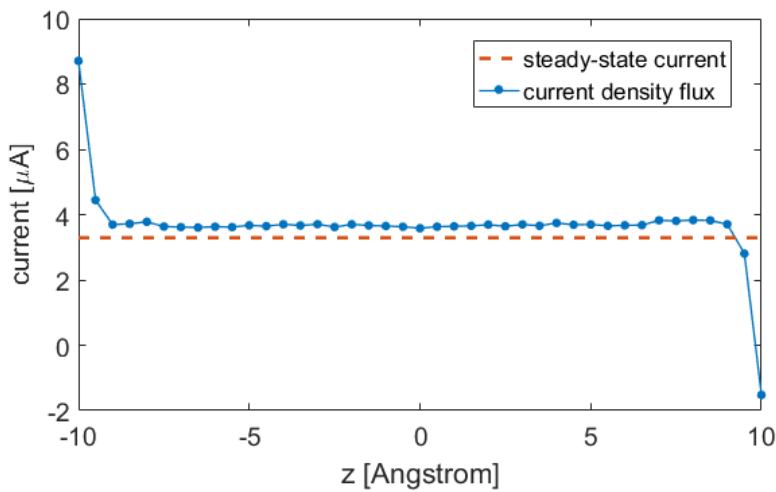

Figure S8: The integrated current density (blue circles) across the perpendicular plane to the main axis of a hydrogen chain (same as that studied in Fig. 3 of the main text) calculated via Eq. 18 of the main text, at different positions along the EM section, under a bias voltage of $V = 0.3$ V, and with a driving rate of $\hbar\Gamma = 0.61$ eV. The steady-state current evaluated directly from Eq. 13 of the main text is shown by the horizontal dashed orange line.

## 8. Total number of electrons for the case of fractional occupations

In Eq. (S60) of SI section 4 above, we used the expression $N = Tr(\mathcal{P} \cdot \mathcal{S})$ for the particle number. Here, we demonstrate that this expression is valid also for the case of non-idempotent density matrices, representing fractionally occupied states, of the following form:

$$\mathcal{P}_{\mu\nu} = \sum_i^M f_i\, C_{\mu i} C_{\nu i}^*. \tag{S99}$$

Here, $M$ is the total number of basis functions, and $C_{\mu i}$ are the expansion coefficients of the orthonormal Kohn-Sham orbitals, $\psi_i(\mathbf{r})$, within the non-orthogonal atom-centered orbital basis $\{\phi_\mu\}$:

$$\psi_i(\mathbf{r}) = \sum_\mu^M C_{\mu i} \phi_\mu(\mathbf{r}), \tag{S100}$$

and $0 \leq f_i \leq 1$ are the occupation numbers. For brevity of the presentation, spin indices have been omitted herein.

The density at a point $\mathbf{r}$ in space can be written as:

$$n(\mathbf{r}) = \sum_i^M f_i |\psi_i(\mathbf{r})|^2 = \sum_i^M \sum_{\mu\nu}^M f_i C_{\mu i} \phi_\mu(\mathbf{r}) C_{\nu i}^* \phi_\nu^*(\mathbf{r}). \tag{S101}$$

Spatially integrating the density $n(\mathbf{r})$, we obtain the total number of electrons:

$$N = \sum_{\mu\nu}^M \mathcal{S}_{\mu\nu} \left( \sum_i^M f_i\, C_{\mu i} C_{\nu i}^* \right), \tag{S102}$$

where the overlap matrix elements in the atom-centered orbital basis are given by $\mathcal{S}_{\mu\nu} = \int \phi_\mu(\mathbf{r})\, \phi_\nu^*(\mathbf{r}) d^3 r$. Using Eq. (S99) equation (S102), we finally obtain:

$$N = \sum_{\mu\nu}^M \mathcal{P}_{\mu\nu} \mathcal{S}_{\mu\nu} = Tr(\mathcal{P} \cdot \mathcal{S}), \tag{S103}$$

which has the same form as the standard expression for the total number of electrons of a closed ground state system with integer state occupations.

# 9. Cartesian atomic coordinates of the GNR/Benzene/GNR junction

All atomic coordinates are given in Å.

```
C -6.57401 14.9863 0.01103
C -6.14147 19.2275 -0.00037
C -5.70864 23.4702 -0.01177
C -5.60033 12.7631 0.01674
C -5.16742 17.0046 0.00534
C -4.73516 21.2455 -0.00605
C -4.15623 14.7624 0.01106
C -3.72362 19.003 -0.00034
C -3.29063 23.2418 -0.01173
C -1.70817 14.5122 0.01109
C -1.2756 18.7514 -0.00031
C -0.842592 22.9913 -0.0117
C 0.739796 14.2625 0.01112
C 1.1721 18.5017 -0.00028
C 1.60409 22.7417 -0.01167
C 3.18798 14.0133 0.01115
C 3.62049 18.2538 -0.00024
C 4.05229 22.4927 -0.01164
C -4.30118 13.346 0.01486
C -3.86808 17.5865 0.00346
C -3.43523 21.8259 -0.00794
C -1.85247 13.0991 0.01489
C -1.41986 17.3382 0.00349
C -0.987224 21.5782 -0.0079
C -0.556851 25.8226 -0.0193
C 0.595791 12.8494 0.01492
C 1.02805 17.0885 0.00352
C 1.46043 21.3285 -0.00787
C 1.89592 25.5724 -0.01927
C 3.04395 12.5968 0.01495
C 3.47598 16.8374 0.00356
C 3.90802 21.0768 -0.00784
C 5.46128 12.3714 0.01498
C 5.89377 16.6127 0.00359
C 6.32622 20.8551 -0.00781
C -5.31148 15.5944 0.00914
C -4.87881 19.8352 -0.00225
C -4.44695 24.0767 -0.01365
C -2.86004 15.3433 0.00917
C -2.42714 19.583 -0.00222
C -1.99228 23.8248 -0.01362
C -0.412129 15.0937 0.00921
C 0.0203384 19.3333 -0.00219
C 0.452809 23.5729 -0.01359
C 2.03577 14.8439 0.00924
C 2.46783 19.0837 -0.00216
C 2.89841 23.3259 -0.01355
C 4.48742 14.5948 0.00927
C 4.91994 18.8357 -0.00213
C 5.35338 23.077 -0.01352
C -3.14813 12.517 0.01677
C -2.71568 16.7563 0.00538
C -2.28284 20.9959 -0.00602
C -1.85023 25.2414 -0.01742
C -0.700356 12.2682 0.0168
C -0.26797 16.5069 0.00541
C 0.164502 20.7465 -0.00599
C 0.596896 24.9854 -0.01738
C 1.74724 12.0177 0.01683
C 2.17968 16.2569 0.00544
C 2.61178 20.4966 -0.00596
C 3.04531 24.742 -0.01735
C 4.19861 11.7635 0.01687
C 4.63106 16.0051 0.00547
C 5.06401 20.246 -0.00593
```

```
H -7.46139 15.6201 0.00950414
H -7.02877 19.8614 -0.00190323
H -6.59624 24.1036 -0.0133055
H 7.21389 20.2217 -0.00627613
H 6.78102 15.9787 0.00512481
H 6.34865 11.7376 0.0165072
C 6.46661 22.2282 -0.01161
H 7.46383 22.6694 -0.0131194
C 6.03388 17.9856 -0.00021
H 7.0309 18.4272 -0.00171507
C 5.60115 13.7443 0.01118
H 6.59823 14.1859 0.00966866
C -6.71425 13.6134 0.01482
H -7.71132 13.1718 0.0163203
C -6.28143 17.8547 0.00343
H -7.27842 17.4129 0.00493961
C -5.84839 22.097 -0.00797
H -6.84567 21.6558 -0.00646443
C -4.26825 25.4651 -0.01745
H -5.14886 26.1083 -0.0189645
C -3.00883 26.03 -0.01933
H -2.90404 27.1155 -0.0223383
C -0.380245 27.2142 -0.0231
H -1.26132 27.8567 -0.0245901
C 0.881558 27.776 -0.02498
H 0.992192 28.8606 -0.0279657
C 2.00395 26.971 -0.02307
H 2.99659 27.4223 -0.0245354
C 4.33929 25.2805 -0.01924
H 4.45582 26.3647 -0.0222592
C 5.4587 24.4729 -0.01732
H 6.45103 24.925 -0.0188068
C -7.0056 10.7417 0.02242
C -6.03095 8.51998 0.02813
C -4.58765 10.5222 0.02245
C -2.56748 6.02954 0.03388
C -2.14046 10.2737 0.02248
C -0.131647 5.77861 0.03391
C 0.30741 10.0241 0.02252
C 2.75429 9.77322 0.02255
C -4.73071 9.10588 0.02625
C -2.28412 8.86111 0.02628
C 0.162977 8.61167 0.02631
C 2.60836 8.3572 0.02635
C 5.02563 8.12672 0.02638
C -5.74413 11.3525 0.02054
C -3.72005 6.85606 0.03196
C -3.2923 11.1043 0.02057
C -1.27592 6.62218 0.032
C -0.844501 10.8552 0.0206
C 1.16416 6.35785 0.03203
C 1.60322 10.605 0.02063
C 4.05461 10.353 0.02066
C -3.57648 8.27312 0.02817
C -1.13226 8.03245 0.0282
C 1.30973 7.77464 0.02823
C 3.76354 7.52085 0.02826
H -7.89405 11.374 0.0208867
H 5.9128 7.49266 0.0279143
C 5.16675 9.49997 0.02258
H 6.16459 9.93986 0.0210758
C -7.1447 9.36822 0.02622
H -8.14164 8.92637 0.0277289
C -5.01368 6.31669 0.03385
H -5.12959 5.23239 0.0368687
C 2.32235 5.56838 0.03394
H 2.21691 4.48301 0.036948
C -0.30562 4.38327 0.03771
H 0.596048 3.79425 -0.0186168
C -2.65672 4.63222 0.03768
H -3.65047 4.18197 0.0342575
C -1.54922 3.78574 0.03959
C -6.13382 7.1237 0.03193
```

```
H -7.12555 6.67027 0.0334184
C 3.58237 6.13257 0.03206
H 4.46212 5.48823 0.0335753
C 1.82579 -2.28092 0.03103
H 2.46633 -2.07473 -0.832772
H 2.45568 -2.13974 0.91549
C 0.769673 -1.16669 0.06603
C 1.29883 0.123429 0.11955
C -0.606622 -1.25715 0.09045
C 0.517671 1.26655 0.19779
H 2.38391 0.235449 0.107119
C -1.38988 -0.114994 0.16918
H -1.12839 -2.19988 0.0526206
C -0.860722 1.17593 0.22319
H 1.03802 2.20865 0.26908
H -2.475 -0.227062 0.164841
C -1.89189 2.29088 -0.01959
H -2.32171 2.10685 -1.00957
H -2.70242 2.12653 0.697825
C -5.06377 -8.15663 -0.01349
C -4.0803 -10.3774 -0.00778
C -2.64526 -8.37354 -0.01346
C -0.198544 -8.61433 -0.01343
C 2.24993 -8.84999 -0.01339
C 4.69782 -9.08088 -0.01336
C -2.78325 -9.79037 -0.00966
C -0.335018 -10.0276 -0.00963
C 0.0803285 -5.7796 -0.02102
C 2.11422 -10.2634 -0.0096
C 2.51714 -6.01686 -0.02099
C 4.56277 -10.498 -0.00956
C 6.98191 -10.7038 -0.00953
C -3.80511 -7.54368 -0.01537
C -1.34992 -7.78369 -0.01534
C 1.0934 -8.02788 -0.01531
C 3.53891 -8.25464 -0.01528
C 5.99472 -8.48763 -0.01525
C -1.62754 -10.6157 -0.00774
C -1.21225 -6.3661 -0.01914
C 0.821559 -10.8522 -0.00771
C 1.22911 -6.61687 -0.01911
C 3.27073 -11.0874 -0.00768
C 3.67442 -6.83677 -0.01908
C 5.72391 -11.3218 -0.00765
H -5.95449 -7.52756 -0.0150227
H 7.87392 -11.3311 -0.0079951
C 7.11324 -9.32957 -0.01333
H 8.10767 -8.8821 -0.0148335
C -5.19719 -9.53066 -0.00969
H -6.19255 -9.97612 -0.00818048
C -2.37486 -5.58315 -0.02106
H -2.27551 -4.4972 -0.024078
C 4.96498 -6.29011 -0.02096
H 5.07477 -5.20517 -0.0239718
C 0.246236 -4.38327 -0.02482
H -0.660376 -3.80045 -0.0362669
C 2.59803 -4.61909 -0.02479
H 3.58923 -4.16323 -0.0214967
C 1.48593 -3.77793 -0.02671
C 6.08966 -7.09079 -0.01904
H 7.07882 -6.63175 -0.0205214
C -3.63169 -6.1544 -0.01917
H -4.51504 -5.515 -0.0206822
C -1.83595 -25.5845 0.03216
C 0.618191 -25.8209 0.03219
C -2.99 -24.7606 0.03024
C -0.540251 -24.9902 0.03027
C 1.90828 -25.2324 0.0303
C -6.29274 -20.8922 0.0207
C -5.88419 -16.6474 0.0093
C -5.47559 -12.4037 -0.00209
C -5.30741 -23.1086 0.02641
C -4.89786 -18.8649 0.01501
```

```
C -4.48923 -14.6217 0.00362
C -3.87334 -21.1003 0.02073
C -3.46518 -16.8585 0.00933
C -3.05704 -12.6155 -0.00206
C -1.42436 -21.3382 0.02076
C -1.01587 -17.0959 0.00936
C -0.607488 -12.8544 -0.00203
C 1.02466 -21.5741 0.02079
C 1.43342 -17.3318 0.0094
C 1.84215 -13.0903 -0.002
C 3.47402 -21.808 0.02082
C 3.883 -17.5663 0.00943
C 4.29223 -13.3234 -0.00197
C -4.00963 -22.517 0.02453
C -3.6017 -18.2757 0.01313
C -3.19308 -14.0328 0.00174
C -1.56007 -22.7522 0.02456
C -1.15196 -18.5098 0.01316
C -0.743531 -14.2682 0.00177
C 0.887987 -22.988 0.02459
C 1.29712 -18.7457 0.01319
C 1.70581 -14.5042 0.0018
C 3.3374 -23.2247 0.02462
C 3.74652 -18.9835 0.01323
C 4.15525 -14.7406 0.00183
C 5.75665 -23.4395 0.02465
C 6.16559 -19.1944 0.01326
C 6.57425 -14.9508 0.00186
C -5.03398 -20.276 0.01881
C -4.62493 -16.0327 0.00742
C -4.21636 -11.7887 -0.00398
C -2.58038 -20.5128 0.01884
C -2.17216 -16.2708 0.00745
C -1.7636 -12.0291 -0.00395
C -0.13173 -20.7489 0.01888
C 0.276862 -16.507 0.00748
C 0.685373 -12.266 -0.00391
C 2.31698 -20.9845 0.01891
C 2.72595 -16.7426 0.00751
C 3.13451 -12.5009 -0.00388
C 4.77067 -21.2204 0.01894
C 5.17905 -16.977 0.00754
C 5.58807 -12.7331 -0.00385
C -2.85107 -23.3437 0.02644
C -2.44439 -19.0991 0.01505
C -2.03621 -14.857 0.00365
C -0.404119 -23.5769 0.02647
C 0.00447323 -19.3349 0.01508
C 0.413063 -15.093 0.00368
C 2.04235 -23.815 0.0265
C 2.45332 -19.5709 0.01511
C 2.86236 -15.3288 0.00371
C 4.4984 -24.0531 0.02654
C 4.90637 -19.8093 0.01514
C 5.31517 -15.566 0.00375
H -7.18397 -20.2638 0.0191737
H -6.775 -16.0185 0.0077667
H -6.36651 -11.7749 -0.00361496
H 7.46519 -15.5796 0.00338807
H 7.05645 -19.8233 0.0147948
H 6.6478 -24.0679 0.0261787
C 6.70676 -13.5771 -0.00194
H 7.70132 -13.13 -0.00345047
C 6.29783 -17.8208 0.00946
H 7.29231 -17.3735 0.00795574
C 5.88867 -22.0655 0.02085
H 6.88345 -21.6188 0.0193383
C -6.4254 -22.2661 0.02449
H -7.42012 -22.7129 0.0259908
C -6.01657 -18.021 0.0131
H -7.01109 -18.4683 0.0146095
C -5.60773 -13.7774 0.0017
H -6.60231 -14.2246 0.0032012
```

```
C -5.40487 -24.5051 0.03021
H -6.39464 -24.9627 0.0316989
C -4.28092 -25.3063 0.03212
H -4.39134 -26.3912 0.0351291
C -1.9361 -26.9837 0.03595
H -2.92618 -27.4406 0.0374054
C -0.809196 -27.7823 0.03787
H -0.913722 -28.8675 0.0408587
C 0.449424 -27.2135 0.03599
H 1.3341 -27.8509 0.0374806
C 3.0713 -26.0145 0.03222
H 2.97262 -27.1005 0.0352344
C 4.32752 -25.4425 0.03034
H 5.21174 -26.0807 0.0318553
```